\documentclass[%
 prb,
 amsmath,amssymb,
 aps,
twocolumn]{revtex4-2}

\usepackage[latin1]{inputenc}
\usepackage{mathtools}
\usepackage{slashed}
\usepackage{physics}
\usepackage{array}
\usepackage{relsize}
\usepackage[compat=1.1.0]{tikz-feynman}
\usepackage{makecell}

\usepackage{tabu}
\usepackage{graphicx}
\usepackage[caption=false]{subfig}
\usepackage{epsfig}

\usepackage{hyperref}
\hypersetup{
	linkcolor=blue,
	citecolor=blue,
	filecolor=blue,
	urlcolor=blue,
	colorlinks=true
}

\usepackage{comment}

\usepackage{tabularx}

\usepackage{euscript}
\usepackage{amsmath}
\usepackage{amsfonts}
\usepackage{amssymb}
\usepackage{mathrsfs}
\usepackage[scr=boondox]{mathalfa}
\usepackage{exscale}
\usepackage{amsbsy}
\usepackage{textcomp}
\usepackage{bm}
\usepackage[normalem]{ulem}

\usepackage{babel}
\usepackage[autostyle, english = american]{csquotes}
\MakeOuterQuote{"}

\usepackage{tikz}
\usepackage{soul}

\newcommand{\beq}{\begin{eqnarray}}
\newcommand{\eeq}{\end{eqnarray}}

\begin{document}

\def\ppnumber{\vbox{\baselineskip14pt
}}

\def\ppdate{
} \date{\today}

\title{\bf Topological Field Theories of Three-Dimensional Rotation Symmetric Insulators: Coupling Curvature and Electromagnetism}
\author{Julian May-Mann}
\email{Maymann2@illinois.edu }
\author{Mark R. Hirsbrunner}
\author{Xuchen Cao}
\author{Taylor L. Hughes}
\affiliation{ \it Department of Physics and Institute for Condensed Matter Theory,\\  \it University of Illinois at Urbana-Champaign, \\  \it 1110 West Green Street, Urbana, Illinois 61801-3080, USA}

\begin{abstract}
Quantized responses are important tools for understanding and characterizing the universal features of topological phases of matter. In this work, we consider a class of topological crystalline insulators in $3$D with $C_n$ lattice rotation symmetry along a fixed axis, in addition to either mirror symmetry or particle-hole symmetry. These insulators can realize  quantized mixed geometry-charge responses.  When the surface of these insulators is gapped, disclinations on the surface carry a fractional charge that is half the minimal amount that can occur in purely $2$D systems. Similarly, disclination lines in the bulk carry a fractionally quantized electric polarization. These effects, and other related phenomena, are captured by a $3$D topological response term that couples the lattice curvature to the electromagnetic field strength. Additionally,  mirror symmetric insulators with this response can be smoothly deformed into a higher-order octopole insulator with quantized corner charges.  We also construct a symmetry indicator form for the topological invariant that describes the quantized response of the mirror symmetric topological crystalline insulators, and discuss an unusual response quantization in time-reversal breaking systems.
\end{abstract}

\maketitle

\bigskip
\newpage

\section{Introduction}

In modern condensed matter physics, it has been well established that for a given symmetry class, there can be multiple insulating phases of matter that are topologically distinct from one another~\cite{schnyder2008classification,kitaev2009periodic}. These topologically inequivalent insulators are denoted as symmetry protected topological phases (SPTs)~\cite{gu2009tensor,hasan2010colloquium, pollmann2012symmetry,chen2013symmetry,bernevig2013topological,sato2017topological} and they have been a central area of condensed matter research for more than a decade. Concretely, SPTs are defined as symmetric insulators that cannot be smoothly deformed into a trivial insulator without either breaking the symmetry or closing the energy gap. SPTs also display a bulk-boundary correspondence, where the topologically non-trivial bulk is accompanied by gapless degrees of freedom on symmetry preserving surfaces~\cite{hatsugai1993chern,prodan2016bulk}.

One reason that topological phases of matter have attracted so much attention is that they can exhibit quantized responses in the presence of external gauge fields. These responses arise from the underlying topology of the SPT and are robust to symmetry preserving disorder and perturbations. In experimental contexts, these quantized responses serve as smoking-gun characteristics of topological insulators~\cite{bernevig2006quantum,fu2007topologicalIn,konig2007quantum,hsieh2008topological,xia2009observation}. The first observed, and most famous, topological response is the quantized Hall-conductance of $2$D (two spatial dimensions) insulators~\cite{klitzing1980new,thouless1982quantized,girvin1987quantum,haldane1988model}. Similarly, $1$D insulators with particle hole symmetry (PHS) have a quantized polarization~\cite{su1979solitons,su1983erratum,king1993theory,ortiz1994macroscopic}, and  $3$D insulators with time-reversal symmetry (TRS) display quantized axion electrodynamics~\cite{fu2007topological,qi2008topological,qi2013axion,li2010dynamical}. All of these effects have a topological field theory description that captures the quantized responses: the integer Hall conductance corresponds to a Chern-Simons term~\cite{girvin1987quantum}, electric polarization corresponds to a $1+1$-d $\Theta$-term/Goldstone-Wilczek response term~\cite{goldstone1981fractional}, and axion electrodynamics corresponds to a $3+1$-d $\Theta$-term~\cite{qi2008topological}. 

More recently, the topological responses of topological crystalline insulators (TCIs)--SPTs that are protected by crystalline symmetries--have also gained attention~\cite{fu2011topological,song2017topological,cornfeld2019classification,  huang2022effective}. Notably, it has been shown that certain TCIs can host mixed geometry-charge responses, where charge fluctuations are driven by lattice effects, e.g., shears, strains, or defects~\cite{chaikin1995principles, ran2009one, teo2010topological, jurivcic2012universal, barkeshli2012topological, asahi2012topological, chung2016dislocation, teo2017topological,ramamurthy2017electromagnetic, roy2021dislocation, gioia2021unquantized, teo2013existence,gopalakrishnan2013disclination, benalcazar2014classification,li2020fractional}. A well known example of such a mixed geometry-charge response occurs in $2$D, where charge is bound to disclination defects in TCIs with $C_n$ lattice rotation symmetry~\cite{ruegg2013bound, ruegg2013corner, liu2019shift,li2020fractional,may2021crystalline,zhang2022fractional}. This effect also has an associated field theory description, i.e.,  the Wen-Zee term. This term couples the electromagnetic gauge field to the spin-connection gauge field, the latter of which represents the geometric distortions~\cite{wen1992shift,han2019generalized,manjunath2021crystalline, manjunath2020classification}. 

In this work, we consider the mixed geometry-charge responses of $3$D systems with $C_n$ lattice rotation symmetry around a fixed axis. We show that such a system can display a novel mixed geometric-charge response, where line-like disclination defects have an electric polarization. This response is described by a topological field theory term that directly couples the lattice curvature $(R)$ to electromagnetic field strength $(F)$. We denote this topological response term as the $RF$-term. The coefficient of the $RF$-term is quantized for systems with either particle-hole symmetry (PHS) or mirror symmetry along the z-direction. This quantized response defines a new class of rotation-invariant topological crystalline insulators (rTCIs).

In the main body of this article, we provide microscopic lattice models that realize three important classes of rTCIs. First, a spinless rTCI with PHS and additional TRS as an illustrative example. Second, a spin-1/2 rTCI with mirror symmetry and additional TRS, which is closer to a more realistic lattice model. Third, a spinless rTCI with mirror symmetry and broken TRS. We relegate similar analysis of other rTCIs (including the spin-1/2 rTCI with TRS and PHS and the spinless rTCI with TRS and mirror symmetry) to the appendix. 

We analyze these rTCIs using the $RF$-field theory description, microscopic lattice models, continuum theories and numerics, and show that they display a number of remarkable topological features. For example, when the surface of a rTCI with a non-trivial $RF$-term is gapped (without electron-electron interactions), the resulting gapped surface theory contains a Wen-Zee response term. This indicates that disclinations on the rTCI surface bind charge (equivalently, intersections of bulk disclination lines and the rTCI surface bind charge). The coefficient of the surface Wen-Zee term is half the value that is allowed for a $2$D system with the same symmetries (i.e., surface disclinations of the rTCI bind half the minimal amount of charge that can be bound to disclinations of a $2$D system with the same symmetries). 

Let us also briefly mention some key features of the bulk-boundary correspondence for rTCIs with various symmetries. For non-interacting rTCIs with PHS and TRS, the surfaces host an even number of Dirac fermions, which cannot be gapped without breaking PHS. However, if interactions are included, the rTCI with PHS can also host a symmetric gapped surface with symmetry-enriched topological order. This symmetry-enriched surface topological order is anomalous and cannot be realized in a purely $2$D system with PHS. For rTCIs with mirror symmetry and TRS, the surfaces also contain an even number of Dirac fermions. However, in this case, the surface Dirac fermions can be gapped without breaking symmetries or adding additional interactions. Such symmetrically gapped surfaces host quantized corner charges, indicating that rTCI with mirror symmetry constitutes a third-order topological insulator with an octopole configuration of charges~\cite{benalcazar2017electric,benalcazar2017quantized}. For the spinless rTCI with mirror symmetry that break TRS, the surface hosts a single Dirac fermion. This two-dimensional Dirac fermions can be fully gapped up to a single one-dimensional chiral mode that circulates in the mirror invariant plane. Additionally, following the approach of Ref. \onlinecite{song2017topological}, we show that a rTCI with PHS or mirror symmetry can be dimensionally reduced to a $1$D insulator that is in the same topological class as the Su-Schrieffer-Heeger (SSH) chain~\cite{su1979solitons} with PHS or mirror symmetry, respectively.

The remainder of this paper is organized as follows. In Sec.~\ref{sec:ResponseTh} we present the $RF$-response term, discuss its physical properties, and show that it defines a  class of rTCIs.  In Sec.~\ref{sec:PHSrTCI} we analyze a lattice model for a spinless rTCI with TRS and PHS. We show that the effective response theory of this rTCI contains a quantized $RF$-term. In Sec.~\ref{sec:MirrorrTCI} we present a lattice model for a spin-1/2 rTCI with TRS and mirror symmetry, and similarly show that its effective response theory contains a quantized $RF$-term. In Sec.~\ref{sec:GenbRespose} we present a symmetry indicator form of the topological invariant for rTCIs that have mirror and inversion symmetry and relate this invariant to the coefficient of the $RF$-term. In Sec.~\ref{sec:RFwoTRS} we consider the $RF$ response of insulators with broken TRS. We conclude in Sec. ~\ref{sec:Con} and discuss possible extensions. We also provide several appendices that contain technical details and analyses of related models.

\section{Response Theory}\label{sec:ResponseTh}
In this section, we consider the effective field theory description of a $3$D fermionic insulator with U$(1)$ charge conservation and $C_n$ lattice rotation symmetry along a fixed axis--which we take to be the z-axis. Our main interest is in the following mixed geometry-charge response term:
\begin{equation}\begin{split}
&\mathcal{L}_{RF} = \frac{\Phi}{4\pi^2} \epsilon^{\mu\nu\rho\kappa} \partial_\mu \omega_\nu \partial_\rho A_\kappa,
\label{eq:BulkResponses}\end{split}\end{equation}
where $A_\mu$ is the electromagnetic gauge field and $\omega_\mu$ is the $C_n$ symmetry gauge field, both of which should be regarded as background probe fields. Physically, fluxes of $\omega_\mu$ correspond to lattice disclinations with Frank-vector parallel to the z-axis~\cite{kleman2008disclinations}. Here and throughout we set the electron charge $e = 1$ as well as $\hbar = 1$. 

 As we show in the following subsections, the response term in Eq.~\ref{eq:BulkResponses} describes a coupling between the lattice curvature $(R)$ and the electromagnetic field strength $(F)$, leading us to refer to it as the ``$RF$-term''. The $RF$-term is a total derivative, but nevertheless leads to a number of non-trivial responses. Furthermore, we show that the coefficient $\Phi$ is quantized for insulators with particle hole symmetry (PHS) or mirror symmetry along the $z$-direction (which we simply refer to as ``mirror symmetry'' unless otherwise noted). 


\subsection{Lattice Geometry in the Continuum Limit}\label{sec:LatticeGeoCont}
The $RF$-term in Eq.~\ref{eq:BulkResponses} is defined in continuous space-time. Because of this, it is worthwhile to discuss how lattice effects, which are inherently discrete, can be described in the continuum limit. Here, for simplicity, we consider the case of a cubic lattice, although this analysis applies to general lattices. 

To begin, take a cubic lattice embedded in a $3$D spatial manifold of a $3+1$D spacetime. In the continuum limit the lattice constant is taken to zero and the lattice points become a dense set of points on the manifold. The spatial metric of the manifold, $g_{ij}$ ($i =x,y,z$), should be consistent with the underlying lattice in the continuum limit. To this end, we introduce the frame-fields (AKA vielbeins or tetrads)~\cite{de1992relativity} $e^A_i$ $(A = x,y,z)$ such that $g_{ij} = e^A_i e^{B}_j \delta_{AB}$, where $\delta_{AB}$ is the Kronecker delta, and the inverse frame-fields $E^i_A$, satisfying $E^i_A e^A_j  =\delta^i_{j}$. The frame-fields and inverses are not constants, and in general can be functions of space and time. In order for the metric of the manifold to be consistent with the lattice, the inverse frame-fields $E^i_A$ should be identified with the primitive lattice vector in the $A$-direction, in units of the lattice constant~\cite{dzyaloshinskii1980poisson}\footnote{Compared to the definition of the frame-fields in Ref. \onlinecite{dzyaloshinskii1980poisson}, we have included a factor of the lattice constant in our definition of the frame-fields so they are dimensionless.}. For a perfect lattice that is free of defects, we can take  $E^i_A = \delta^i_A$. In principle, we can also introduce a fourth time-like frame-field (and corresponding inverse frame-field), but since the temporal direction is fixed for lattice systems this will not be necessary here. 

In this work, we are primarily interested in the continuum interpretation of lattice disclinations with Frank-vector parallel to the z-axis. In $3$D these defects are line-like fluxes of the $C_n$ lattice rotation symmetry around the z-axis ($n =4$ for the cubic lattice). With this in mind, consider a lattice where the only defects are disclinations with Frank-vector parallel to the z-axis (which we shall refer to simply as disclinations from now on). Since the disclinations only rotate the lattice vectors that span the xy-plane, the inverse frame-fields for a generic lattice with disclinations can be defined as
\begin{equation}\begin{split}
E^i_z &= \delta^i_z\\
E^i_x &= \cos(\varphi)\delta^i_x + \sin(\varphi)\delta^i_y,\\
E^i_y &= \cos(\varphi)\delta^i_x - \sin(\varphi)\delta^i_x,
\label{eq:FramefieldDef}\end{split}\end{equation}
for some spatially varying angle $\varphi$. For these inverse frame-fields, the metric $g_{ij}$ is flat everywhere, which is a consequence of us only considering rotation symmetry fluxes. It should be noted that if non-trivial disclinations are present, a global definition of $E_x$ and $E_y$ is not possible, and it is necessary to work in coordinate patches where $E_x$ and $E_y$ can be consistently defined.

For the inverse frame-fields defined in Eq.~\ref{eq:FramefieldDef} the only non-vanishing-components of the spin connection~\cite{de1992relativity} are: 
\begin{equation}\begin{split}
\omega_\mu \equiv \omega^{x}_{y\mu} = - \omega^{y}_{x\mu} =e^x_{i} \partial_\mu E^i_y = -e^y_{i} \partial_\mu E^i_x,
\label{eq:SpinConnectionDef}\end{split}\end{equation}
For brevity, we refer to $\omega_\mu$ simply as the spin connection. Physically, $\omega_\mu$ measures how much the lattice vectors that span the xy-plane rotate as we move along the $\mu$-direction. The spin connection has a $C_n$ gauge ambiguity, which corresponds to a local redefinition of the primitive lattice vectors that span the xy-plane. Under this $C_n$ gauge symmetry, the frame-fields $E_x$ and $E_y$ transform as,
\begin{equation}\begin{split}
E^i_x &\rightarrow \cos(\theta) E^i_x + \sin(\theta)E^i_y, \\
E^i_y &\rightarrow\cos(\theta) E^i_y - \sin(\theta)E^i_x,
\label{eq:FFGauge}\end{split}\end{equation}
where $\theta$ is a function that takes on values in $\{0, \frac{2\pi}{n}, \frac{4\pi}{n}, .... \frac{(n-1)2\pi}{n}\}$. It is straightforward to show that $C_n$ gauge transformations do not change the metric. Using Eq.~\ref{eq:SpinConnectionDef}, the $C_n$ gauge transformation acts on the spin connection as
\begin{equation}\begin{split}
\omega_\mu \rightarrow \omega_\mu - \partial_\mu \theta.
\label{eq:SpinConnectionGauge}
\end{split}\end{equation}
For the theories we consider in the following sections, this $C_n$ gauge symmetry is actually part of a larger SO$(2)$=U$(1)$ gauge symmetry that emerges within the continuum limit of the lattice model. The U$(1)$ gauge symmetry transforms the frame-fields and spin connections as in Eq.~\ref{eq:FFGauge} and~\ref{eq:SpinConnectionGauge} but with $\theta$ taking continuous values in $[0,2\pi)$. 

Based on the gauge transformation defined in Eq.~\ref{eq:SpinConnectionGauge}, we define the gauge invariant lattice curvature tensor $R_{\mu\nu} = \partial_\mu \omega_\nu - \partial_\nu \omega_\mu$.  This curvature is related to the full curvature tensor of the $3$D spacetime as $R_{\mu\nu} \equiv R_{\mu\nu y}^{x}$~\cite{de1992relativity}. The $RF$-term in Eq.~\ref{eq:BulkResponses} therefore describes a coupling between the effective lattice curvature $R_{\mu\nu}$ and the dual electromagnetic field strength $F^*_{\mu\nu} = \frac{1}{2}\epsilon^{\mu\nu\rho\kappa} F_{\rho\kappa}$. We note that, in principle, it is also possible to define torsion for the lattice system. However, for the frame-fields in Eq.~\ref{eq:FramefieldDef} the torsion vanishes. At the level of the lattice, the absence of torsion is the result of the assumption that the lattice only has disclinations and is free of dislocations~\cite{hughes2011torsional}. 

Disclinations of the underlying lattice correspond to fluxes of $\omega_\mu$ and are singular points of the curvature $R$. For example, consider a disclination line on the $z$-axis located at $x=y=0$ with Frank angle $\Theta_F$. Away from the disclination core, the lattice vectors that span the $xy$-plane are rotated by $\Theta_F$ upon encircling the disclination. Using our previous identification of the inverse frame-fields with the lattice vectors, we find that such a disclination corresponds to the inverse frame-fields defined in Eq.~\ref{eq:FramefieldDef}, where $\varphi$ winds by $\Theta_F$ on any loop that encircles $x=y=0$. To be explicit, we choose  $\varphi(x,y) = \frac{\Theta_F}{2\pi}\tan^{-1}(x/y)$. Then, using Eq.~\ref{eq:SpinConnectionDef} we find $\omega_\mu = \partial_\mu \varphi$, and $\oint \omega = -\Theta_F$, where the loop integral is defined on a loop that encircles the disclination line. This confirms that fluxes of the spin connection correspond to lattice disclinations. For lattice systems with $C_n$ symmetry, the Frank angles are necessarily multiples of $2\pi/n$, and the physical fluxes of $\omega_\mu$ are quantized in multiples of $2\pi/n$.

\subsection{Physical Implications of the Response Theory}\label{ssec:PhysicalImp}
To set the stage for a discussion of the physical implications of the $RF$-term in Eq.~\ref{eq:BulkResponses} it is useful to first discuss a related response term, the (axion) $\Theta$-term~\cite{huang1985structure,wilczek1987two,lee1987topological},
\begin{equation}\begin{split}
&\mathcal{L}_{\Theta} = \frac{\Theta}{8\pi^2} \epsilon^{\mu\nu\rho\kappa} \partial_\mu A_\nu \partial_\rho A_\kappa,
\label{eq:ThetaTerm}\end{split}\end{equation}
that describes time-reversal invariant fermionic topological insulators when $\Theta = \pi$~\cite{qi2008topological}, and bosonic topological insulators when $\Theta = 2\pi$~\cite{vishwanath2013physics, metlitski2013bosonic, bi2015classification, xu2015self}. We show below that many features of the $\Theta$-term have direct analogs in the $RF$-term. For a more detailed discussion of the $\Theta$-term in the context of fermionic and bosonic topological phases of matter, see Ref. \onlinecite{qi2008topological} and \onlinecite{vishwanath2013physics}.

The first feature of note is that non-vanishing values of $\Theta$ indicate that magnetic monopoles carry charge $-\Theta/2\pi$. This is known as the Witten effect~\cite{witten1979dyons}. Second, the $\Theta$-term imparts magnetic flux tubes with non-trivial braiding statistics, e.g., linking a pair of $2\pi$ ($\frac{h}{e}$) electromagnetic vortices produces a phase of $e^{i\Theta}$ relative to the unlinked configuration. Third, at domain walls where the value of $\Theta$ changes by $\Delta \Theta$, there is a $2$D Chern-Simons term of the form
\begin{equation}\begin{split}
\mathcal{L}_{\text{CS-DW}} = \frac{\Delta \Theta}{8\pi^2} \epsilon^{\mu\nu\rho} A_\mu \partial_\nu A_\rho. 
\label{eq:DWChernSimons}\end{split}\end{equation}

It is also important to note that $\Theta$ is periodic. For fermionic systems, the period of $\Theta$ is $2\pi$. This is easily demonstrated by considering a domain wall where $\Theta$ changes by $\Delta \Theta$. According to Eq.~\ref{eq:DWChernSimons} this leads to a $2$D domain wall Chern-Simons term with coefficient $\Delta\Theta/8\pi^2$. For a purely $2$D fermionic system without topological order, the Chern-Simons coefficient must be an integer multiple of $1/4\pi$~\cite{tong2016lectures}. So, when $\Delta \Theta$ is an integer multiple of $2\pi$, the domain wall physics can be trivialized by adding a purely $2$D system, indicating that the value of $\Theta$ in Eq.~\ref{eq:ThetaTerm} is only meaningfully defined modulo $2\pi$ in fermionic systems. The periodicity of $\Theta$ in bosonic systems can be found using the same logic: for $2$D bosonic systems without topological order, the Chern-Simons coefficient must be an integer multiple of $1/2\pi$~\cite{senthil2013integer}. Because of this, $\Theta$ is defined modulo $4\pi$ in bosonic systems. 

Having discussed the essential features of the $\Theta$-term in Eq.~\ref{eq:ThetaTerm}, we will now describe the analogous features of the $RF$-term in Eq.~\ref{eq:BulkResponses}. We assume that the $RF$-term describes an insulator and that the disclination defects we discuss do not close the bulk gap (these assumptions will be satisfied in the models we consider below). Since the $RF$-term couples the electromagnetic gauge field $A_\mu$ and the spin connection $\omega_\mu$, it gives rise to mixed geometric-charge effects. As is well known, the U$(1)$ charge conservation symmetry implies a conserved electromagnetic charge 4-current $j^\mu = \delta S /\delta A_\mu$, where $S$ is the minimally coupled action. Similarly, we can define an angular momentum 4-current $j^\mu_{AM} = \delta S /\delta \omega_\mu$, where we have implicitly embedded $C_n$ in U$(1)$. However, since we are considering $C_n$ symmetry, the angular momentum is only defined modulo $n$.

Using these currents we first note that, for a non-vanishing value of $\Phi$, there is a mixed Witten effect where magnetic monopoles carry angular momentum $-\Phi/2\pi$. We can also define a $2\pi/n$ ``disclination monopole'' where a $2\pi/n$ disclination line terminates in the bulk of the insulator. Such a disclination monopole will likely have a high energy cost in an actual crystalline solid, but they are still useful to consider as a theoretical tool. The mixed Witten effect indicates that a $2\pi/n$ disclination monopole carries electromagnetic charge $\Phi/2\pi n$. Since the disclination monopoles can be viewed as the ends of a $1$D disclination line, the surface charge theorem for electric polarization indicates that the $RF$-term binds polarization $\Phi/2\pi n$ to $2\pi/n$ disclination lines. This can also be seen from the fact that inserting a configuration of $\omega$ harboring a $2\pi/n$ disclination line into Eq.~\ref{eq:BulkResponses} generates a $1$D Goldstone-Wilczek response term~\cite{goldstone1981fractional,qi2008topological} with coefficient $\Phi/2\pi n$.

The $RF$-term also indicates that electromagnetic flux lines and disclination lines have non-trivial braiding statistics. Linking a $2\pi$ electromagnetic flux line with a $2\pi/n$ disclination line produces a phase of $e^{i \Phi/n}$ compared to the unlinked configuration. The $RF$-term does not affect the self-statistics of the flux or disclination lines.

\begin{figure}
    \centering
    \subfloat[]{\includegraphics[width=0.45\textwidth]{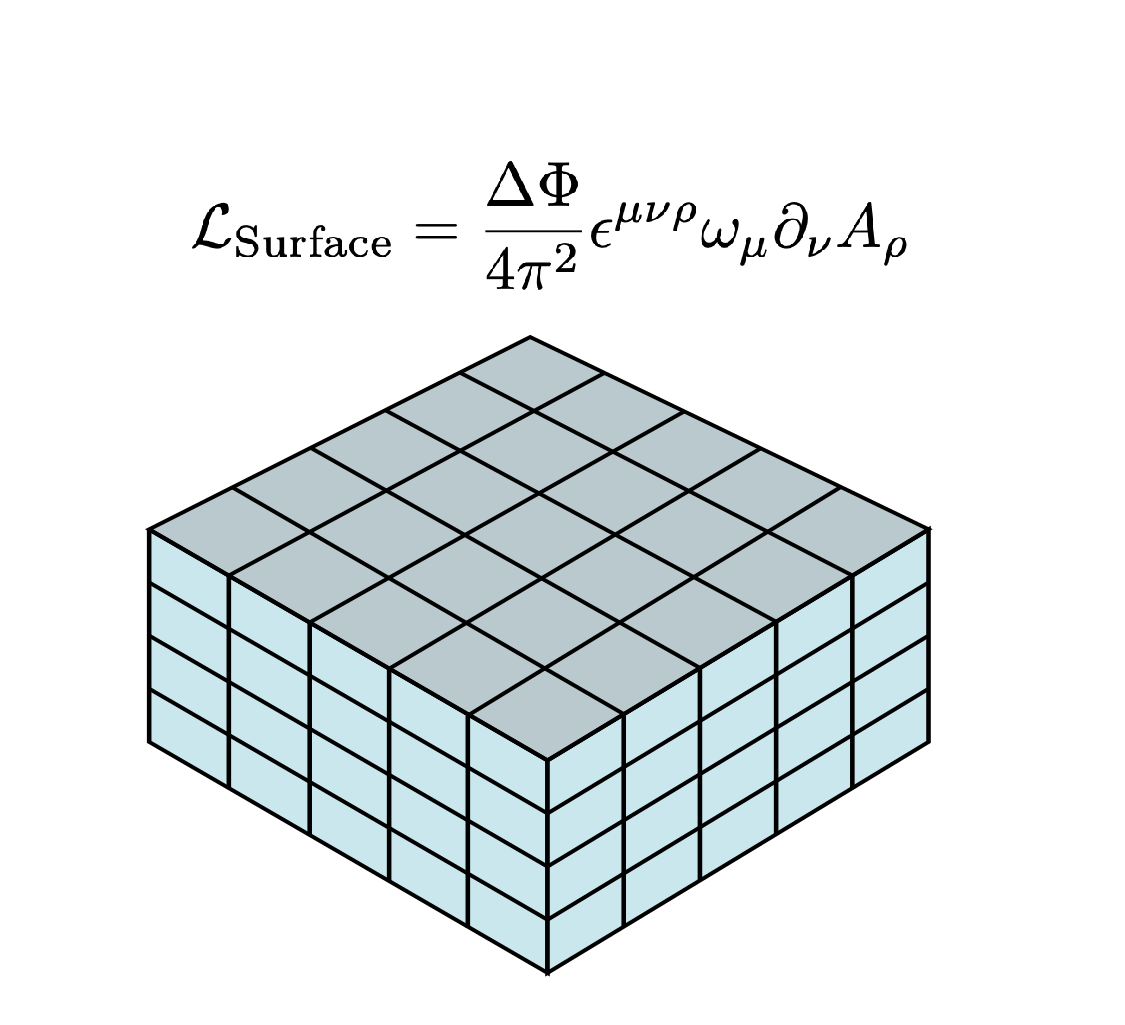}}\\
    \subfloat[]{\includegraphics[width=0.45\textwidth]{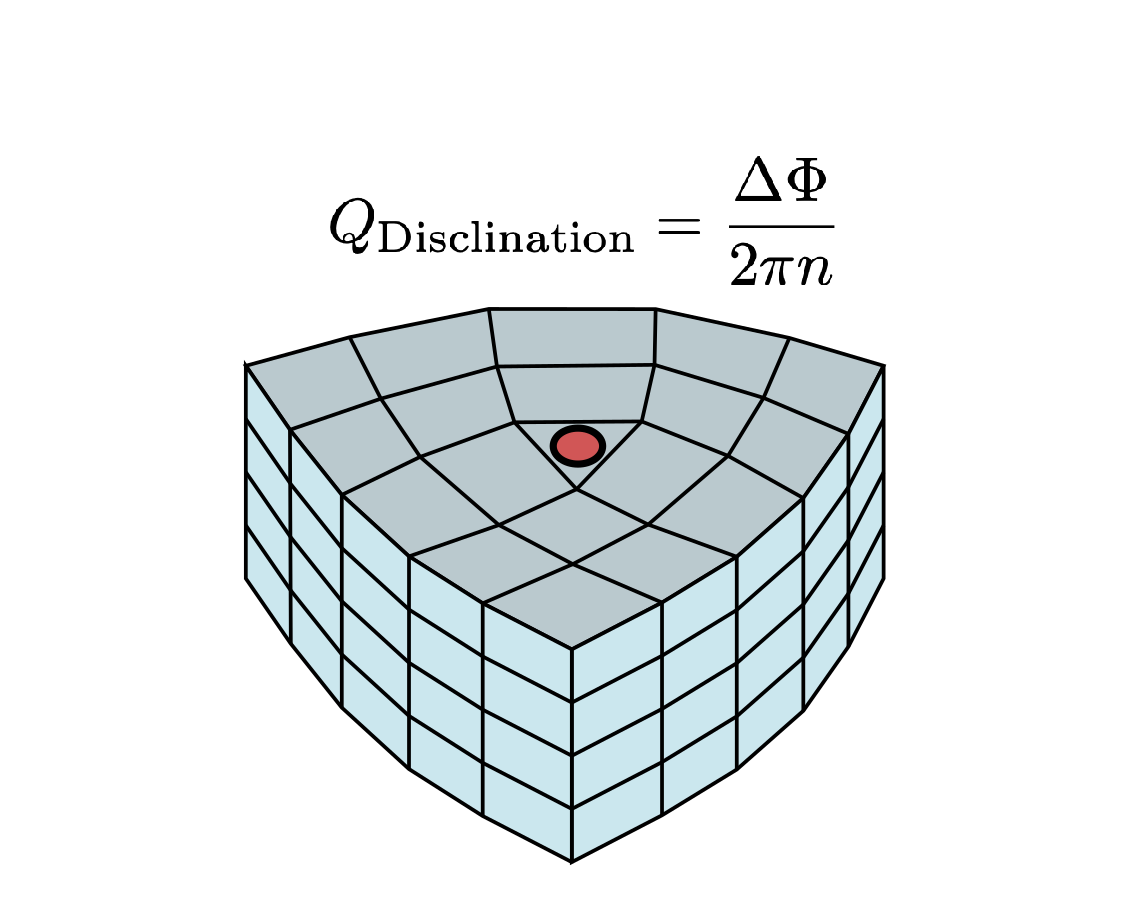}}
    \caption[]{(a) A schematic of a surface where the coefficient of Eq.~\ref{eq:BulkResponses} changes by $\Delta \Phi$. This surface hosts a Wen-Zee term with coefficient $\Delta \Phi/4\pi^2$. (b) A $2\pi/n$ disinclination, shown here for $n = 4$, and the surface charge bound to the disclination, $Q_{\text{Disclination}} = \Delta \Phi/2\pi n$, depicted in red.}\label{fig:DisclinationDiagram}
\end{figure}

Finally, we can consider domain walls, where the value of $\Phi$ changes by $\Delta \Phi$. For a domain wall that preserves $C_n$ symmetry, i.e., a domain wall normal to the z-axis, there will be a $2$D Wen-Zee term of the form~\cite{wen1992shift}:
\begin{equation}\begin{split}
\mathcal{L}_{\text{WZ-DW}} = \frac{\Delta \Phi}{ 4 \pi^2} \epsilon^{\mu\nu\rho} \omega_\mu \partial_\nu A_\rho.
\label{eq:WZDomainWall}\end{split}\end{equation}
At this domain wall, the electromagnetic 3-current is
\begin{equation}\begin{split}
j^\mu =  -\frac{\Delta \Phi}{4\pi^2} \epsilon^{\mu\nu\rho} \partial_\nu \omega_\rho,
\end{split}\end{equation}
indicating that a $2\pi/n$ disclination at the domain wall binds charge $-\Delta \Phi /2\pi n$. This effect is shown schematically in Fig.~\ref{fig:DisclinationDiagram}. Similarly, the domain wall angular momentum 3-current is 
\begin{equation}\begin{split}
j^\mu_{\text{AM}} =  \frac{\Delta \Phi}{4\pi^2} \epsilon^{\mu\nu\rho} \partial_\nu A_\rho,
\end{split}\end{equation}
and hence a $2\pi$ magnetic flux on the surface has angular momentum $\Delta \Phi /2\pi$.

The value of the $\Phi$ coefficient of the $RF$-term is also periodic. However, the period depends on the presence of time-reversal symmetry (TRS), and on the spin of the particles. Here and throughout, we are primarily interested in systems with time-reversal symmetry, as it simplifies our discussions and makes it more applicable to realistic materials. We provide a discussion of the periodicity of $\Phi$ in systems without time-reversal symmetry in Sec.~\ref{sec:RFwoTRS}. 

Similar to our discussion above for $\Theta$, the period of $\Phi$ can be determined by finding the value of $\Delta \Phi$ such that the domain-wall Wen-Zee term in Eq.~\ref{eq:WZDomainWall} can be realized in a purely $2$D system without topological order. For spinless fermions with time-reversal symmetry ($\hat{\mathcal{T}}^2 = +1$), the minimal disclination charge is $1/n$~\cite{li2020fractional}. This response corresponds to a Wen-Zee term with coefficient $1/2\pi$. This indicates that $\Phi$ has period $2\pi$. For spin-1/2 fermions where $\hat{\mathcal{T}}^2 = -1$, the minimal disclination charge is $2/n$ due to Kramers' degeneracy~\cite{li2020fractional}. This response corresponds to a Wen-Zee term with coefficient $1/\pi$. Therefore, for spin-1/2 fermions with TRS, $\Phi$ has period $4\pi$. 

Here, and in the coming sections, we assume that all defects carry trivial quantum numbers in a $\Phi = 0$ insulator. However, this is not always true, as the quantum numbers of a $1$D defect can be locally changed by embedding a $1$D insulator with non-trivial quantum numbers at the defect core. This embedding does not change the $3$D bulk of the insulator. The polarization of defects in an insulator with non-vanishing $\Phi$ should therefore be implicitly understood as the difference in polarization compared to that of the same defect in an insulator with $\Phi = 0$ (see Appendix~\ref{app:RFOrigin} for further discussion).



\subsection{Symmetry Quantization of the $RF$-term}\label{sec:SymQuantRF}
In this subsection we discuss how the coefficient of the $RF$-term, $\Phi$, is quantized by time-reversal symmetry (TRS), particle-hole symmetry (PHS), and mirror symmetry along the $z$-direction. For these symmetries, a non-zero quantized value of $\Phi$ describes a rotation-invariant topological crystalline insulator (rTCI). The essential features of the rTCIs in different symmetry classes are summarized in table~\ref{table:summary}.

\subsubsection{Quantization for Insulators with Particle-Hole and Time-Reversal Symmetry}\label{ssec:SymQuant}
It is well known that the coefficient of the $\Theta$-term in Eq.~\ref{eq:ThetaTerm} is quantized by TRS~\cite{qi2008topological}. This can be seen by noting that Eq.~\ref{eq:ThetaTerm} is odd under TRS, such that $\Theta = -\Theta$ for a time-reversal invariant insulator. Recalling the periodicity of $\Theta$, we find that $\Theta = -\Theta$ has solutions $0$ and $\pi(2\pi)$ for fermions (bosons). In both cases, the former corresponds to a trivial insulator, while the latter corresponds to a time-reversal invariant topological insulator.

Similarly, the $\Phi$ coefficient of the $RF$-term is quantized by particle-hole-symmetry (PHS). Here, we will restrict our attention to systems with additional time-reversal symmetry. Since the $RF$-term is odd under PHS, ($\mathcal{C}:A_\mu \rightarrow -A_\mu$, and $\omega_\mu \rightarrow \omega_\mu$), the $RF$-term of a particle-hole symmetric insulator satisfies $\Phi = -\Phi$ (we take angular momentum to be even under PHS). As noted previously, $\Phi$ is $2\pi$ periodic for spinless insulators with TRS, indicating that $\Phi = 0$ or $\pi$ in spinless insulators with PHS and TRS. For spin-1/2 insulators with TRS, $\Phi$ is $4\pi$ periodic and the $RF$-term has $\Phi = 0$ or $2\pi$. A non-zero value of $\Phi$ indicates that the insulator is an rTCI protected by TRS and PHS. Since $\Phi$ can take only one of two quantized values, both the spinless and spin-1/2 rTCIs have a $\mathbb{Z}_2$ classification.

\begin{figure}[h]
\centering
\includegraphics[width=0.4\textwidth]{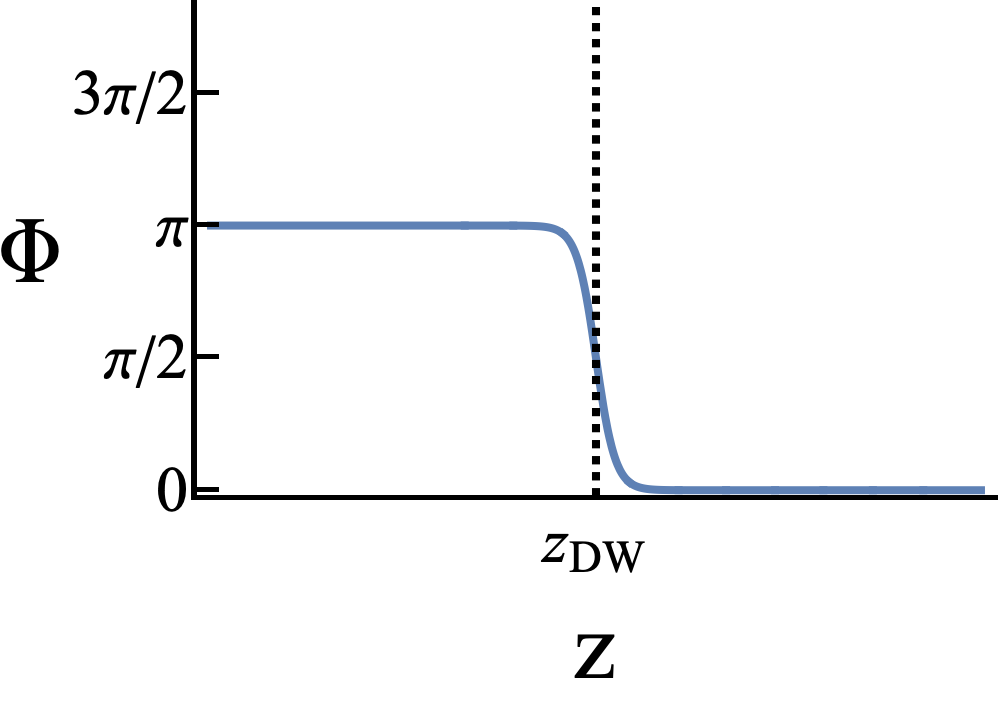}
\caption{The value of $\Phi$ near a PHS-breaking domain wall located at $z = z_\text{DW}$. The domain wall separates a spinless rTCI with $\Phi = \pi$ ($z < z_\text{DW}$), and a trivial insulator with $\Phi = 0$ ($z > z_\text{DW}$).}\label{fig:PHSDW}
\end{figure}

As discussed in the previous subsection, $2\pi/n$ disclinations of the rTCI carry polarization $1/2n$. For the spinless rTCIs (with PHS and TRS), a domain wall where the value of $\Phi$ changes from $\Phi = \pi$ (rTCI) to $\Phi = 0$ (trivial) corresponds to a surface where PHS is explicitly broken. Such a domain wall is shown in Fig.~\ref{fig:PHSDW}. Based on Eq.~\ref{eq:WZDomainWall}, this PHS breaking surface hosts a Wen-Zee term with coefficient $1/4\pi + m/2\pi$ ($m \in \mathbb{Z}$), where $m/2\pi$ is determined from purely $2$D surface effects. The coefficient of this Wen-Zee term is exactly half of what is allowed for a purely $2$D system of spinless fermions with TRS and without topological order. The surface disclination charge modulo $1/n$ is therefore a quantized signature of the bulk topology of the spinless rTCI. 

Similarly for the spin-1/2 rTCIs, $2\pi/n$ disclinations carry charge $1/n$ and domain walls where the value of $\Phi$ changes from $\Phi = 2\pi$ (rTCI) to $\Phi = 0$ (trivial) correspond to PHS broken surfaces and host Wen-Zee terms with coefficient $1/2\pi + m/\pi$. A $2\pi/n$ surface disclination on this type of surface binds charge $1/n + 2m/n$. So for spin-1/2 rTCIs, the surface disclination bound charge modulo $2/n$ is a quantized signature of the bulk topology. As before, the coefficient of the surface Wen-Zee term is half of what is allowed in a purely $2$D system of spin-1/2 fermions with TRS and without topological order.


\subsubsection{Quantization for Insulators with Mirror and Time-Reversal Symmetry}\label{ssec:SymQuantMirror}
In addition to PHS, the $RF$-term is also quantized by mirror symmetry along the $z$-direction. As before, this arises from the fact that the $RF$-term is odd under mirror symmetry ($\mathcal{M}_z:(A_0,A_x,A_y,A_z) \rightarrow (A_0,A_x,A_y,-A_z)$, and $(\omega_0,\omega_x, \omega_y, \omega_z) \rightarrow (\omega_0,\omega_x, \omega_y, -\omega_z)$). Hence, for spinless insulators with TRS and mirror symmetry, $\Phi = 0$ or $\pi$, and, for spin-1/2 insulators with TRS and mirror symmetry, $\Phi = 0$ or $2\pi$. Again, the non-trivial values of $\Phi$ correspond to a non-trivial rTCI with mirror symmetry, and the rTCIs with mirror symmetry have a $\mathbb{Z}_2$ classification. These insulators have both $C_n$ symmetry and mirror symmetry, leading to the total spatial symmetry group of $C_{nh}$. However, since we only gauge the $C_n$ symmetry, we will refer to the mirror and rotation symmetries separately throughout this work. 

The bulk physics is much the same for rTCIs with PHS and rTCIs with mirror symmetry. However, these two classes of rTCIs have different surface physics, since PHS is an on-site symmetry, while mirror symmetry is a spatial symmetry that exchanges surfaces. In particular, a non-trivial $RF$-term does not necessarily lead to symmetry protected surface modes for mirror symmetric insulators. To see this, we note that if $\Phi$ is a function of the $z$ coordinate, mirror symmetry requires that $\Phi(z) = - \Phi(-z)$, while PHS requires that $\Phi(z) = - \Phi(z)$.

\begin{figure}
    \centering
    \subfloat[]{\label{fig:MDW}\includegraphics[width=0.4\textwidth]{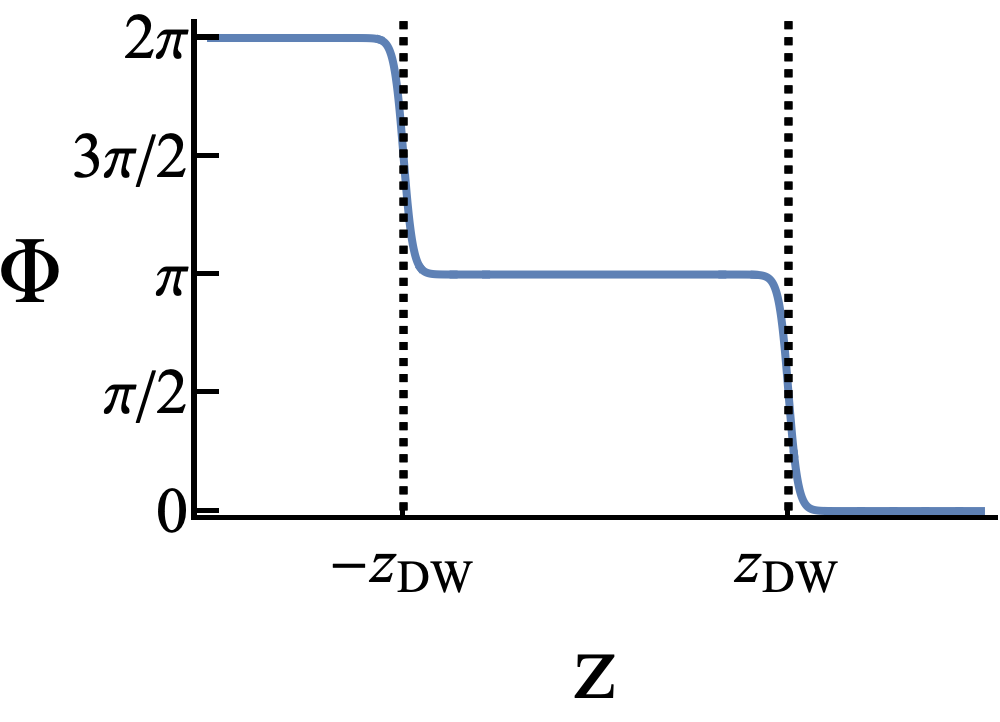}}
    \\
    \subfloat[]{\label{fig:MBDW}\includegraphics[width=0.4\textwidth]{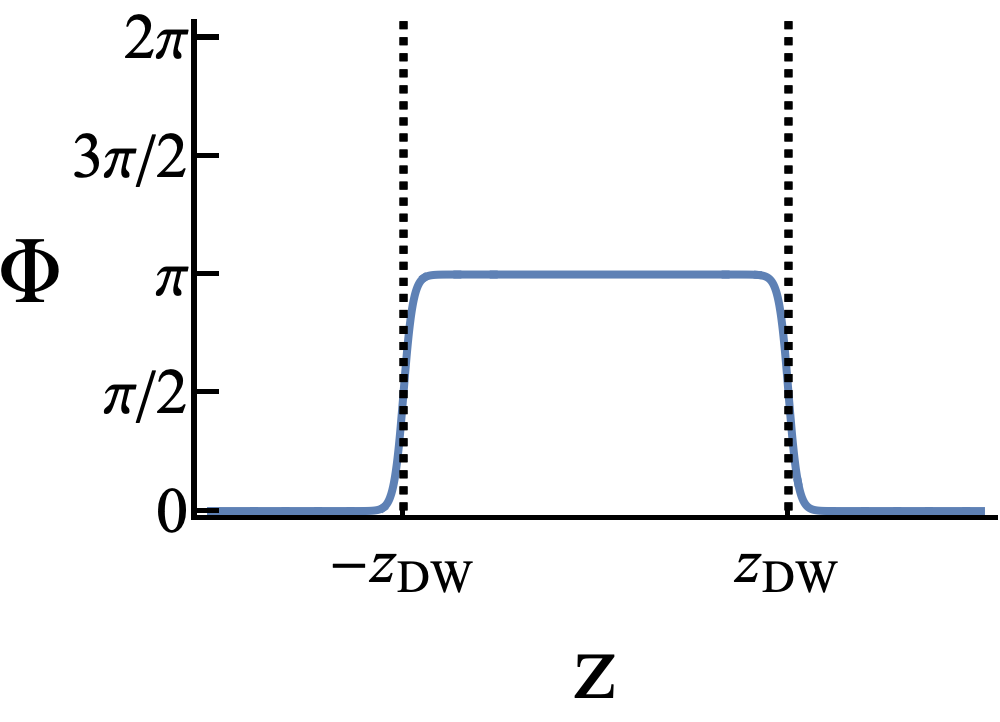}}
    \caption[]{The value of $\Phi$ near a pair of (a) mirror symmetry preserving and (b) mirror symmetry breaking domain walls located at $z = \pm z_\text{DW}$. The domain walls separate the spinless rTCI with $\Phi = \pi$ ($-z_{\text{DW}}<z<z_{\text{dwDW}}$) from two trivial insulators with  $\Phi = 0$ mod$(2\pi)$ ($z < -z_{\text{DW}}$ and $z>z_{\text{DW}}$).}
    \label{fig:MirrorDWs}
\end{figure}

With this in mind, consider a spinless insulator with mirror symmetry, TRS, and a $\Phi = \pi$ $RF$-term that is separated from two trivial insulators ($\Phi = 0$ mod$(2\pi)$) by two domain walls. These domain walls can be gapped while preserving mirror symmetry and TRS if $\Phi$ winds by $\pi + 2\pi q$ ($q \in \mathbb{Z}$) at both domain walls. A pair of mirror symmetry preserving domain walls is shown in Fig.~\ref{fig:MDW}. Similarly, for a spin-1/2 insulator with a $\Phi = 2\pi$ $RF$-term, mirror symmetry is preserved when $\Phi$ winds by $2\pi + 4\pi q$ at both domain walls. For both the spinless and spin-1/2 insulators, domain wall configurations that do not satisfy $\Phi(z) = - \Phi(-z)$ correspond to mirror symmetry breaking surfaces, as shown in Fig.~\ref{fig:MBDW}. 

For the mirror symmetry preserving surfaces, each surface theory consists of a Wen-Zee term with the same coefficient, and a disclination binds the same amount of charge on both mirror related surfaces. A $2\pi/n$ disclination line that connects two mirror symmetry preserving surfaces therefore carries a net charge $1/n + 2q/n$ for spinless fermions ($2/n + 4q/n$ for spin-1/2 fermions), with half the charge localized on each end of the disclination line.

For mirror symmetry breaking surfaces, each surface consists of a Wen-Zee term with opposite coefficients (modulo local surface terms). Disclinations of the two mirror symmetry breaking surfaces will bind opposite amounts of charge, up to local contributions. 

The above analysis also indicates the coefficient of the $RF$-term is quantized by inversion symmetry. However, since inversion symmetry is the product of mirror symmetry and a $C_2$ rotation, we primarily focus on mirror symmetry here.


\subsubsection{Quantization for Insulators with Broken Time-Reversal Symmetry}\label{sssec:QuantizedBrokenTRS}

For insulators with broken TRS, the coefficient of the $RF$-term, $\Phi$, can still take on symmetry-quantized values. However, these quantized values are, in general, different from those found in insulators with TRS. Here we briefly go over the quantization of $\Phi$ in systems with broken TRS. More details are provided in Sec.~\ref{sec:RFwoTRS}.

For spin-1/2 insulators with broken TRS, $\Phi$ has period $2\pi$ (see Sec.~\ref{ssec:periodRFwoTRS}). Therefore, $\Phi = 0$ or $\pi$ for insulators with either PHS or mirror symmetry. As noted previously, spinless insulators with TRS and either PHS or mirror symmetry also have $\Phi = 0$ or $\pi$. This indicates that spin-1/2 insulators with either mirror symmetry or PHS can, at most, realize the same geometry-charge responses as spinless insulators with additional TRS. 

For spinless insulators with broken TRS, $\Phi$ does not have a well-defined periodicity. Rather, $\Phi$ and the coefficient of the $\Theta$-term have a \textit{shared} periodicity, where (see Sec.~\ref{ssec:periodRFwoTRS})
\begin{equation}\begin{split}
(\Phi,\Theta)&\equiv  (\Phi+\pi,\Theta+2\pi) \equiv (\Phi+\pi,\Theta-2\pi).
\label{eq:PhiandThetaPeriod}\end{split}\end{equation}
Since $\Phi$ and $\Theta$ have a shared periodicity, a quantized $\Phi$ must be accompanied by a quantized $\Theta$ (we shall discuss this in more detail in Sec.~\ref{sec:RFwoTRS}). As a result, only mirror symmetry, not PHS, can quantize $\Phi$ (and $\Theta$) in spinless insulators with TRS.

Under mirror symmetry, $(\Phi,\Theta)\rightarrow (-\Phi,-\Theta)$, and a mirror symmetric spinless insulator can therefore have $(\Phi,\Theta) = (0,0)$,  $(\pi/2,\pm \pi)$,  $(\pi,0)$ or $(0,2\pi)$. The $(\Phi,\Theta) = (0,0)$ insulator is clearly a trivial insulator. The $(\Phi,\Theta) = (\pi/2,\pm \pi)$ insulators are a new class of rTCI that has both a non-trivial $RF$-term and s non-trivial $\Theta$-term (the $\Theta = +\pi$ and $-\pi$ insulators are related by time-reversal and do not need to be considered separately). The $(\Phi,\Theta) = (\pi,0)$ and $(0,2\pi)$ insulators are simply the sum and difference of $(\Phi,\Theta) = (\pi/2,\pm \pi)$ insulators, respectively. Based on our previous discussion, insulators with TRS cannot realize a $\Phi = \pi/2$ $RF$-term. Breaking TRS in mirror symmetric insulators can therefore lead to new geometry-charge responses that are not observed in insulators with TRS.

\begin{table*}\begin{center}
\addtolength{\tabcolsep}{2pt}
\renewcommand{\arraystretch}{2.5}
\begin{tabular}{||c | c | c  | c||} 
 \hline
\textbf{System and symmetries} & \textbf{Value of $\Phi$} & \textbf{Surface Disclination Charge} & \textbf{Symmetry Preserving Surfaces} \\[0.5ex]
\hline
 \makecell{Spinless rTCI with $C_n$ symmetry,\\ TRS, and PHS} & $\Phi = \pi$ & $1/2n$ mod($1/n$) & \makecell{Gapless or gapped with\\anomalous topological order} \\ [0.5ex]
 \hline
  \makecell{Spin-1/2 rTCI with $C_n$ symmetry,\\ TRS, and PHS} & $\Phi = 2\pi$ & $1/n$ mod($2/n$) &  \makecell{Gapless or gapped with\\anomalous topological order} \\ [0.5ex]
 \hline
  \makecell{Spinless rTCI with $C_n$ symmetry,\\TRS, and mirror symmetry} & $\Phi = \pi$ & $1/2n$ mod($1/n$) & Gapped with a filling anomaly \\ [0.5ex]
 \hline
   \makecell{Spin-1/2 rTCI with $C_n$ symmetry,\\ TRS, and mirror symmetry} & $\Phi = 2\pi$ & $1/n$ mod($2/n$) & Gapped with a filling anomaly \\ [0.5ex]
 \hline
  \makecell{Spinless rTCI with $C_n$ symmetry\\and mirror symmetry} & $\Phi = \pi/2$ & $1/4n$ mod($1/2n$) & \makecell{Gapped up to a single\\ one-dimensional chiral surface mode} \\ [0.5ex]
 \hline
\end{tabular}
\caption{Summary of the properties and quantized responses of the systems in this paper. The first column lists the system and symmetries. The second column lists the coefficient of the $RF$-term (Eq.~\ref{eq:BulkResponses}) that describes the system. The third column lists the charge that is bound to $2\pi/n$ disclinations of surfaces that preserve $C_n$ symmetry (i.e. surfaces normal to the $\pm$z-direction). The fourth column lists the properties of symmetry-preserving states on surfaces normal to the $\pm$z-direction. }\label{table:summary}
\end{center}
\end{table*}

\section{rotation-invariant Topological Crystalline Insulators with Particle-Hole Symmetry}\label{sec:PHSrTCI}
In this section we will analyze time-reversal invariant rTCIs that have an $RF$-term that is quantized by PHS. We restrict our attention to spinless fermions here, for which the rTCIs is described by $RF$-terms with $\Phi = \pi$. We extend this analysis to spin-1/2 fermions in Appendix~\ref{app:Spin1/2PHS}.

It is worth noting that PHS does not occur as an exact symmetry in realistic electronic insulators~\footnote{PHS does occur in superconductors, but here we are interested in insulators with unbroken U$(1)$ symmetry}. Nevertheless, it is useful to discuss rTCIs with PHS as theoretical constructions to better understand the physical consequences of the $RF$-term. The rTCIs with PHS also serve as a primer for our discussion of rTCIs with mirror symmetry in Sec.~\ref{sec:MirrorrTCI}. Since mirror symmetry is common in electronic insulators, the mirror symmetric rTCIs are more likely to be realized in real materials. However, the mirror symmetric rTCIs have a more complex theoretical structure, since mirror symmetry is also a spatial symmetry. For these reasons, we consider rTCIs with PHS first to gain intuition about the mixed geometry-charge responses that are described by the $RF$-term. Furthermore, there exist materials for which PHS is approximately obeyed near the Fermi-level, as well as large classes of engineered metamaterials, e.g., photonic or acoustic crystals, that can have approximate PHS symmetry.

\subsection{Lattice Model with Spinless Fermions}\label{sec:SpinlessLattice}
In this subsection we present a lattice model for the spinless rTCI with TRS and PHS. To be concrete, we construct a model with $C_4$ rotation symmetry. The minimal model for the rTCI is an $8$-band model with Bloch Hamiltonian
\begin{equation}\begin{split}
\mathcal{H}(\bm{k}) &= \sin(k_x)\Gamma^x \sigma^0 + \sin(k_y)\Gamma^y\sigma^0 + \sin(k_z)\Gamma^z\sigma^0\\ 
&+\sin(k_x)\sin(k_z)\Gamma^0\sigma^x + \sin(k_y)\sin(k_z)\Gamma^0\sigma^y\\
&+(M+\cos(k_x)+\cos(k_y)+\cos(k_z))\Gamma^0\sigma^z,
\label{eq:LatticeHam}\end{split}\end{equation} where $\sigma^{x,y,z,0}$ are the Pauli matrices and the $2\times 2$ identity, and the $4\times 4$ $\Gamma$ matrices are defined as 
\begin{equation}\begin{split}
&\Gamma^x = \sigma^x \sigma^0,\phantom{=} \Gamma^y = \sigma^y\sigma^0,\phantom{=}\Gamma^z = \sigma^z\sigma^z,\\ &\Gamma^0 = \sigma^z \sigma^x, \phantom{=}\Gamma^5 = \sigma^z\sigma^y.
\end{split}\end{equation}
Here and throughout, the Kronecker products are implicit in our definitions. 

The model in Eq.~\ref{eq:LatticeHam} has U$(1)$ charge conservation and is invariant under TRS, PHS, and $C_4$ rotation symmetry. The on-site TRS and PHS operators are
\begin{equation}\begin{split}
&\hat{\mathcal{T}} = \Gamma^y \sigma^y K = \sigma^y \sigma^0 \sigma^y \mathcal{K}, \\
&\hat{\mathcal{C}} = \Gamma^{5y}\sigma^y K = \sigma^x \sigma^y \sigma^y \mathcal{K},
\end{split}\end{equation}
where $\Gamma^{ab}\equiv -i \Gamma^a \Gamma^b$ ($a,b = 0,x,y,z,5$) and $\mathcal{K}$ is complex conjugation. Since the fermions are spinless, $\hat{\mathcal{T}}^2 = + 1$. The model also possesses chiral symmetry, defined as $\hat{\Pi} = \hat{\mathcal{T}}\hat{\mathcal{C}}$. The $C_4$ symmetry operation is
\begin{equation}\begin{split}
\hat{U}_4 &= \exp(i \frac{\pi}{4}[\Gamma^{yx}\sigma^0+ \text{I}\sigma^z ]),
\label{eq:C4Def}\end{split}\end{equation}
and the Hamiltonian satisfies the relation $\hat{U}_4^{-1} \mathcal{H}(\bm{k}) \hat{U}_4 = \mathcal{H}(R^z_4 \bm{k})$, where $R^z_4$ rotates the lattice momentum by $\pi/2$ around the $z$-axis. Since we are considering spinless fermions, $(\hat{U}_4)^4 = +1$. Without $C_4$ symmetry this model is in symmetry class BDI, which is trivial in $3$D~\cite{schnyder2008classification,kitaev2009periodic}. 

\begin{figure}[h]
\centering
\includegraphics[width=0.5\textwidth]{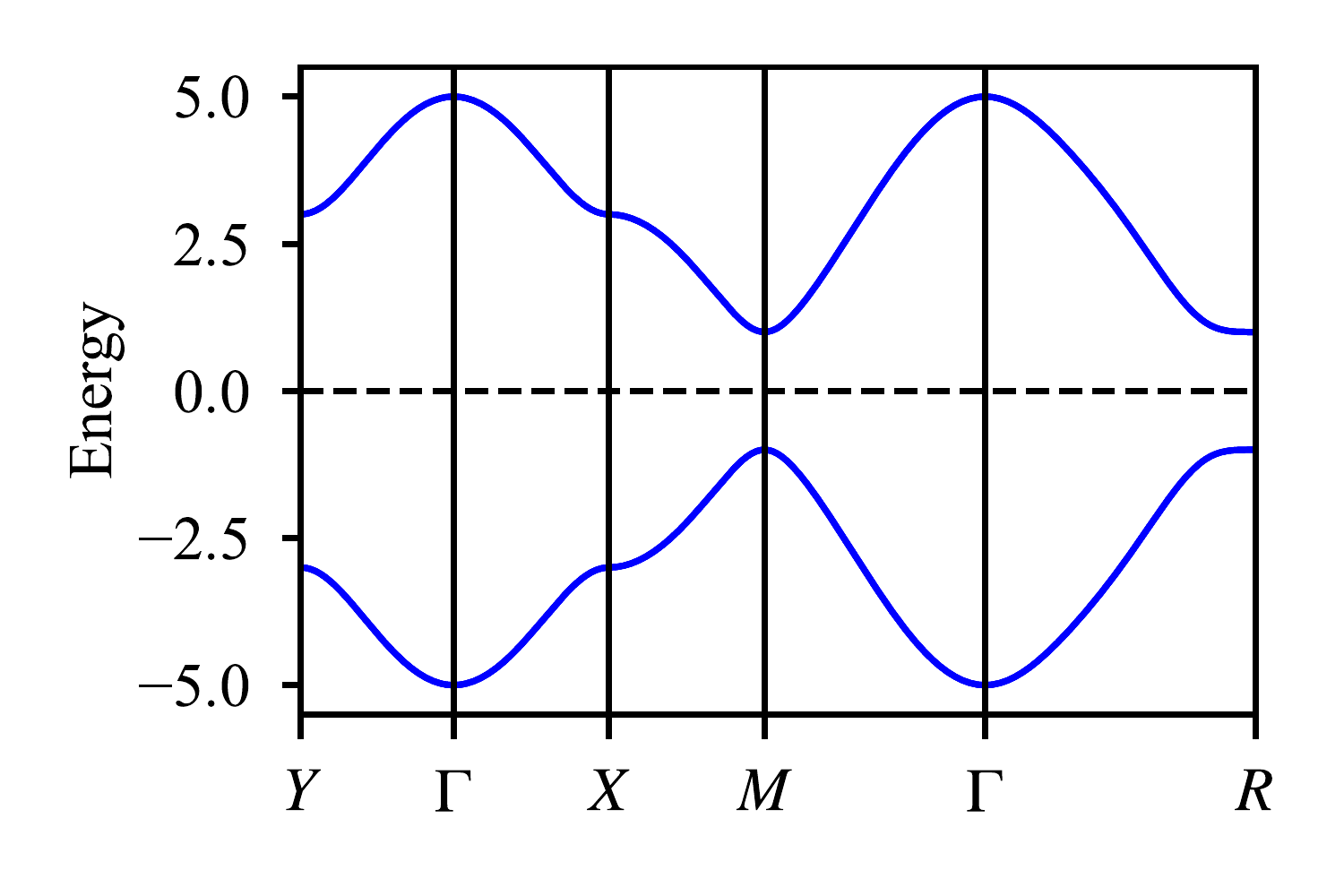}
\caption{The band structure of the Hamiltonian in Eq.~\ref{eq:LatticeHam} along high-symmetry lines with $M=2$.}\label{fig:band_structure}
\end{figure}

The bulk spectrum of Eq.~\ref{eq:LatticeHam} has four-fold degeneracy with energy bands
\begin{equation}\begin{split}
E_{\pm}(k) &= \pm\left[\sin^2(k_x) + \sin^2(k_y) + \sin^2(k_z)\right.\\ 
&+\sin^2(k_x)\sin^2(k_z) + \sin^2(k_y)\sin^2(k_z)\\
&+ \left. (M+\cos(k_x)+\cos(k_y)+\cos(k_z))^2\right]^{1/2}.
\label{eq:latticeSpectrum}\end{split}\end{equation}
The bulk spectrum for $M = 2$ is plotted in Fig.~\ref{fig:band_structure}. The spectrum has a bulk gap unless $|M| = 1,3$.  For $|M|>3$, the lattice model is a trivial insulator, and is adiabatically connected to the atomic limit ($M\rightarrow \pm \infty$). In the following subsection we show that the lattice model is an rTCI that exhibits an $RF$-term where $\Phi = \pi$ for $1<|M|<3$, and the $RF$-term vanishes again for $|M|<1$. 

\subsection{Response Theory}\label{ssec:ResponseSpinless}
Here we calculate the $RF$-term (Eq.~\ref{eq:BulkResponses}) in the response theory of the lattice model (Eq.~\ref{eq:LatticeHam}) for different values of the parameter $M$. To determine the coefficient of the $RF$-term, $\Phi$, our strategy is to first determine the \textit{change} induced in $\Phi$ by the band crossings that occur at $|M| = 1,3$. Combining this information with the fact that $\Phi$ is quantized for gapped insulators with PHS and TRS, and that $\Phi = 0$ for a trivial insulator, we determine the value of $\Phi$ as a function of $M$.

With this in mind, we consider Eq.~\ref{eq:LatticeHam} close to the band crossing at $M = -3$ where two 4-component Dirac fermions form at $\bm{k} = (0,0,0)$. To proceed, we need to consider only the low-energy physics of the lattice model. Expanding to leading order around $\bm{k} = (0,0,0)$ we arrive at the continuum Dirac Hamiltonian
\begin{equation}\begin{split}
\hat{\mathcal{H}} &= \Psi^\dagger \mathcal{H}\Psi\\
\mathcal{H} &= \Gamma^x \sigma^0 i \partial_x + \Gamma^y \sigma^0 i\partial_y +\Gamma^z \sigma^0 i \partial_z+m \Gamma^0 \sigma^z,
\label{eq:LatticeHamLowEnergy}\end{split}\end{equation}
where $m \sim M + 3$, and $\Psi$ is an 8-component spinor. Eq.~\ref{eq:LatticeHamLowEnergy} describes the lattice model when $M \sim -3$. The mass term $m$ controls the transition between the trivial phase of the lattice model with $M<-3$, and the regime where $-3<M<-1$. As we show below, this transition generates an $RF$-term with $\Phi = \pi$ in the effective response theory, indicating that the lattice model is an rTCI when $-3<M<-1$.

To determine the response theory of Eq.~\ref{eq:LatticeHamLowEnergy}, we gauge the U$(1)$ and $C_4$ symmetries and introduce the electromagnetic gauge field $A_\mu$ and spin connection $\omega_\mu$ respectively. As shown in Appendix~\ref{app:SpinConnectCoup}, the spin connection minimally couples to the Dirac fermions in the continuum limit via a term proportional to the angular momentum in the covariant derivative:
\begin{equation}\begin{split}
D_\mu =   \partial_\mu - i A_\mu - i \frac{1}{2}\omega_\mu [\Gamma^{yx} \sigma^0 + \text{I}\sigma^z]. 
\label{eq:CovariantDeriv}\end{split}\end{equation}
The Lagrangian for the minimally coupled Dirac fermions in curved space is given by~\cite{lawrie2012unified}
\begin{equation}\begin{split}
\mathcal{L} = \bar{\Psi} [i \bar{\Gamma}^0\sigma^z D_0  + i E^i_A \bar{\Gamma}^A\sigma^z D_i - m \text{I}\sigma^0]\Psi, 
\label{eq:LagMin}\end{split}\end{equation}
where $E^i_A$ are the inverse frame-fields introduced in Sec.~\ref{sec:LatticeGeoCont}, $\bar{\Psi} = \Psi^\dagger \bar{\Gamma}^0\sigma^z$, and the $4\times 4$ $\bar{\Gamma}$ matrices are $\bar\Gamma^x = \sigma^y \sigma^x$, $\bar\Gamma^y = \sigma^x \sigma^x$, $\bar\Gamma^z = \sigma^0 \sigma^y$, $\bar\Gamma^0 = \sigma^z \sigma^x$, $\bar\Gamma^5 = \sigma^0 \sigma^z$. Under a $C_4$ gauge transformation, the inverse frame-fields, spin connection, and continuum fermions transform as
\begin{equation}\begin{split}
&E^i_x \rightarrow \cos(\theta) E^i_x+ \sin(\theta)E^i_y, \\
&E^i_y \rightarrow\cos(\theta) E^i_y - \sin(\theta)E^i_x,\\
&\omega_\mu \rightarrow \omega_\mu - \partial_\mu \theta,\\
&\Psi \rightarrow e^{i \theta \frac{1}{2}[\Gamma^{xy}\sigma^0 + \text{I}\sigma^z]}\Psi,
\label{eq:ContGaugeSym}\end{split}\end{equation}
where $\theta$ is a function of $x_\mu$ that takes values in $\{0, \pi/2, \pi , 3\pi/2\}$. Here, the $C_4$ gauge symmetry is actually part of a larger SO$(2)$=U$(1)$ gauge symmetry that continuously rotates the Dirac fermions. The U$(1)$ rotation gauge symmetry is defined as in Eq.~\ref{eq:ContGaugeSym}, but with $\theta$ taking continuous values in $[0,2\pi)$. The original lattice model does not have this U$(1)$ rotation gauge symmetry. Rather, it is a feature that emerges in the continuum limit.

In addition to the gauge fields, we include an additional PHS breaking perturbation to Eq.~\ref{eq:LagMin},
\begin{equation}\begin{split}
\mathcal{L}' = \bar\Psi m'\bar{\Gamma}^5\sigma^0 \Psi,
\label{eq:ResponsePert}\end{split}\end{equation}
and set $m = -\bar{m}\cos(\phi)$ and $m' = -\bar{m}\sin(\phi)$, where $\phi$ is a background field. We keep $\bar{m}>0$ fixed, and treat $\phi$ as a new parameter for the theory, such that $\phi = 0$ corresponds to $m<0$ phase and $\phi = \pi$ corresponds to the $m>0$ phase~\cite{qi2008topological}. Physically, non-constant values of $\phi$ encode either domain walls or adiabatic evolutions of the Hamiltonian. For example, $\phi = \frac{\pi}{2}[1-\tanh(z/\xi)]$ corresponds to a PHS breaking domain wall between the $m<0$ and $m>0$ phases that is located near $z = 0$ and has width $\xi$. Similarly, $\phi = \pi t/T$ corresponds to a PHS breaking adiabatic evolution from the $m<0$ phase at $t = 0$ to the $m>0$ phase at $t = T$.

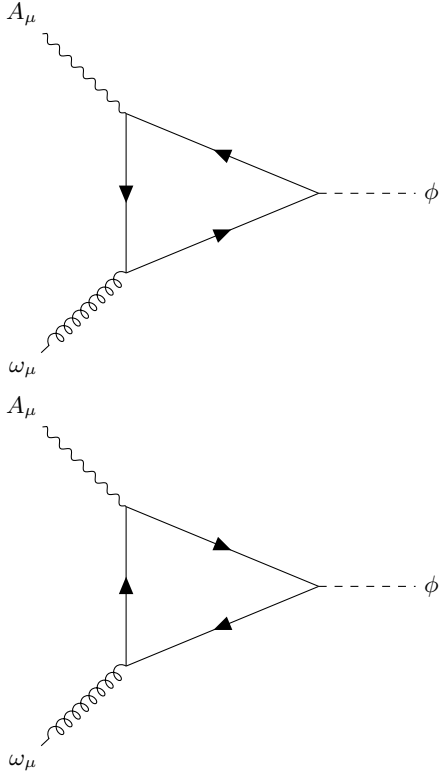
\begin{figure}[h]
\centering
\begin{tikzpicture}
  \begin{feynman}
    \vertex (p) {\(\phi\)};
    \vertex [left=of p] (a);
    \vertex [above left=of a] (e);
    \vertex [below left=of a] (f);
    \vertex [left=of e] (b);
    \vertex [left=of f] (c);
    \vertex [above left=of b] (em) {\(A_\mu\)};
    \vertex [below left=of c] (w) {\(\omega_\mu\)};
    \diagram* {
      (a) -- [fermion] (b),
      (b) -- [fermion] (c),
      (c) -- [fermion] (a),
      (a) --[dashed]  (p),
      (b) --[photon]  (em),
      (c) --[gluon]  (w),
    };
  \end{feynman}
\end{tikzpicture}
\begin{tikzpicture}
  \begin{feynman}
    \vertex (p) {\(\phi\)};
    \vertex [left=of p] (a);
    \vertex [above left=of a] (e);
    \vertex [below left=of a] (f);
    \vertex [left=of e] (b);
    \vertex [left=of f] (c);
    \vertex [above left=of b] (em) {\(A_\mu\)};
    \vertex [below left=of c] (w) {\(\omega_\mu\)};
    \diagram* {
      (b) -- [fermion] (a),
      (c) -- [fermion] (b),
      (a) -- [fermion] (c),
      (a) --[dashed]  (p),
      (b) --[photon]  (em),
      (c) --[gluon]  (w),
    };
  \end{feynman}
\end{tikzpicture}
\caption{The relevant Feynman diagrams for calculating the response theory in Eq.~\ref{eq:EffDerivedphi}. The solid lines indicate fermion propagators.}\label{fig:TDia}
\end{figure}

Now we are ready to obtain the effective topological response theory in terms of $A_\mu$, $\omega_\mu$, and $\phi$ by integrating out the massive fermions via a diagrammatic expansion. For our purposes, we are only interested in the topological contributions from the triangle diagrams in Fig~\ref{fig:TDia}~\footnote{There are additional diagrams that contribute Maxwell-like terms in the effective response theory, but these terms are not topological and will not be considered here.}. These diagrams evaluate to
\begin{equation}\begin{split}
\mathcal{L}_{\text{eff}} = \frac{\phi}{4\pi^2} \epsilon^{\mu\nu\rho\kappa} \partial_\mu \omega_\nu \partial_\rho A_\kappa. 
\label{eq:EffDerivedphi}\end{split}\end{equation}
For $m<0$ (i.e., $\phi = 0$) the effective response vanishes, in agreement with the fact that the lattice model is trivial for $M<-3$. For $m>0$ (i.e., $\phi = \pi$) the effective response is
\begin{equation}\begin{split}
\mathcal{L}_{\text{eff}} = \frac{1}{4\pi} \epsilon^{\mu\nu\rho\kappa} \partial_\mu \omega_\nu \partial_\rho A_\kappa,
\label{eq:EffDerived}\end{split}\end{equation}
which is the $RF$-term from Eq.~\ref{eq:BulkResponses} with $\Phi = \pi$. We only considered the low-energy Lagrangian and leading order diagrams here. However, corrections to the continuum Lagrangian and higher-order diagrams will not change these results since the value of $\Phi$ is quantized by PHS and TRS.

Strictly speaking, we have only confirmed in this subsection that $\Phi$ \textit{changes} by $\pi$ at the $ M = -3$ ($m=0$) band crossing. This ambiguity is related to the fact that the $RF$-term is a total derivative. However, because the $M<-3$ phase corresponds to a trivial insulator where $\Phi$ must vanish, we conclude that the $-3<M<-1$ phase is an rTCI with a $\Phi = \pi$ $RF$-term. Similar calculations show that $\Phi$ vanishes for $|M|<1$ and $|M|>3$, and that $\Phi = \pi$ for $1<|M|<3$ as well. 

As a final point, we also note that increasing $\phi$ from $0$ to $2\pi$, in Eq.~\ref{eq:LagMin} is a periodic process that takes the trivial insulator back to itself. Based on the diagrammatic calculation in Eq.~\ref{eq:EffDerivedphi}, an $RF$-term with $\Phi = 2\pi$ is generated during this process. This agrees with our earlier conclusion that $\Phi$ is $2\pi$ periodic for spinless fermions with TRS.

\subsection{Surface Theory}\label{ssec:SurfaceTh}
In this subsection we analyze the surface theory of the rTCI. This is accomplished by considering a domain wall where $-3< M < -1$ for $z<0$, and $M < -3$ for $z>0$, i.e. a domain wall between the rTCI and a trivial insulator. This mass configuration generates a pair of gapless 2-component Dirac fermions that are localized at the $2$D domain wall~\cite{shen2012topological}. The Hamiltonian for the surface Dirac fermions is
\begin{equation}\begin{split}
\hat{\mathcal{H}}_{\text{surf}} &= \psi^\dagger \mathcal{H}_{\text{surf}} \psi,\\
\mathcal{H}_{\text{surf}} &= \left[\sigma^x i\partial_x - \sigma^y i\partial_y\right]\sigma^0,
\label{eq:SurfHam}\end{split}\end{equation}
where $\psi$ is a 4-component spinor. The TRS, PHS, and $C_4$ symmetry act on the surface Hamiltonian as
\begin{equation}\begin{split}
&\hat{\mathcal{T}}_{\text{surf}} = \sigma^y  \sigma^y K,\\
&\hat{\mathcal{C}}_{\text{surf}} = \sigma^x  \sigma^x K,\\
&\hat{U}_{4-\text{surf}} = \exp(i\frac{\pi}{4}[-\sigma^z  \sigma^0 + \sigma^0  \sigma^z] ).
\label{eq:SurfSym}\end{split}\end{equation}
Consistent with the bulk theory, $\hat{\mathcal{T}}^2_{\text{surf}} = (\hat{U}_{4-\text{surf}})^4 = +1$. 
To show that the surface Dirac cones are symmetry protected, we note that a mass term for Eq.~\ref{eq:SurfHam} must be proportional to $\sigma^z  \sigma^{0,x,y,z}$. All of these terms break one of the symmetries in Eq.~\ref{eq:SurfSym}. 
Specifically, the $\sigma^z  \sigma^{0}$ term breaks TRS, the $\sigma^z  \sigma^{z}$ term breaks PHS, and the $\sigma^z  \sigma^{x,y}$ terms break $C_4$ symmetry. As such, all three symmetries are required to protect the pair of gapless Dirac cones.

\begin{figure}[h]
\centering
\begin{tikzpicture}
  \begin{feynman}
    \vertex (em) {\(A_\mu\)};
    \vertex [left=of em] (a);
    \vertex [left=of a] (b);
    \vertex [left=of b] (w) {\(\omega_\mu\)};
    \diagram* {
      (em) --[photon]  (a),
      (a) --[fermion,half left, looseness=1.6] (b),
      (b) --[fermion,half left, looseness=1.6] (a),
      (b) --[gluon]  (w),
    };
  \end{feynman}
\end{tikzpicture}
\caption{The relevant Feynman diagram for calculating the surface Wen-Zee term. The solid lines indicate fermion propagators.}\label{fig:WZDia}
\end{figure}
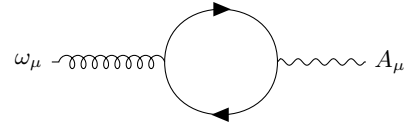

Based on our discussion of the effective field theory in Sec.~\ref{ssec:SymQuant}, we now gap the surface by adding a PHS breaking mass term $m_s {\sigma^z  \sigma^{z}}$. The response theory for the massive symmetry broken surface is found by coupling the fermions to the gauge field $A_\mu$ and spin connection $\omega_\mu$ via the covariant derivative,
\begin{equation}\begin{split}
D_\mu = \partial_\mu -i A_\mu - i \frac{1}{2} \omega_\mu [-\sigma^z  \sigma^0 +  \sigma^0  \sigma^z],
\end{split}\end{equation}
and integrating out the massive fermions via a diagrammatic expansion. The resulting response theory contains a Wen-Zee term,
\begin{equation}\begin{split}
\mathcal{L}_{\text{surf}} = \frac{\text{sgn}(m_s)}{4\pi} \epsilon^{\mu\nu\rho} \omega_\mu \partial_\nu A_\rho,
\label{eq:SurfaceWZspinless}\end{split}\end{equation}
that corresponds to the one-loop diagram in Fig~\ref{fig:WZDia}. This is exactly the anomalous surface term from Eq.~\ref{eq:WZDomainWall} with $\Delta \Phi = \pm \pi$. Local surface effects can shift the coefficient of the surface Wen-Zee term by $1/2\pi$, and, in general, a $\pi/2$ surface disclination binds charge $\frac{1}{8}+\frac{n}{4}$ with $n \in \mathbb{Z}$.

\subsection{Numerics}\label{ssec:spinlessnumerics}
\begin{figure}
    \centering
    \subfloat[]{\label{fig:dos_phs}\includegraphics[width=0.45\textwidth]{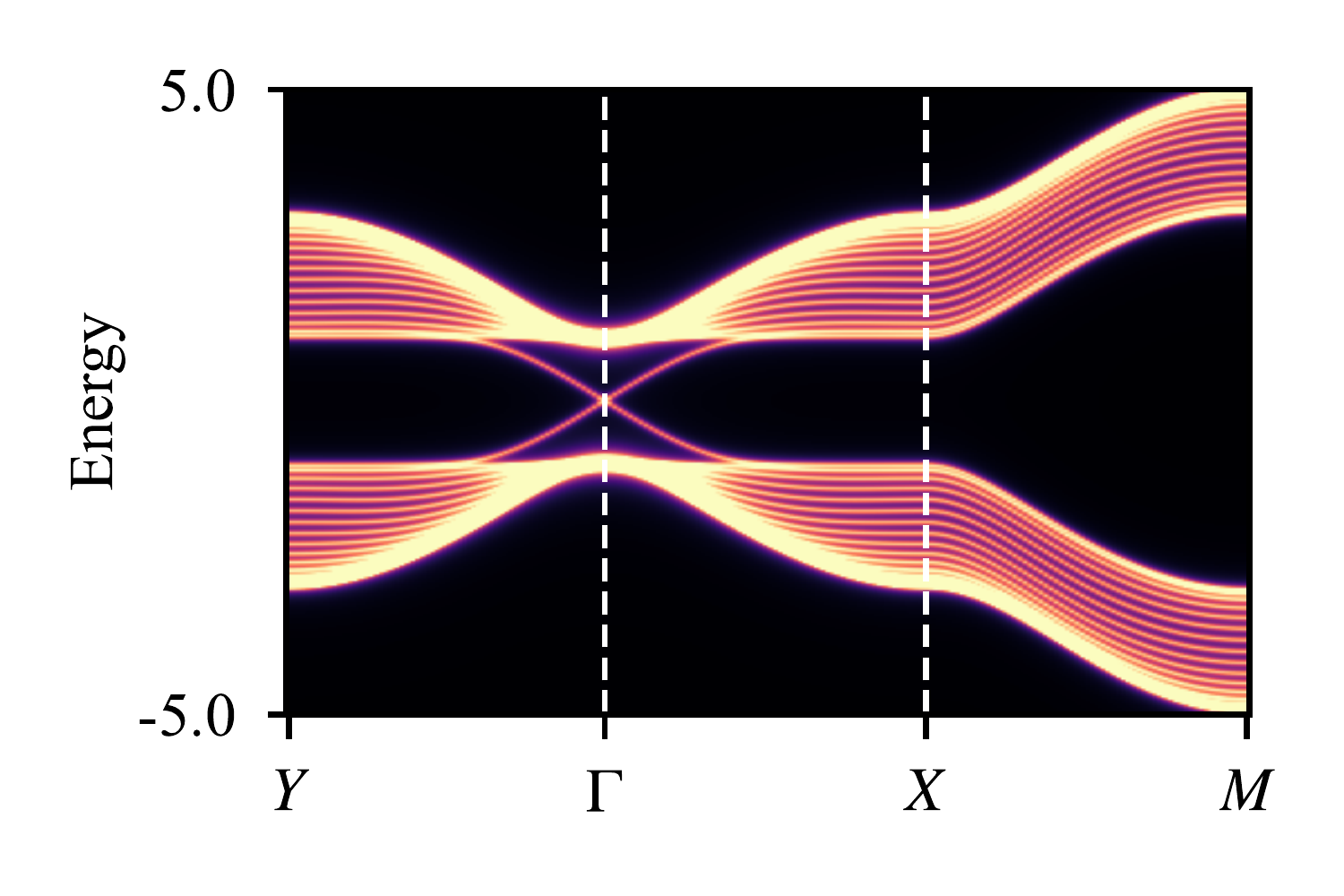}}
    \\
    \subfloat[]{\label{fig:dos_no_phs}\includegraphics[width=0.45\textwidth]{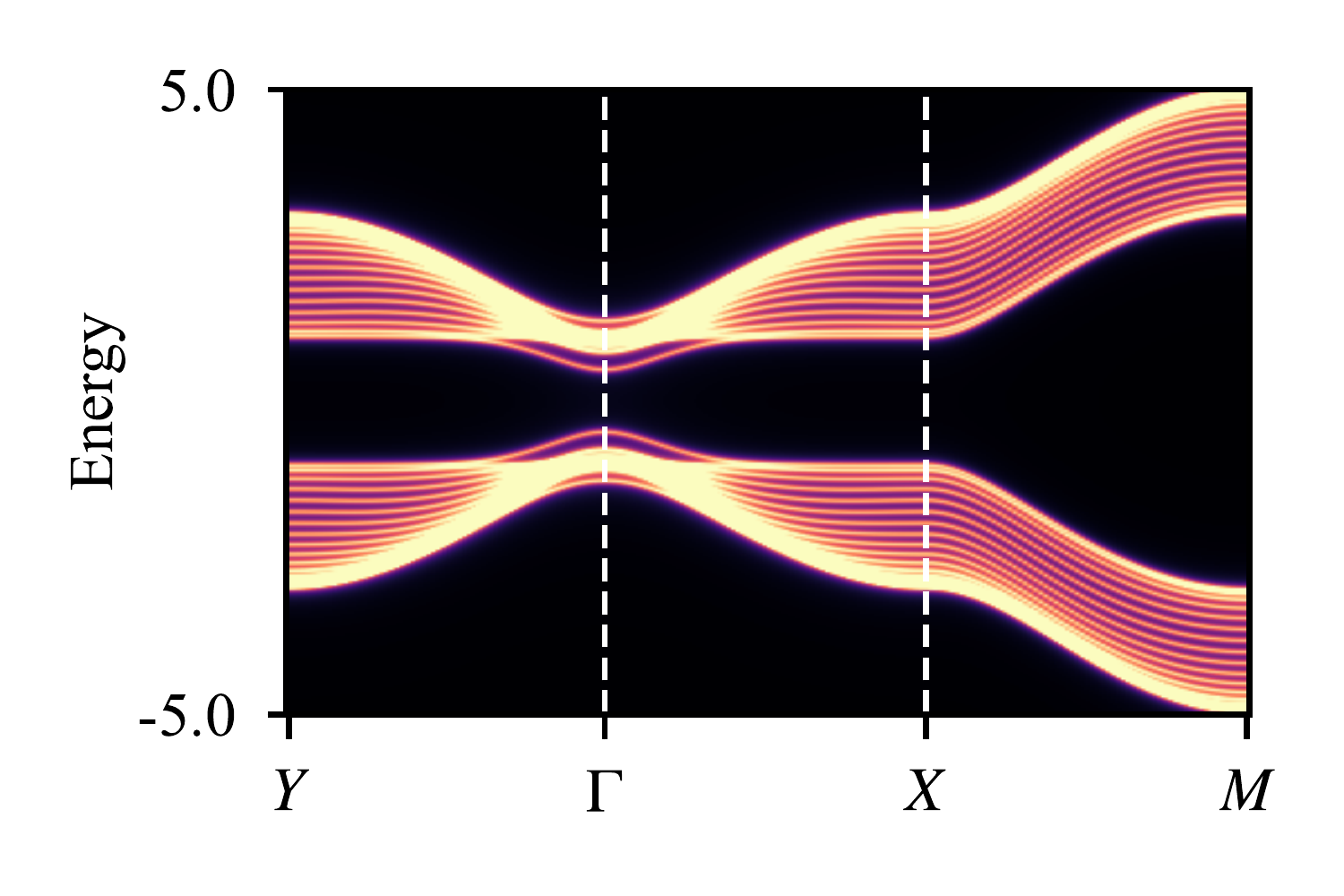}}
    \caption[]{(a) The spectral function along high-symmetry lines of the lattice model in Eq.~\ref{eq:LatticeHam} with $M = -2$,sixteen sites and open boundary conditions in the z-direction, and symmetry preserving surfaces hosting midgap states. (b) The same spectral function with a PHS breaking mass term Eq.~\ref{eq:LatticeSurfMass} with $m_s = 0.5$ that gaps the surface states.}
    \label{fig:DensityofStatesSpinless}
\end{figure}

In this subsection we numerically verify our previous analysis. For a lattice with open boundaries along the z-direction, we find mid-gap states with a Dirac-like dispersion (see Fig.~\ref{fig:dos_phs}). The midgap states correspond to the gapless surface states of the rTCI and can be gapped out by adding an on-site PHS breaking term of the form
\begin{equation}\begin{split}
\mathcal{H}_{s} = m_s \sum_{r\in \text{surface}}c^\dagger(r)\Gamma^5\sigma^0c(r),
\label{eq:LatticeSurfMass}
\end{split}\end{equation}
where the sum is taken over the sites on the open surfaces of the lattice and $c^\dagger(r)$ is the $8$-component fermion creation operator at site $r$. The density of states of the system with the PHS breaking surface perturbation shows no midgap states (see Fig.~\ref{fig:dos_no_phs}).

\begin{figure}
    \centering
    \includegraphics[width=0.4\textwidth]{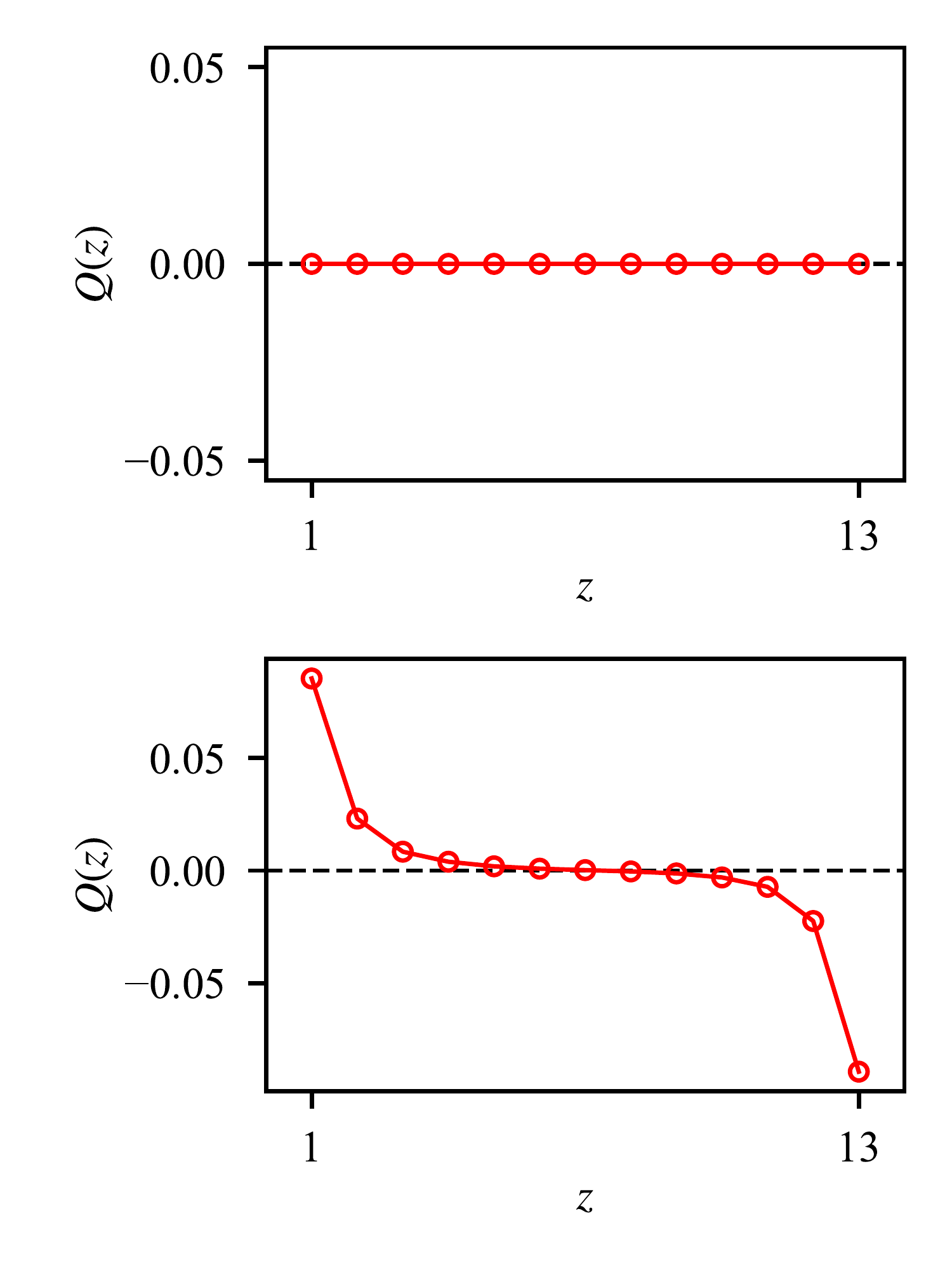}
    \caption[]{The charge per layer, $Q(z)$, for the Hamiltonian in Eq.~\ref{eq:LatticeHam} on a $13\times13\times13$ lattice with $M=-2$, a PHS breaking mass term Eq.~\ref{eq:LatticeSurfMass} with $m_s = 0.25$, and either no disclination (top) or a $\pi/2$ site-centered disclination (bottom). The background charge of $-4$ per site is added to obtain charge neutrality at half-filling.}
    \label{fig:ChargePerLayerSpinless}
\end{figure}

\begin{figure}
    \centering
    \includegraphics[width=0.45\textwidth]{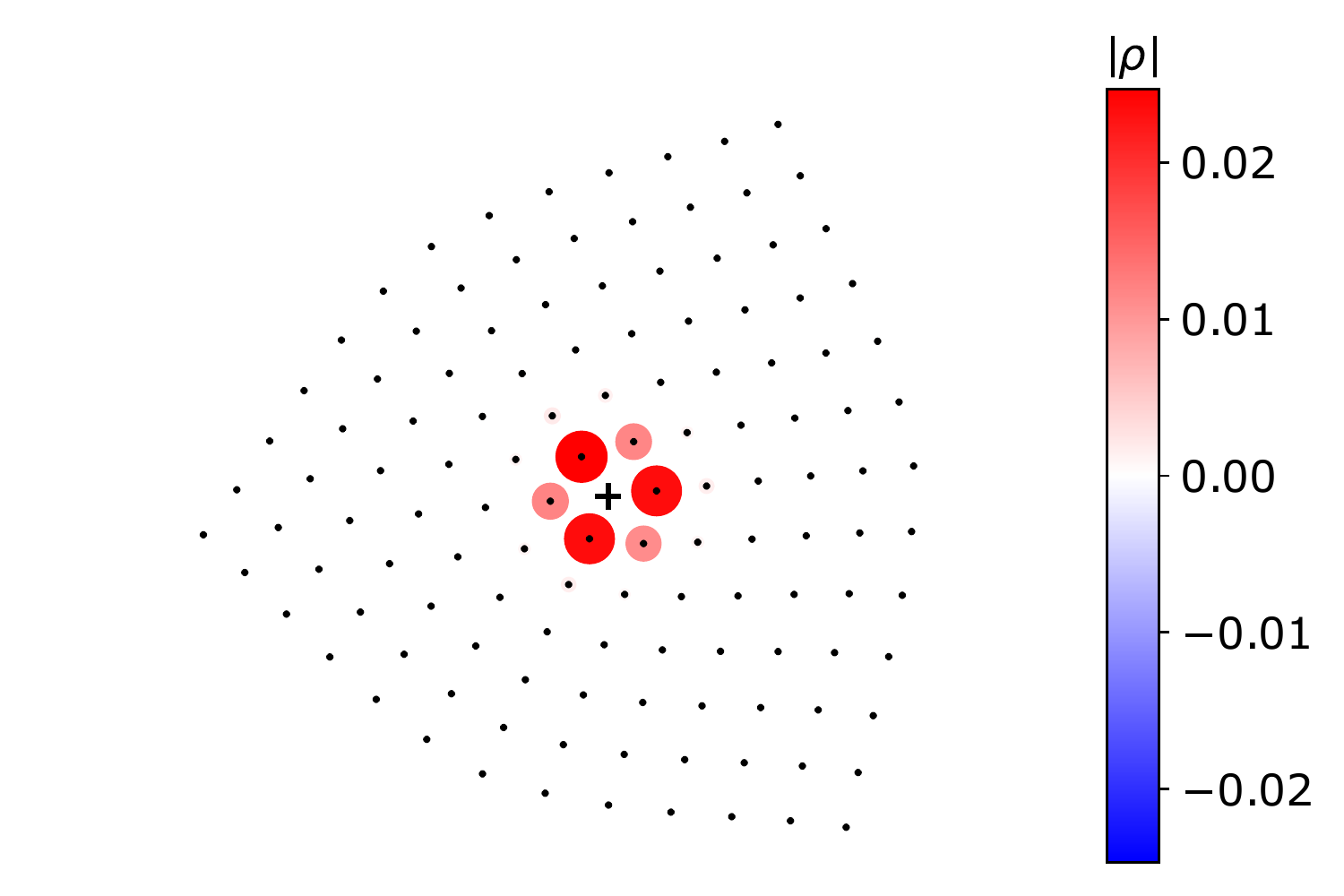}
    \caption[]{The charge density summed over the top half of the $z$-layers for the Hamiltonian in Eq.~\ref{eq:LatticeHam} on a $13\times13\times13$ lattice with $M=-2$, a PHS breaking mass term Eq.~\ref{eq:LatticeSurfMass} with $m_s = 1$, and a $\pi/2$ disclination. The background charge of $-4$ per site is added to obtain charge neutrality at half-filling. The cross ($\bm{+}$) marks the disclination core.}
    \label{fig:disclination_charge_density}
\end{figure}

\begin{figure}
    \centering
    \includegraphics[width=0.45\textwidth]{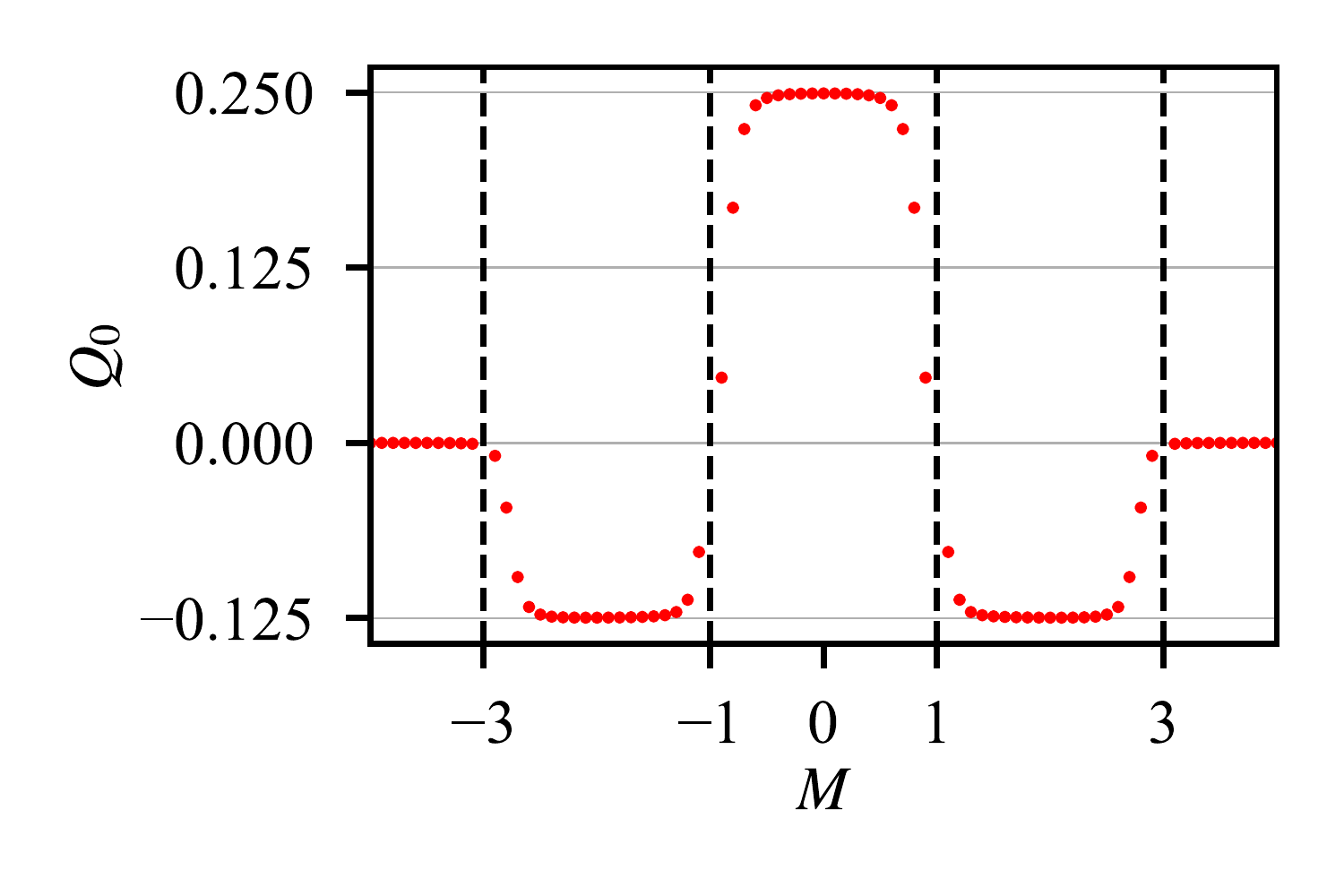}
    \caption{The surface charge $Q_0$ bound to a $\pi/2$ site-centered disclination as a function of the mass $M$. The charge is calculated using Eqs.~\ref{eq:LatticeHam} and~\ref{eq:LatticeSurfMass} on a $15 \times 15 \times 15$ lattice with open boundaries and a PHS breaking mass $m_s$ equal to the bulk mass gap. We attribute the deviation of the charge from its quantized value near $|M| = 1,3$ (vertical dashed lines) to finite size effects and the closing of the bulk gap.}\label{fig:surfaceChargeVsM}
\end{figure}

We now calculate the charge distribution of the rTCI when PHS symmetry is broken at the surface by Eq.~\ref{eq:LatticeSurfMass}. We include negative background charges at each lattice site such that the system is charge neutral at half-filling. Physically, these negative charges correspond to the ions that form a crystalline solid. The charge distribution is uniform when the lattice is free of disclinations, as shown in Fig.~\ref{fig:ChargePerLayerSpinless}.  To probe the mixed geometry-charge response of the rTCI with PHS breaking surfaces, we add a $\pi/2$ disclination-line to the lattice (detail of the disclianted lattice are given in Appendix~\ref{app:latticeDisc}).  As shown in Fig.~\ref{fig:ChargePerLayerSpinless}, excess charge is localized on the top and bottom surfaces of the rTCI when there is a disclination. This excess surface charge is localized around the core of the disclination (see Fig.~\ref{fig:disclination_charge_density}). Fig~\ref{fig:surfaceChargeVsM} shows the net charge that is localized on the top surface of the disclinated lattice model for various values of $M$. When $1<|M|<3$ the disclinated surface has surface charge $\frac{1}{8}$ (mod $\frac{1}{4}$)  (up to finite size corrections), indicating that the lattice model is an rTCI with a non-trivial $RF$-term is these regimes. When $|M|>3$ and $|M|<1$ the surface charge is $0$ (mod $\frac{1}{4}$), indicating that the $RF$-term is trivial in these regimes. These results are in full agreement with our previous analysis.

\subsection{Dimensional reduction to a 1+1D SPT}\label{sec:DimRed}
In Ref. \onlinecite{song2017topological}, Song et al. showed that a TCI protected by a crystalline symmetry is adiabatically connected to a lower-dimensional SPT and that the crystalline symmetry of the higher-dimensional TCI becomes an on-site symmetry of the lower-dimensional SPT. In this subsection we use this logic to dimensionally reduce the $3$D rTCI to a $1$D SPT with an on-site U$(1)$ symmetry, PHS, TRS, and $\mathbb{Z}_4$ symmetry, the latter of which is inherited from the $C_4$ rotation symmetry of the rTCI. This $1$D SPT is equivalent to the topological phase of the well known Su-Schrieffer-Heeger (SSH)~\cite{su1979solitons} chain with an additional trivial $\mathbb{Z}_4$ symmetry.

This connection is established as follows. Using the rTCI surface theory from Sec.~\ref{ssec:SurfaceTh}, we show that there exists a symmetry preserving deformation that trivializes the entire rTCI surface, except for the $C_4$ rotation center of the surface. Since the rotation center of the surface is single point, this deformation reduces the effective dimension of the surface from $2$D to $0$D. Treated as a $0$D system, the rotation center of the deformed surface has a zero-energy mode that is protected by PHS, and has a net half-integer of charge. These are exactly the same topological features that are found at the $0$D edge of a $1$D SSH chain. The equivalence of the deformed rTCI and the SSH chain therefore follows from the bulk-boundary correspondence. 

To this end, we consider the surface theory for a single domain wall oriented normal to the z-direction in Eq.~\ref{eq:SurfHam}. We add the following generic mass deformation term to the surface Hamiltonian,
\begin{equation}\begin{split}
\mathcal{H}_{\text{surf-mass}} =  m_x \sigma^z\sigma^x + m_y \sigma^z\sigma^y + m_z \sigma^z \sigma^z.
\label{eq:DimRedMassTerm}\end{split}\end{equation}
Here we set $m_z = 0$ and $m_x + im_y =  m_s(r) \exp(i\theta)$, where $(r,\theta)$ are polar coordinates on the surface and $m_s(r) \geq 0$ is a function of the radial coordinate that vanishes at $r = 0$ and goes to a non-zero constant $\bar{m}_s$ as $r\rightarrow \infty$. Due to the dependence on the radial angle $\theta$, this term is invariant under $C_4$ rotations as well as PHS and TRS. The single particle spectrum of the surface theory can be explicitly solved. In Appendix~\ref{app:ZeroEnergy} we show that the deformed surface has a single zero-energy mode $\psi_0$ that is localized at the rotation center, $r=0$. This mode is protected by PHS, and transforms trivially under $C_4$ rotations. 

Additionally, a net half-integer of charge is localized at $r = 0$. To show this, we take the zero-energy mode to be empty, and integrate out the remaining massive fermions. The effective theory for the massive fermions can be written in terms of the fluctuations of the mass terms $m_{x,y,z}$~\cite{jaroszewicz1984induced,abanov2000theta},
\begin{equation}\begin{split}
\mathcal{L}_{\text{surf-mass}} &= \frac{\epsilon^{\mu\nu\rho}}{8\pi} \bm{n}\cdot(\partial_\mu \bm{n}\times \partial_\nu \bm{n})A_\rho\\
&+ \frac{n_z}{4\pi}\epsilon^{\mu\nu\rho} \omega_\mu \partial_\nu  A_\rho ,\\
\bm{n} &= \frac{\bm{m}}{|\bm{m}|}, \phantom{=} \bm{m} = (m_x,m_y,m_z).
\label{eq:SurfaceEffectiveSky}\end{split}\end{equation}
The first term $\mathcal{L}_{\text{surf-mass}}$ arises from fluctuations of the mass terms, while the second term is the half-quantized Wen-Zee term that occurs on PHS breaking surfaces. Note that for the constant PHS breaking mass term $m_x = m_y = 0$, $m_z \neq 0$, Eq.~\ref{eq:SurfaceEffectiveSky} reduces to Eq.~\ref{eq:SurfaceWZspinless}, while for PHS preserving surfaces mass terms with $m_z = 0$ the Wen-Zee term vanishes. 

Using Eq.~\ref{eq:SurfaceEffectiveSky}, we find that charge $Q= -\frac{1}{2}$ is localized at $r = 0$ for the mass configuration $m_z = 0$, $m_x + im_y =  m_s(r) \exp(i\theta)$~\cite{chamon2008irrational}. Due to the aforementioned zero-energy mode, this charge is only meaningfully defined modulo $1$. 

In total, we find that the surface of the rTCI can be symmetrically gapped except for a single zero-energy mode that is localized at the rotation center and protected by PHS. The rotation center also binds charge $\frac{1}{2}\mod(1)$. Treated as a $0$D system, the topological features of the rotation center match those of the $0$D surface of a $1$D SSH chain with additional trivial $\mathbb{Z}_4$ symmetry. Since the surface physics of the deformed rTCI and SSH chain are equivalent, the bulks of the two systems are also equivalent due to the bulk-boundary correspondence. We can also understand the disclination bound charge from this picture. For example, a $\pi/2$ disclination will remove one quarter of the $1/2$ charge, i.e., it removes $\tfrac{1}{4}\cdot\tfrac{1}{2}=\tfrac{1}{8}$ charge at the rotation center leaving $3/8.$ However, this would paradoxically imply the existence of a $1$D insulator with PHS and a charge polarization $3/8$. The resolution to this is that the disclination itself must also bind an additional $1/8$ charge. The total charge therefore remains as $1/2$, as expected for a $1$D insulator with PHS.


\subsection{Surface Topological Order}\label{ssec:SurfTopo}
If interactions are not present, the spinless rTCI does not support a fully gapped symmetric surface. Here we show that if interactions are included, the rTCI can support a fully gapped symmetric surface with topological order. This is similar to the topological orders that can be found on the surfaces of time-reversal invariant topological insulators~\cite{vishwanath2013physics,wang2013gapped,wang2014interacting,metlitski2015symmetry}.

We show here that the rTCI surface topological order is Abelian with anyon content $\{1,e,e^2,e^3,m,m^2,m^3,e^am^b\}\times\{1,f\}$ for $a,b = 1,2,3$. The $f$ particle is a fermion and the $e$ and $m$ anyons are self-bosons with $\pi/2$ mutual statistics. The surface topological order is enriched by $C_4$ rotation symmetry~\cite{lu2016classification} such that the anyons carry both charge and angular momentum. Specifically, the $e$ anyon has charge $\frac{1}{2}$ and angular momentum $0$ and the $m$ anyon has charge $0$ and angular momentum $\frac{1}{4}$. This topological order is anomalous for $2$D systems with PHS but can be realized on the surface of the $3$D rTCI with PHS. 

We use a vortex proliferation argument to construct this topologically ordered surface state~\cite{wang2013gapped}. The first step of this argument is to gap out the surface fermions by adding superconducting terms that break U$(1)$ charge conservation, $C_4$, rotation symmetry, TRS, PHS, and chiral symmetry. To accomplish this it is useful to rewrite the 4-component surface spinor $\psi$ from Eq.~\ref{eq:SurfHam} as $\psi = (\psi_1, \psi_2)$, where $\psi_{1,2}$ are two-component Dirac spinors. In terms of these spinors, the superconducting terms are
\begin{equation}\begin{split}
    \hat{\mathcal{H}}_{\text{SC}} = i \Delta_1 \psi_1 \sigma^y \psi_1 + i \Delta_2 \psi_2 \sigma^y \psi_2 + \text{H.c.}
\end{split}\end{equation}
Under the U$(1)$ and $C_4$ symmetries, $\Delta_{1,2}$ transform as
\begin{equation}\begin{split}
 &\text{U}(1): (\Delta_1,\Delta_2 )\rightarrow (\Delta_1 e^{i 2\theta},\Delta_2 e^{i 2\theta})\\
 &C_4:  (\Delta_1,\Delta_2 )\rightarrow (\Delta_1 e^{i \pi/2},\Delta_2 e^{-i \pi/2}).
\label{eq:SCSymmetryTrans} \end{split}\end{equation}
 The surface superconductivity therefore consists of a condensate of Cooper pairs with charge $2$ and angular momentum $\pm 1$. TRS and PHS act as $\mathcal{T}:\Delta_{1,2} \rightarrow \Delta^*_{2,1}$, and $\mathcal{C}:\Delta_{1,2} \rightarrow -\Delta^*_{2,1}$.

To restore the surface symmetries, we follow the procedure of vortex proliferation and disorder the superconducting terms, ($\langle \Delta_i\rangle = 0$). There are two types of vortices that we must consider. First are vortices where $\Delta_1$ and $\Delta_2$ both wind by $2\pi n $ ($n \in \mathbb{Z}$). Preempting our later identification of these vortices with the $m$ anyons of the theory, we refer to them as $2\pi n$ $m$-vortices. Second, are vortices where $\Delta_1$ winds by $-2\pi n$ and $\Delta_2$ winds by $2\pi n$, which we refer to as $2\pi n$ $e$-vortices. To understand why we must consider both $m$- and $e$-vortices, we note that if we proliferate only $m$-vortices the composite operator $\Delta_1 \Delta^*_2$ is not disordered (see Eq.~\ref{eq:SCSymmetryTrans}). This composite operator breaks $C_4$ symmetry, so the resulting surface would have U$(1)$ symmetry but not $C_4$ symmetry. A similar argument shows that only proliferating the $e$-vortices results in a surface state with unbroken $C_4$ symmetry and broken U$(1)$ charge conservation. 

With this in mind, we now ask what vortices can be condensed to restore the surface symmetry. Following the usual logic, the condensable vortices must be commuting bosons with vanishing quantum numbers. Additionally, in order for the resulting surface to be gapped, the condensed vortices must not have any protected zero-modes. To determine the quantum numbers and statistics of the $e$ and $m$-vortices, we first note that due to the symmetry transformations in Eq.~\ref{eq:SCSymmetryTrans}, a $2\pi$ $m$-vortex is created by a $\pi$ U$(1)$ flux~\footnote{This is a $\frac{hc}{2e}$ vortex in dimensionful units}, and a $2\pi$ $e$-vortex is generated by a $-2\pi$ disclination. Let us now imagine tunneling an electromagnetic monopole into the bulk of the rTCI. This process leaves behind a $2\pi$ U$(1)$ flux on the surface, equivalent to a $4\pi$ $m$-vortex. Based on our discussion of the RF term in Sec.~\ref{ssec:PhysicalImp} a magnetic monopole in the bulk of the spinless rTCI carries angular momentum $-\frac{1}{2}$. From conservation of angular momentum, the $4\pi$ $m$-vortex must have angular momentum $\frac{1}{2}$, and the $2\pi$ $m$-vortex must have angular momentum $\frac{1}{4}$. Similarly, a $-2\pi$ disclination monopole in the bulk carries charge $\frac{1}{2}$, and so a $2\pi$ $e$-vortex has charge $\frac{1}{2}$. Additionally, from the bulk braiding statistics of the flux and disclination lines, we conclude that both types of vortices are self-bosons and that a $2\pi$ $e$-vortex and a $2\pi$ $m$-vortex have $\pi/2$ mutual statistics. 

We will now determine the fate of the zero modes of the vortices. The $2\pi$ $e$ and the $2\pi$ $m$-vortices both host a single \textit{complex} fermion zero-mode. This comes from the fact that a $2\pi$ $m$-vortex is a combination of a $2\pi$ vortex of $\Delta_1$ and a $2\pi$ vortex of $\Delta_2$, each of which host a single Majorana zero mode~\cite{fu2008superconducting}. This complex zero-mode is protected from acquiring a gap by PHS. The same logic indicates that a $2\pi$ $e$-vortex also has a complex zero-mode that is protected by PHS. 

Based on these observations, the surface symmetry can be restored by simultaneously condensing the following two combinations of vortices: first, the combination of an $-8\pi$ $e$-vortex, an $-8\pi$ $m$-vortex, and a Cooper pair with charge $2$ and angular momentum $1$, and second, the combination of an $-8\pi$ $e$-vortex, a $8\pi$ $m$-vortex, and a Cooper pair with charge $2$ and angular momentum $-1$. Both of these combinations have vanishing charge, vanishing angular momentum, and trivial mutual statistics. Additionally, both vortex combinations have a total of $8$ complex fermions, and these fermions can be gapped out while preserving PHS. Under the first combination of vortices, $\Delta_1$ winds by $16\pi$, while $\Delta_2$ winds by $0$, and under the second combination $\Delta_1$ winds by $0$, while $\Delta_2$ winds by $16\pi$. Condensing both combinations of vortices therefore disorders $\Delta_1$, $\Delta_2$, and any composite operator descendants. The resulting surface state is therefore both gapped and symmetry-preserving. 

The resulting surface has several non-trivial deconfined excitations (anyons). First, there are the fermionic excitations that are the remnant of the gapped complex fermion zero modes. We label these excitations as $f$. The rest of the anyons correspond to vortices that have trivial statistics with the condensed vortex combinations. Here, both $2\pi n$ $e$-vortices and $2\pi n$ $m$-vortices (as well as the fusions of the two) are deconfined. We label the $2\pi$ $e$-vortex with an unoccupied complex zero-mode as the $e$ anyon (the $2\pi$ $e$-vortex with an occupied complex zero-mode as $e\times f$). Similarly, we label the $2\pi$ $m$-vortex with an unoccupied zero-mode as the $m$ anyon. Based on our earlier observations, the $e$ and $m$ anyons are self-bosons with $\pi/2$ mutual statistics. The $e$ particle has charge $\frac{1}{2}$ and angular momentum $0$, while the $m$ particle has charge $0$ and angular momentum $\frac{1}{4}$. Since the $e^4$ and $m^4$ anyons have trivial statistics with all other anyons and have non-fractionalized quantum numbers, they should be regarded as local particles and do not enter into the anyonic data of the theory.

Having established the existence of the surface topological order, we will now show that this surface topological order is anomalous with respect to PHS. First, consider a purely $2$D theory with the same anyon content as the surface topological order we constructed. The bosonic part of the topological order can be represented in the K-matrix formalism as~\cite{wen1992shift, wen2004quantum}
\begin{equation}\begin{split}
\mathcal{L}_{2\text{D-top}} &= K_{IJ}\frac{\epsilon^{\mu\nu\rho}}{4\pi} a^I_\mu \partial_\nu a^J_\rho + \frac{\epsilon^{\mu\nu\rho}}{4\pi} q_I a^I_\mu \partial_\nu A_\rho\\
 &+ \frac{\epsilon^{\mu\nu\rho}}{4\pi} s_I a^I_\mu \partial_\nu \omega_\rho,
\end{split}\end{equation}
where $K = 4\sigma^x$, $q_I = (2,0)$, and $s_I = (0,1)$. This $2$D theory is not consistent with PHS, as can be seen from the fact that integrating out the dynamic gauge fields $a^I_\mu$ produces a Wen-Zee term
\begin{equation}\begin{split}
\mathcal{L}_{2\text{D-top}} = \frac{\epsilon^{\mu\nu\rho}}{4\pi} \omega_\mu \partial_\nu A_\rho,
\end{split}\end{equation}
which breaks PHS. 

For an alternative perspective, let us assume that there is a purely $2$D lattice system with the same topological order as the gapped rTCI surface. For such a system, we can consider the instanton process where a $2\pi$ U$(1)$ flux is adiabatically inserted in a local region~\cite{metlitski2015symmetry}. This instanton event is a local process for lattice systems. However, the $e$ anyons have charge $\frac{1}{2}$ and hence pick up an Aharonov-Bohm phase of $-1$ upon encircling the flux. The resolution to this seeming paradox is that the instanton event must bind an anyon that has $\pi$ mutual statistics with the $e$ anyon. This anyon must be the $m^2$ anyon. However, the $m^2$ anyon has angular momentum $\frac{1}{2}$. If angular momentum is conserved, the instanton event must therefore be accompanied by a flow of angular momentum current. The fact that inserting an electromagnetic flux drives an angular momentum current indicates that the $2$D lattice system must necessarily break PHS (e.g., this is exactly what the Wen-Zee term would generate).

Now, consider the same instanton event when the topological order is defined on the surface of the rTCI. Here, the flux insertion on the surface is accompanied by a monopole tunneling event in the bulk of the rTCI. As before, the flux insertion on the surface binds the $m^2$ anyon, which has angular momentum $\frac{1}{2}$. Additionally, due to the mixed Witten effect in the bulk of the rTCI, the monopole has angular momentum $-\frac{1}{2}$. So, the angular momentum of the full $3$D system is conserved during the instanton event and there is no need for an accompanying angular momentum current on the surface. The topological order is therefore consistent with PHS when placed on the surface of the rTCI.

\section{rotation-invariant Topological Crystalline Insulators with Mirror Symmetry}\label{sec:MirrorrTCI}
In this section, we discuss time-reversal invariant rTCIs where the $RF$-term is quantized by mirror symmetry instead of PHS. Much of the bulk physics of the rTCIs with mirror symmetry is the same as that of the rTCIs with particle-hole symmetry, which we discussed in detail in Sec.~\ref{sec:PHSrTCI}. However, as noted in Sec.~\ref{ssec:SymQuantMirror}, unlike the rTCI with PHS, the surfaces of the rTCI with mirror symmetry can be gapped while preserving mirror symmetry, and without introducing surface topological order. However, these symmetric gapped surfaces still carry an anomaly, as a $2\pi/n$ surface disclination binds charge $1/2n$ in spinless systems, and charge $1/n$ in spin-1/2 systems (half the amount that is allowed in purely $2$D systems). In this section, we will only consider spin-1/2 insulators, since they are more relevant to real materials. We include the analysis for spinless fermions with mirror symmetry in Appendix~\ref{app:spinlessMirrorrTCI}. 

Before we begin, we note that rTCIs with mirror symmetry along the z-axis and  $C_2$ rotation symmetry along the z-axis also have inversion symmetry. However, the models we study are trivial in terms of the classification of inversion symmetric topological insulators~\cite{fu2007topologicalIn} and higher-order topological insulators with hinge modes~\cite{benalcazar2017electric,schindler2018higher,khalaf2018higher}. This is because the topological properties of these systems are due to the rotation and mirror symmetries \textit{separately}, not inversion symmetry. In addition to the analysis of a specific model, in  Sec.~\ref{sec:GenbRespose} we construct appropriate topological invariants for more generic mirror symmetric rTCIs.

\subsection{Lattice Model with Spin-1/2 Fermions} 
In this subsection we present a lattice model for the spin-1/2 rTCI with TRS and mirror symmetry $C_4$ rotation symmetry. The spin-1/2 rTCI is realized by the following a $16$-band model ($8$-bands per spin):
\begin{equation}\begin{split}
\mathcal{H}(\bm{k}) &= \Bigl[\sin(k_x)\Gamma^x\sigma^0 + \sin(k_y)\Gamma^y\sigma^0 + \sin(k_z)\Gamma^z\sigma^0\\
&+\sin(k_x)\sin(k_z) \Gamma^{0}\sigma^x+ \sin(k_y)\sin(k_z) \Gamma^{0} \sigma^y\\
  &+(M+\cos(k_x)+\cos(k_y)+\cos(k_z))\Gamma^0\sigma^z\Bigr]\sigma^0.
\label{eq:LatticeHamSpin1/2Mirror}\end{split}\end{equation}
The spin of the fermions is given by $S^z = \frac{1}{2}\text{I}\sigma^0\sigma^z$. Eq.~\ref{eq:LatticeHamSpin1/2Mirror} conserves charge and is invariant under TRS, mirror symmetry and $C_4$ rotation symmetry. The TRS operator is given by
\begin{equation}\begin{split}
&\hat{\mathcal{T}} = i\Gamma^y \sigma^y \sigma^y K,
\end{split}\end{equation}
mirror reflection is defined as
\begin{equation}\begin{split}
\hat{\mathcal{M}}_z = i\Gamma^{5z} \sigma^0\sigma^z,
\end{split}\end{equation}
where $\hat{\mathcal{M}}_z^{-1} \mathcal{H}(k_x,k_y,k_z) \hat{\mathcal{M}}_z = \mathcal{H}(k_x,k_y,-k_z)$, and $C_4$ rotation is defined as
\begin{equation}\begin{split}
\hat{U}_4 &= \exp(i \frac{\pi}{4}[ \Gamma^{yx}\sigma^0\sigma^0 + \text{I}\sigma^z\sigma^0 + \text{I}\sigma^0\sigma^z]).
\label{eq:C4DefSpin1/2}\end{split}\end{equation}
Here, $\hat{\mathcal{T}}^2 = (\hat{\mathcal{M}}_z)^2 = (\hat{U}_4)^4= - 1 $, since the fermions have spin-1/2. 

The lattice model in Eq.~\ref{eq:LatticeHamSpin1/2Mirror} also has PHS given by
\begin{equation}\begin{split}
\hat{\mathcal{C}} = \Gamma^{5y} \sigma^y \sigma^y \mathcal{K}.
\end{split}\end{equation}
However, this PHS is not relevant to our discussion and should be regarded as an ``accidental'' symmetry of the lattice model. 


The spectrum of the spin-1/2 lattice model is 8-fold degenerate but otherwise the same as in Eq.~\ref{eq:latticeSpectrum}, and hence is gapped for $|M|\neq 1,3$. Below we show that this lattice model realizes a spin-1/2 rTCI with a $\Phi = 2\pi$ $RF$-term for $1<|M|<3$.


\subsection{Response Theory}\label{ssec:ResponseThSpin1/2Mirror}
To derive the response theory for the spin-1/2 model, we follow the methodology used in Sec.~\ref{ssec:ResponseSpinless} and consider the system close to the band crossing at $M = -3.$ Near this point in the phase diagram  the low-energy degrees of freedom have a Dirac-like form:
\begin{equation}\begin{split}
\mathcal{H} &= \left[\Gamma^x  \sigma^0  i \partial_x + \Gamma^y  \sigma^0  i\partial_y +\Gamma^z  \sigma^0  i \partial_z +m \Gamma^0  \sigma^z \right] \sigma^0, 
\label{eq:LatticeHamLowEnergySpin1/2}\end{split}\end{equation}
where $m \sim M + 3$ controls the transition between the $M<-3$ trivial phase, and the $-3<M<-1$ phase of the lattice model. To determine the effective response theory, we gauge the U$(1)$-charge and $C_4$-rotation symmetries and couple the fermions to the gauge field $A_\mu$ and spin connection $\omega_\mu$ via the covariant derivative (see Appendix~\ref{app:SpinConnectCoup}):
\begin{equation}\begin{split}
D_\mu &=   \partial_\mu - i A_\mu \\
&- i \frac{1}{2}\omega_\mu [\Gamma^{xy}\sigma^0\sigma^0 + \text{I}\sigma^z\sigma^0 + \text{I}\sigma^0\sigma^z ]. 
\label{eq:CovariantDerivSpin1/2}\end{split}\end{equation}
Similar to before, the $C_4$ rotation symmetry of Eq.~\ref{eq:LatticeHamLowEnergySpin1/2} is actually part of an enlarged U$(1)$ rotation symmetry. In addition to the gauge fields, we also include a perturbation
\begin{equation}\begin{split}
\mathcal{H}' =  m' \Gamma^5 \sigma^0\sigma^0,
\label{eq:mass2spin1/2}\end{split}\end{equation}
and set $m = -\bar{m}\cos(\phi)$, and $m' = -\bar{m}\sin(\phi)$, with $\bar{m}>0$, such that $m<0$ when $\phi = 0$, and $m>0$ when $\phi = \pi$. If $\phi$ is promoted to be a function of $z$, mirror symmetry is preserved only if $\phi(z) = -\phi(-z)$ mod$(2\pi)$. If $\phi$ is constant (as it should be in the interior of an insulator) mirror symmetry requires that $\phi = 0$, or $\pi$.

The effective response theory is obtained by a diagrammatic expansion in terms of $A_\mu$, $\omega_\mu$, and $\phi$, and we are again primarily interested in the triangle diagrams depicted in Fig.~\ref{fig:TDia}. The contribution from the triangle diagrams is
\begin{equation}\begin{split}
\mathcal{L}_{\text{eff}} = \frac{\phi}{2\pi^2} \epsilon^{\mu\nu\rho\kappa} \partial_\mu \omega_\nu \partial_\rho A_\kappa, 
\label{eq:EffDerivedPhiSpin1/2}\end{split}\end{equation}
which differs by a factor of $2$ from the effective response of the spinless lattice model, Eq.~\ref{eq:EffDerivedphi}. For $\phi = \pi$ (constant), the effective response is,
\begin{equation}\begin{split}
\mathcal{L}_{\text{eff}}= \frac{1}{2\pi} \epsilon^{\mu\nu\rho\kappa} \partial_\mu \omega_\nu \partial_\rho A_\kappa,
\label{eq:EffDerivedPhiSpin2}\end{split}\end{equation}
which is exactly the $RF$-term in Eq.~\ref{eq:BulkResponses} with $\Phi = 2\pi$. Following the same logic from Sec.~\ref{sec:ResponseTh}, the $RF$-term vanishes for $M<-3$ and the coefficient of the $RF$-term is $\Phi = 2\pi$, for $-3<M<-1$. Repeating this procedure for the band crossing at $M = \pm 1,3$ we conclude that $\Phi = 2\pi$ for $1<|M|<3$ and vanishes otherwise. 

 As noted before, when the coefficient of the $RF$-term is non-constant, mirror symmetry requires that $\Phi(z) = -\Phi(-z)$ mod$(4\pi)$. Because of this, it is possible to have mirror symmetry preserving domain walls between the rTCI and a trivial insulator, as we show in the next subsection. We also note that for Eq.~\ref{eq:LatticeHamLowEnergySpin1/2} and~\ref{eq:mass2spin1/2}, increasing $\phi$ from $0$ to $2\pi$ is a periodic process. During this process a $\Phi = 4\pi$ $RF$-term is generated, which agrees with our earlier conclusion that for spin-1/2 systems with TRS, $\Phi$ is $4\pi$ periodic. 

\subsection{Surface Theory}
To analyze the surface theory of the rTCI with mirror symmetry, we will use a pair of domain walls that are related to one another by mirror symmetry. Specifically, consider a geometry where $-3 <M < 1$ for $|z| < z_{\text{dw}}$ and $M < -3$ for $|z| > z_{\text{dw}}$, which corresponds to a pair of symmetry related domain walls at $z = \pm z_{\text{dw}}$ ($z_{\text{dw}}$ is taken to be large compared to the correlation length of the insulators). The Hamiltonian for the two surfaces is
\begin{equation}\begin{split}
&\mathcal{H}_{\text{t}} = \left[\sigma^x i\partial_x - \sigma^y i\partial_y\right]\sigma^0\sigma^0,\\ &\mathcal{H}_{\text{b}} = \left[\sigma^x i\partial_x - \sigma^y i\partial_y\right]\sigma^0\sigma^0,
\label{eq:SurfHamMirrorSpin1/2}\end{split}\end{equation} or, equivalently,
\begin{equation}\begin{split}
\mathcal{H}_{\text{t-b}} &= \left[\sigma^x i\partial_x - \sigma^y i\partial_y\right]\sigma^0\sigma^0\sigma^0,
\label{eq:SurfHamMirror2Spin1/2}\end{split}\end{equation}
where the two domain walls are indexed by $\sigma^0\sigma^0\sigma^0\sigma^z$. 
The mirror symmetry acts on Eq.~\ref{eq:SurfHamMirror2Spin1/2} as
\begin{equation}\begin{split}
&\hat{M}_{z-\text{surf}} = \sigma^0 \sigma^0 \sigma^z \sigma^x,
\end{split}\end{equation}
while TRS and $C_4$ symmetry act on Eq.~\ref{eq:SurfHamMirror2Spin1/2} as
\begin{equation}\begin{split}
&\hat{\mathcal{T}}_{\text{surf}} = \sigma^y  \sigma^y \sigma^y \sigma^0 K,\\
&\hat{U}_{4-\text{surf}} = \exp\left[i\frac{\pi}{4}\left(-\sigma^z  \sigma^0\sigma^0  + \sigma^0 \sigma^z \sigma^0 + \sigma^0 \sigma^0  \sigma^z \right)\right]\sigma^0.
\label{eq:spin1/2sym}\end{split}\end{equation}
Eq.~\ref{eq:SurfHamMirror2Spin1/2} has two surface mass terms of note: (i) a mass term proportional to $\sigma^z \sigma^z \sigma^0\sigma^z$ that preserves TRS and breaks mirror symmetry, and (ii) a mass term proportional to $\sigma^z \sigma^z \sigma^0\sigma^0$ that preserves TRS and mirror symmetry. Hence, in agreement with our discussion from Sec.~\ref{ssec:SymQuantMirror}, we find that the surface Dirac fermions are not protected by mirror symmetry since they can be gapped with the second mass term while preserving mirror. 

The response theory of the gapped surfaces is found by coupling the surface Dirac fermions to the U$(1)$ gauge field $A_\mu$ and spin connection $\omega_\mu,$ and then integrating out the massive fermions using a diagrammatic expansion. The expansion contains a topological Wen-Zee term, which corresponds to the one-loop diagram in Fig.~\ref{fig:WZDia}. For the mirror symmetry \emph{breaking} surface mass, one surface hosts a Wen-Zee term having coefficient $1/2\pi$ and the other hosts a Wen-Zee term having coefficient $-1/2\pi$. A $\pi/2$ disclination of the rTCI with mirror symmetry therefore binds charge $\frac{1}{4}$ on one surface, and charge $-\frac{1}{4}$ on the other surface (modulo local contributions). 

For the mirror symmetry \emph{preserving} surface mass, both surfaces host a Wen-Zee term with the same coefficient $\pm 1/2\pi$, and a $\pi/2$ disclination binds charge $\pm \frac{1}{4}$ on \textit{both} surfaces. Mirror symmetry preserving surface effects can shift the coefficient of the Wen-Zee term by $1/\pi$ on \textit{both} surfaces. Fractional charges are therefore bound to disclinations of the massive surfaces of the rTCI regardless of whether the surfaces break or preserve mirror symmetry, and the charge bound to the surface disclination is half the amount that is allowed in purely $2$D systems. Hence, the fractional charge bound to surface disclinations is a robust indicator of the topology of the rTCI with TRS and mirror symmetry, even in the absence of symmetry protected surface states.

It should be noted that since the rTCI with mirror symmetry hosts symmetric, non-interacting gapped surfaces, it is not necessary to include additional interactions to generate gapped topologically ordered surfaces, as we did in Sec.~\ref{ssec:SurfTopo}. 

\subsection{Numerics}

In this subsection we numerically analyze the lattice model of the spin-1/2 rTCI with mirror symmetry. As noted previously, the mirror symmetric rTCI admits a mirror symmetry preserving gapped surface. For a lattice model with periodic boundaries along the x and y-directions and open boundaries along the z-direction, the mirror symmetric mass terms are
\begin{equation}\begin{split}
\mathcal{H}_{s} &= \sum_{r\in \text{+z surface}}m_s c^\dagger(r)\Gamma^5\sigma^0\sigma^0c(r) \\&- \sum_{r\in \text{-z surface}}m_s c^\dagger(r)\Gamma^5\sigma^0\sigma^0c(r).
\label{eq:LatticeSurfMassSpin1/2Mirror}
\end{split}\end{equation}
Here, $c^\dagger(r)$ is the $8$-component fermion creation operator at site $r$, and the first (second) sum is over sites on the top (bottom) surface of the lattice. The density of states of the system with the mirror symmetric surface perturbation shows no midgap states, and is the same as the density of states in shown Fig.~\ref{fig:dos_no_phs}, up to a factor of 2 because of the spin of fermions. As we discuss in Sec.~\ref{ssec:octopole}, when the lattice has open boundaries in all directions, the mirror symmetric gapped surface has corner-vertex charges and a filling anomaly~\cite{benalcazar2019quantization, khalaf2021boundary}. 

It is also possible to have gapped surfaces without a filling anomaly if we instead use a mirror symmetry breaking surface mass term of the form. 
\begin{equation}\begin{split}{}
\mathcal{H}_{s} = m_s \sum_{r\in \text{surface}}c^\dagger(r)\Gamma^5\sigma^0\sigma^0c(r),
\label{eq:LatticeSurfMassSpin1/2MB}
\end{split}\end{equation}
where $m_s$ is constant, and the sum is taken over all boundaries of the system. This surface mass term produces a system with an identical spectrum to that of Eq.~\ref{eq:LatticeSurfMassSpin1/2Mirror}.


We now turn our attention to the charge that is bound to the gapped surfaces of the spin-1/2 rTCI with a disclination. As discussed, the spin-1/2 rTCI admits both mirror symmetry preserving and mirror symmetry breaking gapped surfaces. However, the mirror symmetry preserving gapped surfaces lead to a filling anomaly~\cite{benalcazar2019quantization} such that the insulator is not charge neutral (including the negative ionic contribution) when the chemical potential is in the gap. In contrast, when mirror symmetry is broken at the surface, the insulator can be charge neutral and free from the filling anomaly when the chemical potential is in the gap. Since we are only interested in charges that arise from disclinations, it is convenient to use mirror symmetry breaking surface mass terms here.

\begin{figure}
    \includegraphics[width=0.5\textwidth]{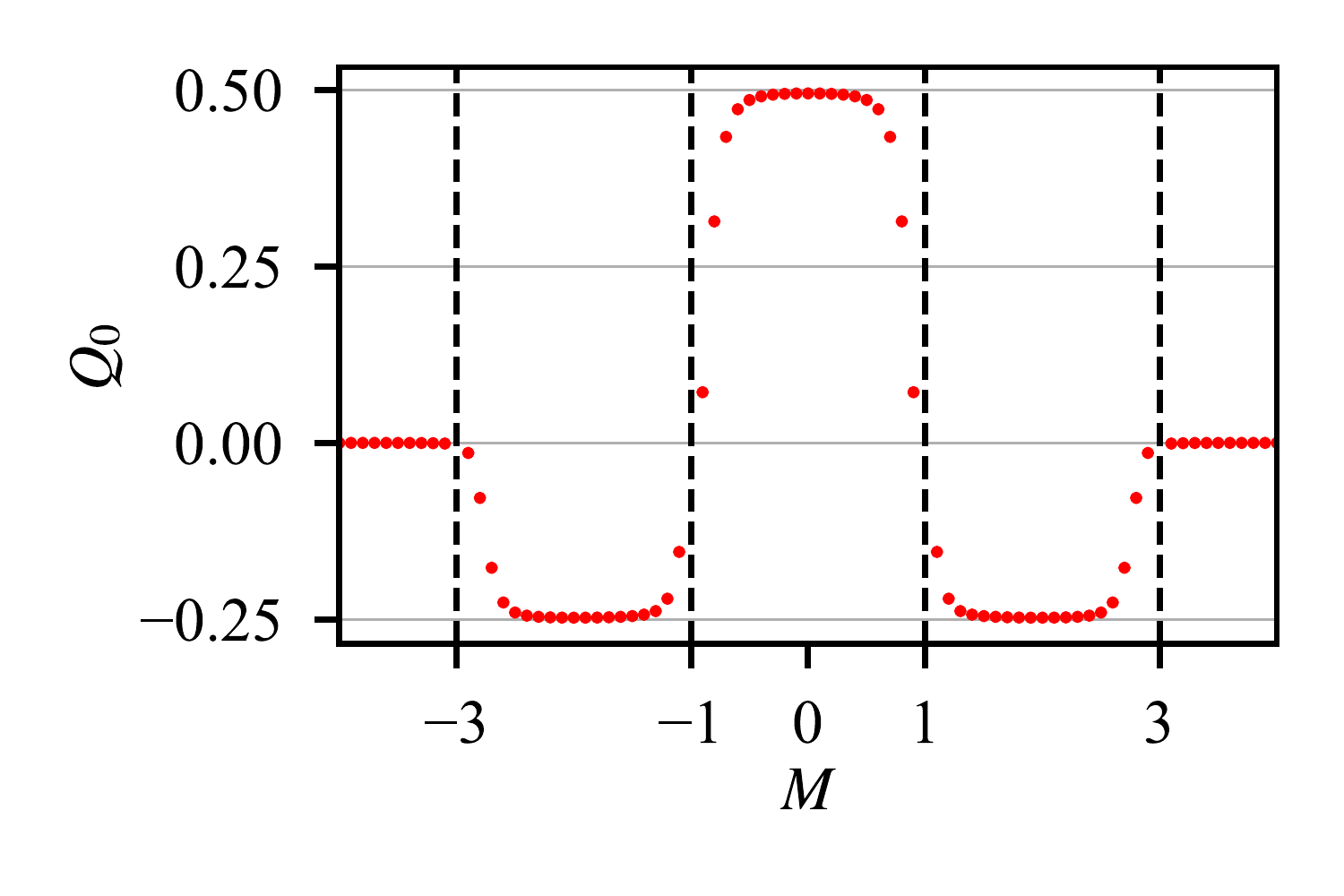}
    \caption{The surface charge $Q_0$ bound to a $\pi/2$ site-centered disclination as a function of the mass $M$. The charge is calculated using the Hamiltonian in Eq.~\ref{eq:LatticeHamSpin1/2Mirror} and~\ref{eq:LatticeSurfMassSpin1/2MB} on a $15\times15\times15$ lattice with open boundaries and a mirror symmetry breaking mass $m_s$ equal to the bulk mass gap. As in the spinless case, the deviations of the charge from the quantized values around the gap closing points $|M|=1, 3$ arise from finite size effects.}\label{fig:surfaceChargeVsMSpin1/2M}
\end{figure}

The surface charge of the spin-1/2 lattice model with a $\pi/2$ disclination and mirror symmetry breaking surface mass term (Eq.~\ref{eq:LatticeSurfMassSpin1/2MB}) is shown in Fig.~\ref{fig:surfaceChargeVsMSpin1/2M}. When $1<|M|<3,$ the disclinated surface has an extra $\frac{1}{4}$ mod ($\frac{1}{2}$) surface charge (up to finite size corrections), indicating that the lattice model is a spin-1/2 rTCI with a non-trivial $RF$-term in this regime. When $|M|>3$ and $|M|<1$, the surface charge is $0$ (mod $\frac{1}{2}$), indicating that the $RF$-term is trivial in these regimes. These results are in full agreement with our previous analysis.

\subsection{Dimensional Reduction to $1$D SPT}\label{ssec:DimRedSpin1/2}
The spin-1/2 rTCI with mirror symmetry can be dimensionally reduced to a $1$D spin-1/2 SSH chain protected by mirror symmetry and an on-site $\mathbb{Z}_4$ spin rotation symmetry. The boundaries of the spin-1/2 SSH chain host a Kramers pair of zero energy modes, and have charge $-1$ ($+1$) when both of the zero modes are empty (filled)~\cite{kivelson2001electron}. Filling only one of the zero modes results in a boundary with spin $\pm 1/2$ and charge $0$.  For a pair of mirror symmetry related boundaries, it is possible to have TRS and mirror symmetry preserving gapped boundaries where the zero energy modes at each boundary are all empty or all filled. The gapped boundaries carry the same charge $\pm 1$, and spin $0$.

To show this, we again take the two surface Hamiltonians in Eq.~\ref{eq:SurfHamMirrorSpin1/2}, and add mass perturbations of the form
\begin{equation}\begin{split}
&\mathcal{H}_{\text{t-mass}} =  \left[m_{x,\text{t}} \sigma^z\sigma^x + m_{y,\text{t}} \sigma^z\sigma^y + m_{z,\text{t}} \sigma^z \sigma^z \right]\sigma^0,\\ &\mathcal{H}_{\text{b-mass}} = \left[m_{x,\text{b}} \sigma^z\sigma^x + m_{y,\text{b}} \sigma^z\sigma^y + m_{z,\text{b}} \sigma^z \sigma^z \right]\sigma^0.
\label{eq:SurfHamMirrorMassSpin1/2}\end{split}\end{equation}
Under mirror symmetry, $m_{i,\text{t}} \rightarrow m_{i,\text{b}}$, for $i = x,y,z$. Following our discussion in Sec.~\ref{sec:DimRed}, the spin-1/2 rTCI can be dimensionally reduced to an SSH chain with gapless edge modes by setting $m_{z,\text{t}} = m_{z,\text{b}} = 0$ and $m_{x,\text{t}} + im_{y,\text{t}} =m_{x,\text{b}} + im_{y,\text{b}}=  m_s(r) \exp(i\theta)$, where $(r,\theta)$ are polar coordinates on the surface and $m_s(r) \geq 0$ is a function of the radial coordinate that vanishes at $r = 0$ and goes to a non-zero constant $\bar{m}_s$, as $r\rightarrow \infty$. This mass configuration preserves TRS, mirror symmetry, and $C_4$ rotation symmetry. 

With this mass configuration there are two localized zero-energy modes, $\psi_{0\uparrow, \text{t/b}}$, and $\psi_{0\downarrow, \text{t/b}}$ , at the rotation center of each surface  (see Appendix~\ref{app:ZeroEnergy}). The zero-energy modes of each surface form a Kramers pair with spin $\pm 1/2.$ The zero-energy subspace is acted upon with the symmetries as:
\begin{equation}\begin{split}
&\mathcal{T}: (\psi_{0\uparrow, \text{t/b}},\psi_{0\downarrow, \text{t/b}}) \rightarrow (\psi_{0\downarrow, \text{t/b}},-\psi_{0\uparrow, \text{t/b}})\\
&C_4: (\psi_{0\uparrow, \text{t/b}},\psi_{0\downarrow, \text{t/b}})\rightarrow (\psi_{0\uparrow, \text{t/b}} e^{i\frac{\pi}{4}},\psi_{0\downarrow, \text{t/b}} e^{-i\frac{\pi}{4}}).
\end{split}\end{equation}
Interestingly, it is possible to gap out the edge modes of the SSH chain by instead setting $m_{z,\text{t}} = m_{z,\text{b}} = \sqrt{\bar{m}^2_s - m_s(r)^2}$, such that $m_{z,\text{t}}$ and $m_{z,\text{b}}$ take on the same non-quantized value near $r = 0$. This perturbation preserves all symmetries of the spin-1/2 rTCI and gaps out the zero modes located at $r = 0$ on each surface. 

To determine the charge that is bound at $r = 0$, we integrate out the massive fermions, leading to the effective response theory
\begin{equation}\begin{split}
\mathcal{L}_{\text{eff-t}} &= \frac{\epsilon^{\mu\nu\rho}}{4\pi} \bm{n}_{\text{t}}\cdot(\partial_\mu \bm{n}_{\text{t}}\times \partial_\nu \bm{n}_{\text{t}})A_\rho+ \frac{n_{z,t}}{4\pi}\epsilon^{\mu\nu\rho} \omega_\mu \partial_\nu  A_\rho ,\\
\mathcal{L}_{\text{eff-b}} &= \frac{\epsilon^{\mu\nu\rho}}{4\pi} \bm{n}_{\text{b}}\cdot(\partial_\mu \bm{n}_{\text{b}}\times \partial_\nu \bm{n}_{\text{b}})A_\rho+ \frac{n_{z,b}}{4\pi}\epsilon^{\mu\nu\rho} \omega_\mu \partial_\nu  A_\rho ,\\
\bm{n}_{\text{t}/\text{b}} &= \frac{\bm{m}_{\text{t}/\text{b}}}{|\bm{m}_{\text{t}/\text{b}}|},\phantom{=} \bm{m}_{\text{t}/\text{b}} = (m_{x,\text{t}/\text{b}},m_{y,\text{t}/\text{b}},m_{z,\text{t}/\text{b}}). 
\end{split}\end{equation}
For the mass configurations discussed above, the response theory indicates that charge $-1$ is localized near $r = 0$ on both the top and bottom surfaces. It is also possible to have a gapped surface with charge $+1$ localized near $r = 0$ by changing the signs of $m_{z,\text{t}}$ and $m_{z,\text{b}}$. Viewed as two $0$D systems, the rotation centers of the top and bottom surfaces each have a Kramers pair of unprotected modes, and carry charge $- 1$ ($+1$) when the zero modes are empty (filled).  These are exactly the characteristic features of the $0$D surfaces of the spin-1/2 SSH chain with mirror symmetry, and, using the bulk-boundary correspondence, we conclude that the spin-1/2 rTCI and spin-1/2 SSH chain with mirror symmetry are equivalent. We can understand the bound disclination charge from this picture as well. For example, a $\pi/2$ disclination will remove one quarter of the $-1$ charge leaving a total charge of $-\tfrac{3}{4}.$ Since such a $1$D chain should have an integer charge polarization we expect the disclination itself to carry the compensating $-\tfrac{1}{4}$ charge.

Since the mirror preserving gapped surfaces \textit{both} carry charge $\pm 1$, the net charge of the rTCI with mirror symmetry preserving gapped surfaces is $\pm 2$. The fact that the symmetrically gapped surface has a net charge indicates that the rTCI with mirror symmetry has a filling anomaly~\cite{benalcazar2019quantization}. This filling anomaly also occurs for the spin-1/2 SSH chain with mirror symmetry~\cite{khalaf2021boundary}.

The spin-1/2 SSH chain with mirror symmetry can also be further dimensionally reduced to a non-trivial $0$D spin-1/2 system with on-site $\mathbb{Z}_2$ symmetry, which is inherited from the mirror symmetry~\cite{khalaf2021boundary}. By extension, the spin-1/2 rTCI can also be further dimensionally reduced to the same $0$D system, with an on-site $\mathbb{Z}_4$ spin rotation symmetry.

\subsection{Octopole Insulator}\label{ssec:octopole}

Interestingly, the rTCI with mirror symmetry can host  a pattern of gapped surfaces that generate corner charges. This surface mass configuration turns the rTCI into an octopole insulator~\cite{benalcazar2017quantized}. Heuristically, we can see this by considering the rTCI defined on a lattice  with open boundaries in all directions. Based on our previous discussion, this system admits a surface mass configuration that fully gaps each surface and that binds the same charge $\pm 1$ to the rotation center on the +z and -z surfaces. To create an octopole insulator, we fractionalize the $\pm 1$ on each surface charge into four charges of $+1/4$ or four charges of $-1/4$ and then we move the charges to the vertices of the cubic lattice surface. This can be done smoothly and without breaking any symmetries. The resulting system is an octopole insulator with the same corner charge of either all $+1/4$ or all $-1/4$ located at each vertex of the surface. The resulting insulator has net charge $\pm 2$ which matches the previously mentioned filling anomaly that is present when mirror-symmetry is preserved. 

\begin{figure}
    \includegraphics[width=0.5\textwidth]{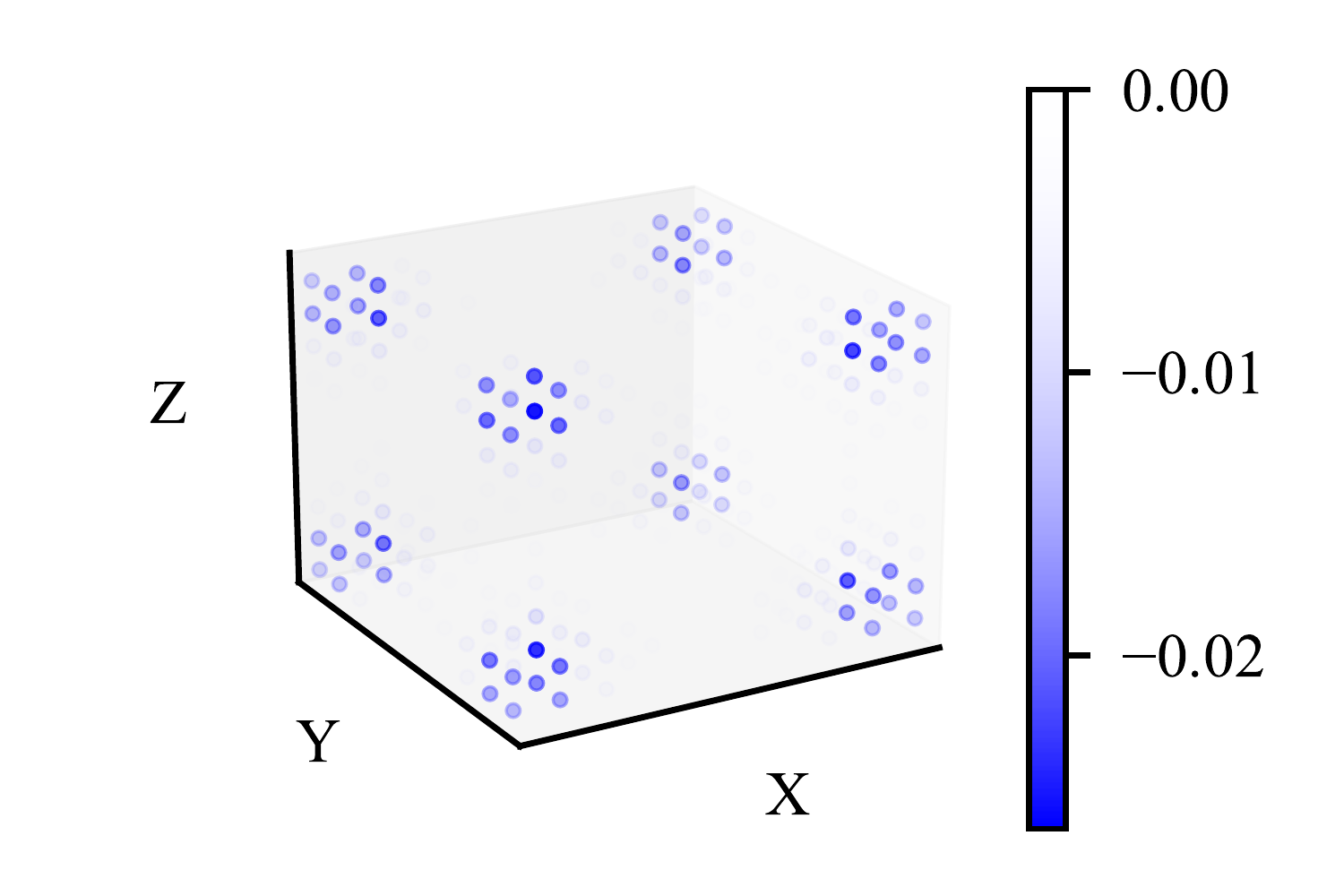}
    \caption{The surface charge configuration for the lattice Hamiltonian in Eq.~\ref{eq:LatticeHamSpin1/2Mirror} with mirror symmetry preserving surface mass terms in Eq.~\ref{eq:OctoSurfMassSpin1/2Mirror} on a $10 \times 10 \times 10$ lattice with $M = -2$, $m_s = 1$. A background charge of $-8$ per site is added such that the bulk of the insulator is charge neutral. A fractional charge $-1/4$ is bound to each corner.}
\label{fig:OctoChargeConfig}
\end{figure}

In terms of the microscopic model, the octopole insulator is found by adding the following terms to the surfaces of the rTCI with mirror symmetry,
\begin{equation}\begin{split}
\mathcal{H}_{s} &= \sum_{r\in \pm\text{z surface}} \pm m_s c^\dagger(r)\Gamma^5\sigma^0\sigma^0c(r) \\
&+\sum_{r\in \pm \text{x surface}} \pm m_s c^\dagger(r)\Gamma^{0}\sigma^x\sigma^0c(r) \\
&+ \sum_{r\in \pm \text{y surface}} \pm m_s c^\dagger(r)\Gamma^{0}\sigma^y\sigma^0c(r),
\label{eq:OctoSurfMassSpin1/2Mirror}
\end{split}\end{equation}
where the sum is taken such that $+m_s$ is used for the fermions on the $+$x,$+$y,$+$z surfaces and $-m_s$ is used for the fermions on the $-$x,$-$y,$-$z surfaces. As shown in Fig.~\ref{fig:OctoChargeConfig} these surface mass terms lead to the octopole insulator with charge $-1/4$ localized on each vertex of the cubic lattice when $m_s > 0$. The charge at each vertex is $+1/4$ when $m_s < 0$. It should be noted that these quantized charges are not fixed to the vertices by the rotation and mirror symmetry alone. For example, it is possible to move the charges such that charge $\pm 1$ is localized at the points where the rotation axis intersects the surface of the rTCI (as in Sec.~\ref{ssec:DimRedSpin1/2}). Equivalently, it is possible to move the charges such that charge $\pm 1/2$ is localized at the corners where the mirror plane intersects the hinges of the surface of the rTCI. 

\section{Symmetry Eigenvalue Formulas for \lowercase{r}TCI Invariants}\label{sec:GenbRespose}

In Sec.~\ref{sec:MirrorrTCI} and Appendix~\ref{app:spinlessMirrorrTCI} we determine the coefficient of the $RF$-term, $\Phi$, using linear response theory for lattice models with $C_4$ symmetry, mirror symmetry, and TRS. In this section we construct a symmetry indicated version of this topological invariant that distinguishes the topological $(\Phi \neq 0)$ and trivial phases $(\Phi = 0)$ phases of mirror symmetric rTCIs with additional inversion symmetry. We consider two generic classes of insulators in the following sections: 
\begin{enumerate} 
  \item Spinless insulators with TRS, $C_n$ symmetry, mirror symmetry, and inversion symmetry.
  \item Spin-1/2 insulators with conserved spin, TRS, $C_n$ symmetry, mirror symmetry and inversion symmetry. 
\end{enumerate}
Notably, this list excludes spin-1/2 insulators with spin-orbit coupling. Determining an analogous formula for such systems remains an open question. It should be noted that the combination of mirror and inversion symmetry leads to $C_2$ symmetry around the same axis as the $C_n$ symmetry. This $C_2$ symmetry is redundant for $n = 2,4,6$, but enlarges the $C_3$ rotation symmetry to $C_6$. So, the above classes effectively consider $C_n$ rotations with only $n = 2,4,6$.

\subsection{Spinless fermions}\label{ssec:InvariantSpinless}
Here we construct an invariant for spinless insulators with TRS ($\mathcal{T}^2 = 1$), $C_n$ symmetry, mirror symmetry, and inversion symmetry. As noted previously, these insulators necessarily have a $C_2$ rotation subgroup. The invariant is constructed in terms of the symmetry eigenvalues of the occupied bands at the time-reversal invariant momenta (TRIM) of the Brillouin zone, i.e., lattice momenta $\bm{k}$ such that $\bm{k} = -\bm{k}$ modulo a reciprocal lattice vector. Since the TRIM are invariant under $C_2$ rotations, we consider only $C_2$ rotations in this section, with the implicit understanding that the $C_2$ rotations may be part of a larger rotation group. In Appendix~\ref{app:InvariantEx} we discuss why it is sufficient to only consider $C_2$ rotations when considering $C_n$ rTCIs with $n = 4$ or $6$. 

Before we define the invariant, it is necessary to go over some preliminary details. Take a generic insulator with the above symmetries. When restricted to the mirror invariant plane $k_z = 0$ or $k_z = \pi$, the occupied bands have an $\hat{\mathcal{M}}_z$ eigenvalue that is independent of the $k_x$- and $k_y$- components of the momentum. We can define the Chern number parity of a band restricted to a mirror invariant plane as~\cite{hughes2011inversion}
\begin{equation}\begin{split}
(-1)^{C_i[k_{z}]} &= \chi_i(0,0,k_{z})\chi_i(0,\pi,k_{z})\chi_i(\pi,0,k_{z})\chi_i(\pi,\pi,k_{z}) \\ &= \zeta_i(0,0,k_{z})\zeta_i(0,\pi,k_{z})\zeta_i(\pi,0,k_{z})\zeta_i(\pi,\pi,k_{z})
\label{eq:ChernNumberC2}\end{split}\end{equation}
where $k_z = 0$ or $\pi$, $i$ is a band index, and $\chi_i = \pm 1$, $\zeta_i = \pm 1$  are the $C_2$  and inversion eigenvalues, respectively, of the $i$-th band at a given high-symmetry point. We can therefore label each band by its mirror eigenvalue and Chern number parity at each mirror invariant plane. 

If we restrict our attention to the $k_z = 0$ ($k_z = \pi$) slice of the Hamiltonian, bands with $(-1)^{C_i[0]} = -1$ ($(-1)^{C_i[\pi]} = -1$) must come in TRS related pairs.  Since mirror symmetry eigenvalues are real for spinless fermions, these pairs must also share the same mirror symmetry eigenvalue at the mirror invariant plane. Additionally, for such a pair of TRS-related bands, the Chern number parity of each band at the mirror invariant plane is separately conserved (only the total Chern number is conserved without $C_2$/inversion symmetry)~\cite{hughes2011inversion}. This follows from Eq.~\ref{eq:ChernNumberC2} and the fact that the $\hat{C}_2$ and inversion eigenvalues of a pair of TRS-related bands must be the same at high symmetry points. For example, take the following 4-band model for a pair of 2D insulators with Chern number $\pm 1$,
\begin{equation}\begin{split}
\mathcal{H}(k_x,k_y) &= (m+\cos(k_x)+\cos(k_y))\sigma^z\sigma^z \\ 
&+ \sin(k_x)\sigma^x\sigma^0 + \sin(k_y)\sigma^y\sigma^0,
\label{eq:2DHamwithTRS}\end{split}\end{equation}
with time reversal symmetry $\hat{\mathcal{T}} = \sigma^y\sigma^y \mathcal{K}$ and $C_2$ symmetry $\hat{C}_2 = \sigma^z\sigma^z$. At half filling, this model has two TRS related bands with Chern number $\pm 1$ for $0<|m|<2$ and two bands with Chern number $0$ for $2<|m|$. Without $C_2$ symmetry, it is possible to adiabatically transition between these two phases by adding a second mass term, e.g., $\sigma^z\sigma^{x}$ or $\sigma^z\sigma^{y}$.

With this in mind, we proceed to construct the invariant for the coefficient of the $RF$-term. Consider the bands with odd Chern number parity for a fixed mirror invariant plane ($k_z = 0$ or $k_z = \pi$) . Since the fermions are spinless, these bands have mirror eigenvalue $\pm 1$ when restricted to this mirror plane. By our above arguments, the bands with odd Chern number parity come in pairs with the same mirror eigenvalue. We take the number of occupied bands with mirror eigenvalue $\pm 1$ and odd Chern number parity at $k_z = 0/\pi$ to be $N^{\text{odd}}_{0/\pi,\pm 1} \in 2\mathbb{Z}$, and define the following four invariants. 
\begin{equation}\begin{split}
\eta_{0,+1} &= \exp(i\frac{\pi}{2} N^{\text{odd}}_{0,+1}),\\
\eta_{\pi,+1} &=\exp(i\frac{\pi}{2} N^{\text{odd}}_{\pi,+1}),\\
\eta_{0,-1} &=\exp(i\frac{\pi}{2} N^{\text{odd}}_{0,-1}),\\
\eta_{\pi,-1} &= \exp(i\frac{\pi}{2} N^{\text{odd}}_{\pi,+1}).
\label{eq:etaDef}\end{split}\end{equation}
Since $N^{\text{odd}}_{0/\pi,\pm 1}$ is even, these four invariants all take values of $\pm 1$. We can understand $\eta_{0/\pi,\pm 1}$ as the net Chern number parity of \textit{half} of the bands with odd Chern number parity in the appropriate sector.

The $\eta$ terms are not meaningful individually, as a redefinition of the mirror symmetry $\hat{M}_z \rightarrow -\hat{M}_z$ exchanges $\eta_{0/\pi,+ 1} \leftrightarrow \eta_{0/\pi,- 1}$, and shifting momentum $k_z \rightarrow k_z + \pi$ exchanges $\eta_{0,\pm 1} \leftrightarrow \eta_{\pi,\pm 1}$. We therefore first consider the product of all four $\eta$ invariants,
\begin{equation}\begin{split}
(-1)^{\nu} = \eta_{0,+1}\eta_{\pi,+1}\eta_{0,-1}\eta_{\pi,-1}.
\label{eq:InvTopoInvariant}\end{split}\end{equation}
Using Eq.~\ref{eq:ChernNumberC2}, we find that $\nu$ is actually the same invariant that describes the quantized axion electrodynamics $\Theta$-term of 3D inversion symmetric topological insulators ($\nu = \Theta/\pi$ in Eq.~\ref{eq:ThetaTerm})~\cite{hughes2011inversion}. 
Since spinless fermions with TRS ($\hat{T}^2 = +1$) cannot support a non-trivial quantized  $\Theta$-term in 3D~\cite{schnyder2008classification}, it is necessarily the case that $\nu = 0$ for the systems we consider here. 

Here we show that the spinless rTCI with mirror symmetry is described by the invariant,
\begin{equation}\begin{split}
(-1)^{\nu_{RF}} = \eta_{0,+1}\eta_{\pi,+1} = \eta_{0,-1}\eta_{\pi,-1},
\label{eq:NuRF}\end{split}\end{equation}
where the second equality follows from $\nu = 0$. In Appendix~\ref{app:spinlessMirrorrTCI}, we construct a lattice model that realizes the spinless rTCI, and determine the coefficient of the $RF$-term using a linear response formalism analogous to those in Secs.~\ref{ssec:ResponseSpinless} and~\ref{ssec:ResponseThSpin1/2Mirror}. For this lattice model, $\nu_{RF} = 0$, when the lattice model is described by a trivial $RF$-term and $\nu_{RF} = 1$ when the lattice model is an rTCI with non-trivial $RF$-term. Concisely, $\nu_{RF} = \Phi/\pi$, where the left-hand side is calculated using the symmetry eigenvalue formula, and the right-hand side is calculated using linear response. Additionally, as we show in Appendix~\ref{app:InvariantEx}, a linear, Dirac-like band crossing for a system of spinless fermions with TRS, $C_2$ symmetry and mirror symmetry requires a minimum of $8$-bands, and a non-trivial $RF$-term is either generated or removed during such a band crossing, if and only if the value of $\nu_{RF}$ changes as well. More generally, we find that $\nu_{RF} = \Phi/\pi$ for all insulators with these symmetries that have transitions generated by Dirac-like band structures at TRIM. We conjecture that the relation $\nu_{RF} = \Phi/\pi$ holds in general, but additional analysis is required to confirm this.

The invariant $\nu_{RF}$ may be written in a Fu-Kane-like form as follows~\cite{fu2011topological}. For a single TRIM, $\Lambda_n$, there are $N^{\text{odd}}_{-1}[\Lambda_n]$ bands that have odd Chern number in the mirror plane that contains $\Lambda_n$, \text{and} have $C_2$ eigenvalue $-1$ at $\Lambda_n$. Based on our previous arguments, these bands come in pairs, and we can order them such that the $2i$ and $2i-1$ bands have the same inversion eigenvalue at $\Lambda_n$, $\zeta_{2i}[\Lambda_n] = \zeta_{2i-1}[\Lambda_n]$. In terms of these bands, $\nu_{RF}$ is
\begin{equation}\begin{split}
(-1)^{\nu_{RF}} &= \prod_{n} \delta_{-1}[\Lambda_n],\\
\delta_{-1}[\Lambda_n] &= \prod^{\frac{1}{2}N^{\text{odd}}_{-1}[\Lambda_n]}_{i = 1} \zeta_{2i}[\Lambda_n],
\label{eq:NuRFTRIM}\end{split}\end{equation}
where we have used Eq.~\ref{eq:ChernNumberC2} and the fact that inversion symmetry is equal to the product of $C_2$ and mirror symmetries. This leads to an interpretation of $\nu_{RF}$ as the Fu-Kane invariant~\cite{fu2011topological} for the non-trivial bands with $C_2$ eigenvalue $-1$ at a given TRIM. A similar invariant can be constructed in terms of the non-trivial bands with $C_2$ eigenvalue $+1$ at a given TRIM, and since $\nu = 0$, these invariants are equal.

We also note that there is a second independent invariant to consider: 
\begin{equation}\begin{split}
(-1)^{\nu_{WZ,z}} = \eta_{0,+1}\eta_{0,-1} = \eta_{\pi,+1}\eta_{\pi,-1}.
\end{split}\end{equation}
Since $\nu_{WZ,z}$ only involves a single mirror invariant plane, it is a weak topological invariant. As we argue here, a non-zero value of $\nu_{WZ,z}$ indicates that the insulator is described by a 3D Wen-Zee term
\begin{equation}\begin{split}
\mathcal{L}_{WZ,z} = \frac{\mathcal{S}_z}{4\pi^2} G_z \epsilon^{\mu\nu\rho}  \omega_\mu \partial_\nu A_\rho
\label{eq:3DWZ}\end{split}\end{equation}
where $\mathcal{S}_z$ is a constant integer, $G_z$ is the reciprocal lattice vector in the z-direction, and $\mu,\nu,\rho$ run over $x,y,t$. The coefficient of the 3D Wen-Zee term is related to the weak invariant as $\mathcal{S}_z = \nu_{WZ,z}$ mod$(2)$. The anisotropic $3$D Wen-Zee term indicates that disclinations in an xy-plane bind charge, or equivalently, disclination lines carry charge per unit length along the z-direction. 

To show the connection between $\nu_{WZ,z}$ and the $3$D Wen-Zee term, we consider a spinless $2$D insulator with $C_2$ rotation symmetry and TRS. Such an insulator will have $N^{\text{odd}}_{2\text{D}}$ occupied bands with odd Chern number parity, and these bands will come in pairs that are related by TRS with opposite Chern number. 
In Ref. \onlinecite{liu2019shift} it was shown that for such insulators, a filled pair of TRS related bands with Chern numbers $\pm C$ are described by a $2$D Wen-Zee term
\begin{equation}\begin{split}
\mathcal{L}_{WZ,2\text{D}} = \frac{\mathcal{S}}{2\pi} \epsilon^{\mu\nu\rho}  \omega_\mu \partial_\nu A_\rho,
\label{eq:2DWZ}\end{split}\end{equation}
with coefficient $\mathcal{S} = C$. To proceed, we use the fact that the Chern number parity of the $i^{\text{th}}$ band is
\begin{equation}
(-1)^{C_i} = \chi_i(0,0)\chi_i(0,\pi)\chi_i(\pi,0)\chi_i(\pi,\pi)
\end{equation}
where $\chi_i$ is the $C_2$ eigenvalue at a given high symmetry point~\cite{hughes2011inversion}. Taking the number of occupied bands with odd Chern number parity to be $N^{\text{odd}}_{2\text{D}} \in 2\mathbb{Z}$ ($N^{\text{odd}}_{2\text{D}}$ must be even due to TRS) we define the following invariant
\begin{equation}
(-1)^{\nu_{WZ}} = \exp(i \frac{\pi}{2} N^{\text{odd}}_{2\text{D}}).
\end{equation}
Following our above discussion, the $2$D insulator we are considering here is described by a Wen-Zee term with coefficient $\mathcal{S} =  \nu_{WZ}$ mod$(2)$, i.e., $(-1)^{\nu_{WZ}}$ is the parity of the Wen-Zee term.

Returning to the weak invariant $\nu_{WZ,z}$ of a $3$D insulator, if we treat the $k_z = 0$ plane as a $2$D insulator, then $\eta_{0,\pm1}$ is the parity of the Wen-Zee term that describes the $\mathcal{M}_z = \pm 1$ sector of the $k_z = 0$ plane. The product $\eta_{0,+1}\eta_{0,-1}$ is therefore the parity of the full Wen-Zee term of the $k_z = 0$ plane. Since the coefficient of the Wen-Zee term is quantized, it cannot change as a function of $k_z$ as long as the gap and rotation symmetry are maintained. Each constant $k_z$ plane is therefore described by the same Wen-Zee term as the $k_z = 0$ plane. From this, we conclude that $\mathcal{S}_z = \nu_{WZ,z}$ mod$(2)$, i.e., $(-1)^{\nu_{WZ,z}}$ is the parity of the $3$D Wen-Zee term. 

For the rTCIs from Secs.~\ref{sec:PHSrTCI} and~\ref{sec:MirrorrTCI}, $\nu_{WZ,z} = 0$. The minimal model for $\nu_{WZ,z} = 1$ consists of stacking the $2$D Hamiltonian in Eq.~\ref{eq:2DHamwithTRS} along the z-direction. Using linear response theory, we find that for $2<|m|$, $\nu_{WZ,z} = 0$ and the $3$D Wen-Zee response vanishes, while for $0<|m|<2$, $\nu_{WZ,z} = 1$ and there is a $3$D Wen-Zee response with $\mathcal{S}_z = \text{sgn}(m)$ (see Appendix~\ref{app:LinRespWZ} for details of the linear response calculation). 

Our $2$D analysis also gives a new perspective on the ``strong'' invariant $\nu_{RF}$. The invariant $\eta_{0/\pi,+1}$ indicates that the $k_z = 0/\pi$ plane of the $\mathcal{M}_z = +1$ sector is described by a Wen-Zee term (Eq.~\ref{eq:2DWZ}) with $(-1)^\mathcal{S} = \eta_{0/\pi,+1}$. We can therefore interpret $(-1)^{\nu_{RF}} = \eta_{0,+1}\eta_{\pi,+1}$ as the change in the parity of the Wen-Zee term between the $\mathcal{M}_z = +1$ sectors of the $k_z = 0$ and $k_z = \pi$ planes. Since the \textit{total} Wen-Zee term must be constant as a function of $k_z$, there must be a compensating change in the parity of the Wen-Zee term between the $\mathcal{M}_z = -1$ sectors of $k_z = 0$ and $k_z = \pi$ planes. This is reflected in the second equality in Eq.~\ref{eq:NuRF}.

\subsection{Spin-1/2 Fermions with Additional Spin Conservation}\label{ssec:spin1/2invariant}
Having discussed spinless fermions, we now construct an analogous invariant for spin-1/2 insulators with conserved spin, TRS ($\mathcal{T}^2 = -1$), $C_n$ symmetry ($(\hat{C}_n)^n = -1$), mirror symmetry ($(\hat{M}_z)^2 = -1$), and inversion symmetry. Again, we only consider the $C_2$ subgroup of rotations here.

 The analysis is simplified by the observation that for a spin-1/2 insulator with conserved spin, it is possible to decompose the insulator into two blocks with $S^z = \pm 1/2$, and that each of these blocks can be treated as a spinless insulator with a spinless TRS and $C_2$ rotation symmetry, mirror symmetry and inversion symmetry. To show this, we write the block diagonal Hamiltonian for a spin-1/2 insulator as
\begin{equation}\begin{split}
\mathcal{H}(\bm{k}) = \mathcal{H}'(\bm{k}) \otimes \sigma^0,
\end{split}\end{equation}
where $\mathcal{H}$ is a $2n \times 2n$ Bloch Hamiltonian, $ \mathcal{H}'$ is an $n\times n$ matrix, and the fermionic spin is given by $S^i = \frac{1}{2}\text{I}_n \otimes \sigma^i$ for $i = x,y,z$. Here we have made the Kronecker product explicit for clarity. In general, the TRS, $C_2$ rotation symmetry, mirror symmetry, and inversion symmetry operators are
\begin{equation}\begin{split}
&\hat{\mathcal{T}} = \hat{T}' \otimes \sigma^y \mathcal{K},\\
&\hat{C}_2 = \hat{C}'_2 \otimes i \sigma^z,\\
&\hat{P} = \hat{P}' \otimes \sigma^0,\\
&\hat{M}_z = \hat{C}_2\hat{P} = [\hat{C}'_2 \hat{P}'] \otimes i \sigma^z,
\label{eq:SymmetryDecomp}\end{split}\end{equation}
where $\hat{P}$ is the inversion symmetry operator, and the Pauli matrices that act on the spin degrees of freedom are fixed by the transformation properties of fermionic spin under the symmetries. The symmetry operators satisfy
\begin{equation}\begin{split}
&\hat{\mathcal{T}}' \mathcal{H}(\bm{k}) \hat{\mathcal{T}}^{-1} = \mathcal{H}(-\bm{k}),\\
&\hat{C}_2 \mathcal{H}(\bm{k}) \hat{C}^{-1}_2 = \mathcal{H}(R_2 \bm{k}),\\
&\hat{P} \mathcal{H}(\bm{k}) \hat{P}^{-1} = \mathcal{H}(-\bm{k}),\\
&\hat{M}_z \mathcal{H}(\bm{k}) \hat{M}_z^{-1} = \mathcal{H}(-R_2 \bm{k}).
\end{split}\end{equation}

If we consider the Hamiltonian for the $S^z = +1/2$ block, $\mathcal{H}'(\bm{k})$, we find that 
\begin{equation}\begin{split}
&\hat{\mathcal{T}}' \mathcal{H}'(\bm{k}) \hat{\mathcal{T}}^{'-1} = \mathcal{H}'^*(-\bm{k}),\\
&\hat{C}'_2 \mathcal{H}'(\bm{k}) \hat{C}^{'-1}_2 = \mathcal{H}'(R_2 \bm{k}),\\
&\hat{P}' \mathcal{H}'(\bm{k}) \hat{P}^{'-1} = \mathcal{H}'(-\bm{k}),\\
&\hat{M}_z' \mathcal{H}'(\bm{k}) \hat{M}_z^{'-1} = \mathcal{H}'(-R_2 \bm{k}),\\
\end{split}\end{equation}
where $\hat{\mathcal{T}}' = \hat{T}'\mathcal{K}$, and $\hat{C}'_2$, $\hat{P}'$ and $\hat{M}_z'$ are defined as in Eq.~\ref{eq:SymmetryDecomp}. We therefore find that the $S^z = +1/2$ block inherits a TRS, $C_2$, mirror, and inversion symmetry. Importantly, $(\hat{\mathcal{T}}')^2 =  (\hat{C}'_2)^2  = (\hat{\mathcal{M}}'_z)^2 = +1$, and so the $S^z = +1/2$ block can be treated as spinless fermions with appropriate TRS, $C_2$ symmetry, mirror symmetry and inversion symmetry. Physically, $\hat{C}'_2$ is generated by the orbital angular momentum of the fermions. Similar logic also holds for the $S^z = -1/2$ block.

We now consider the invariant constructed in Sec.~\ref{ssec:InvariantSpinless} for the $S^z = + 1/2$ block, $\nu_{RF \uparrow}$. As discussed previously, if we restrict the system to a single mirror invariant plane, bands with odd Chern number parity come in pairs with the same mirror eigenvalue. We take the number of bands with $S^z = +1/2$, and $\hat{\mathcal{M}}'_z$ eigenvalue $\pm 1$ and odd Chern number in the $k_z = 0/\pi$ plane to be $N^{\text{odd},\uparrow}_{0/\pi,\pm 1} \in 2\mathbb{Z}$ and define the following invariants
\begin{equation}\begin{split}
\eta^{\uparrow}_{0,+1} &= \exp(i\frac{\pi}{2}N^{\text{odd}, \uparrow}_{0,+1}),\\
\eta^\uparrow_{\pi,+1} &= \exp(i\frac{\pi}{2}N^{\text{odd}, \uparrow}_{\pi,+1}),\\
\eta^\uparrow_{0,-1} &= \exp(i\frac{\pi}{2}N^{\text{odd}, \uparrow}_{0,-1}),,\\
\eta^\uparrow_{\pi,-1} &= \exp(i\frac{\pi}{2}N^{\text{odd}, \uparrow}_{\pi,-1}),.\\
\end{split}\end{equation}
We can understand $\eta_{0/\pi,\pm 1} = \pm 1$ as the net Chern number parity of \textit{half} of the $S^z = +1/2$ bands with odd Chern number parity and mirror eigenvalue $\pm 1$ at $k_z = 0$ or $kz = \pi$. The $\eta^{\downarrow}$ invariants for the $S^z = -1/2$ fermions are constructed analogously. 

Since the $S^z = \pm1/2$ sector can be mapped onto a system of spinless fermions, all results pertaining to $\eta^{\uparrow}$ follow those established in Sec.~\ref{ssec:InvariantSpinless}. In particular,
\begin{equation}\begin{split}
1 = \eta^{\uparrow}_{0,+1} \eta^{\uparrow}_{\pi,+1}\eta^{\uparrow}_{0,-1}\eta^{\uparrow}_{\pi,-1}.
\end{split}\end{equation}
The invariant $\nu_{RF \uparrow}$ is defined as 
\begin{equation}\begin{split}
(-1)^{\nu_{RF \uparrow}} &= \eta^{\uparrow}_{0,+1} \eta^{\uparrow}_{\pi,+1}=\eta^{\uparrow}_{0,-1}\eta^{\uparrow}_{\pi,-1}.
\label{eq:NuRFUp}\end{split}\end{equation}
The corresponding invariant for the $S_z = -1/2$ fermions, $\nu_{RF \downarrow}$,  is related to $\nu_{RF \uparrow}$ by TRS, and as such $\nu_{RF \uparrow} = \nu_{RF \downarrow}$. For the spin-1/2 lattice model in Sec.~\ref{sec:MirrorrTCI}, we calculate $\nu_{RF \uparrow}$ and find that $\nu_{RF \uparrow} = 1$ for $1<|M|<3$, and $\nu_{RF \uparrow} = 0$ otherwise. Hence, $\nu_{RF \uparrow} = 1$ correctly differentiates the spin-1/2 rTCI from the trivial states, and $\nu_{RF \uparrow} = \Phi/2\pi$. In Appendix~\ref{app:InvariantEx} we also show that $\nu_{RF \uparrow} = \Phi/2\pi$ generically for Hamiltonians that have a Dirac-like band structure at TRIM. We again conjecture that $\nu_{RF \uparrow} = \Phi/2\pi$ for the class of insulators considered here, but an analysis for general lattice bandstructures is required.

We can put Eq.~\ref{eq:NuRFUp} into a form similar to Eq.~\ref{eq:NuRFTRIM} as follows. For a single TRIM $\Lambda_n$, there are $N^{\text{odd}}_{-i,\uparrow}[\Lambda_n]$ bands that have $S^z = +1/2$, odd Chern number in the mirror plane that contains $\Lambda_n$, and $C_2$ eigenvalue $-i$ at $\Lambda_n.$ Such bands come in pairs with the same inversion eigenvalue at $\Lambda_n$. If we organize these bands such that the $2j$ and $2j-1$ bands share the same inversion eigenvalue at $\Lambda_n$, the invariant $\nu_{RF,\uparrow}$ is
\begin{equation}\begin{split}
(-1)^{\nu_{RF,\uparrow}} &= \prod_{n} \delta_{-i,\uparrow }[\Lambda_n],\\
\delta_{-i,\uparrow}[\Lambda_n] &= \prod^{\frac{1}{2}N^{\text{odd}}_{-i,\uparrow}[\Lambda_n]}_{j = 1} \zeta_{2j}[\Lambda_n].
\label{eq:NuRFUpTRIM}\end{split}\end{equation}
This is effectively Eq.~\ref{eq:NuRFTRIM} applied to the $S^z = 1/2$ fermions. The corresponding formula for $\nu_{RF,\downarrow}$ is related to Eq.~\ref{eq:NuRFUpTRIM} by the spin-1/2 TRS of the full system. 

We also note that 
\begin{equation}\begin{split}
(-1)^{\nu_{RF \uparrow,z}} &= \eta^{\uparrow}_{0,+1} \eta^{\uparrow}_{0,-1}=\eta^{\uparrow}_{\pi,+1}\eta^{\uparrow}_{\pi,-1},
\end{split}\end{equation}
defines a weak invariant that describes a system with a 3D Wen-Zee term (as does the TRS related invariant $\nu_{RF \downarrow,z}$). This Wen-Zee term differs by a factor of 2 from that given in Eq.~\ref{eq:3DWZ} due to Kramers degeneracy for spin-1/2 fermions. 

To conclude this section, it is worth reiterating that the expressions derived for spinful fermions are based on the assumptions that spin is conserved and cannot be applied to systems with spin-orbit coupling. Determining a general invariant for spin-1/2 insulators, and relating it to the $RF$-term remains an open question for further research.

\section{The $RF$-term in systems with broken time-reversal symmetry}\label{sec:RFwoTRS}
In this section, we discuss the $RF$-term in systems that break time-reversal symmetry (TRS). As we show below, the mixed geometry-charge responses arising from the $RF$-term are intertwined with the charge response of the axion electrodynamics $\Theta$-term (Eq.~\ref{eq:ThetaTerm}) in spinless systems when TRS is broken. Interestingly, we find that a non-zero quantized $RF$-term naturally arises alongside a quantized $\Theta$-term in spinless mirror symmetric axion insulators~\cite{hughes2011inversion,fang2012bulk, khalaf2018higher}  with additional $C_n$ rotation symmetry. 

\subsection{Disclination Charges in $2$D Systems with Broken TRS}
To demonstrate this intertwining of responses in $3$D, we first analyze the interplay between geometry-charge responses and purely charge responses in $2$D systems with broken TRS. As we show here, the charge bound to disclinations in $2$D systems with broken TRS depends on whether the fermions are spinless or have spin-1/2. For spinless fermions without TRS, the charge bound to $2\pi/n$ disclinations satisfies $Q_{\text{disc}} = C/2n \mod(1/n)$, where $C$ is the Chern number of the insulator. For spin-1/2 fermions without TRS, the disclination charge comes in multiples of $1/n$ ($Q_{\text{disc}} = 0 \mod(1/n)$) regardless of the Chern number of the insulator~\cite{li2020fractional}. 

To establish this, we first note that spinless fermions satisfy $(\hat{U}_n)^n = +1$, and spin-1/2 fermions satisfy $(\hat{U}_n)^n = -1$, where $\hat{U}_n$ is the $2\pi/n$ rotation operator. In systems without TRS, a system of spinless fermions can be mapped onto a system of spin-1/2 fermions, and vice versa, by redefining the rotation operator
\begin{equation}\begin{split}
    \hat{U}_n \rightarrow \hat{U}'_n = \hat{U}_n e^{\pm i \pi/n }.
\label{eq:symmetryRedefinition}\end{split}\end{equation} 
This phase shift of the rotation operator also changes the effective structure of lattice disclinations. In $2$D, changing from a spinless (spin-1/2) insulator with rotation operator $\hat{U}_n$ to a spin-1/2 (spinless) insulator with rotation operator $\hat{U}'_n$, amounts to adding an additional $\pm \pi/n$ U$(1)$ symmetry flux to $2\pi/n$ disclinations. The extra U$(1)$ flux binds charge $\pm C/2n$, where $C$ is the Chern number of the insulator (note that redefinition of the rotation operator does not change the Chern number of the insulator). 

From this relationship we can draw some conclusions. If a spin-1/2 insulator exists where $C = +1$ and the disclination-bound charge $Q_{\text{disc}} = 0,$ it implies the existence of a spinless insulator with $C = +1$ and disclination charge $Q_{\text{disc}} = \pm 1/2n$. Continuing this logic, it must then be true that $Q_{\text{disc}} = 0 \mod(1/n)$ for spin-1/2 insulators regardless of the Chern number, and $Q_{\text{disc}} = C/2n \mod(1/n)$ for spinless insulators with Chern number $C$. 

Conversely, if there exists a spinless insulator with $C = +1$, and $Q_{\text{disc}} = 0$, then there must exist a spin-1/2 insulator with $C = +1$, and disclination-bound charge $Q_{\text{disc}} = \pm 1/2n$. If this is true, then $Q_{\text{disc}} = 0 \mod(1/n)$ for spinless insulators regardless of the Chern number, and $Q_{\text{disc}} = C/2n \mod(1/n)$ for spin-1/2 insulators with Chern number $C$. Importantly, it is not possible to have \textit{both} spinless and spin-1/2 insulators with $C = +1$ and $Q_{\text{disc}} = 0$. If this were true, it would imply that there exist insulators with zero Chern number and $Q_{\text{disc}} = 1/2n$, violating the results of Ref. \onlinecite{li2020fractional}.

To find out which one of these two scenarios plays out we can calculate the disclination-bound charge in lattice models. Indeed, this has already been previously done. The calculations in Ref. \onlinecite{liu2019shift} and \onlinecite{li2020fractional} show that there are spin-1/2 insulators with $C = +1$, and $Q_{\text{disc}} = 0$, and spinless insulators with $C = +1$, and $Q_{\text{disc}} = \pm 1/2n$. We therefore conjecture that $Q_{\text{disc}} = 0 \mod (1/n)$ for spin-1/2 insulators having broken TRS regardless of Chern number, and $Q_{\text{disc}} = C/2n \mod(1/n)$ for spinless insulators with broken TRS and Chern number $C$. In terms of the response theories, this means that a system of spinless fermions can have a Wen-Zee term with coefficient $1/4\pi \mod (1/2\pi)$ if and only if it also has a Chern-Simons term with coefficient $1/4\pi \mod (1/2\pi)$. That is spinless fermions can have the response action
\begin{equation}\begin{split}
\mathcal{L}_{2\text{D},spinless} = \frac{\mathcal{S}}{4\pi} \epsilon^{\mu\nu\rho}  \omega_\mu \partial_\nu A_\rho+\frac{C}{4\pi} \epsilon^{\mu\nu\rho}  A_\mu \partial_\nu A_\rho + ...,\label{eq:response2DSpinless}\end{split}\end{equation}
where $\mathcal{S} = C \mod(2)$.

\subsection{Periodicity of the $RF$-Term in Systems with Broken TRS}\label{ssec:periodRFwoTRS}
To see how the intertwining of mixed geometry-charge and pure charge responses in $2$D insulators affects the $3$D $RF$-term, we determine the periodicity of the coefficient of the $RF$-term $\Phi$ when TRS is broken. For spin-1/2 insulators with broken TRS, the disclination charge comes in multiples of $1/n$, regardless of Chern number, and the same logic used in Sec.~\ref{sec:SymQuantRF} indicates that $\Phi$ has period $2\pi$ for these systems. The situation for spinless insulators with broken TRS is more complex. Here we show that the coefficients of the $RF$-term ($\Phi$) and axion electrodynamics $\Theta$-term have a combined periodicity, where 
\begin{equation}\begin{split}
(\Phi,\Theta)&\equiv  (\Phi+\pi,\Theta+2\pi) \equiv (\Phi+\pi,\Theta-2\pi)\\& \equiv  (\Phi+2\pi,\Theta) \equiv (\Phi,\Theta+4\pi).
\label{eq:AppTRSEquiv}\end{split}\end{equation}

 To show this explicitly, consider a domain wall where the value of $\Phi$ changes by $\Delta \Phi$. If the domain wall response can be cancelled by a purely $2$D insulator without topological order, then $\Phi$ and $\Phi + \Delta \Phi$ are equivalent. As noted in the main text, a domain wall where the value of $\Phi$ changes by $\Delta \Phi$ hosts a Wen-Zee term with coefficient $\Delta \Phi/4\pi^2$. When $\Delta \Phi = \pi$ the domain wall Wen-Zee term is cancelled by the response theory of the $2$D spinless TRS breaking insulator in Eq.~\ref{eq:response2DSpinless} with $\mathcal{S} = -1$.  However, when $\mathcal{S} = -1,$ the Chern-Simons term in Eq.~\ref{eq:response2DSpinless} is non-vanshing. Therefore, if we add such a $2$D insulator to the $RF$-term domain wall with $\Delta \Phi  = \pi$, the domain wall does not host a Wen-Zee term, but instead hosts a Chern-Simons term with coefficient $1/4\pi$ mod($1/2\pi$). Hence, a domain wall where $\Phi$ changes by $\pi$ cannot be \textit{completely} trivialized by adding a purely $2$D system. 

Let us now consider a domain wall of both the $RF$-term and the $\Theta$-term (Eq.~\ref{eq:ThetaTerm}), where $\Phi$ changes by $\Delta \Phi$ and $\Theta$ changes by $\Delta \Theta$. There is a Wen-Zee term at this domain wall with coefficient $\Delta \Phi/4\pi^2$ and a Chern-Simons term with coefficient $\Delta \Theta/8\pi^2$. Based on our previous discussion, when $\Delta \Phi = \pi$ and $\Delta \Theta = \pm 2\pi$, the domain wall can be completely trivialized by a $2$D insulator. The $RF$-term and the $\Theta$ term therefore have a combined periodicity where $\Phi$ is shifted by $\pi,$ and $\Theta$ is shifted by $\pm 2\pi$. The other equivalence relationships in Eq.~\ref{eq:AppTRSEquiv} can be established using similar logic. 

An interesting corollary of this analysis is that  $\Theta = 2\pi$  is not necessarily trivial for spinless fermions when $C_n$ rotation symmetry is present. We can show this using similar logic to before. Consider a domain wall where $\Theta$ changes by $2\pi$. This domain wall will host a Chern-Simons term with coefficient $1/4\pi$. We can cancel this out by adding a $2$D insulator with Chern number $1$ to the domain wall. However, as discussed above, this $2$D insulator will contribute a Wen-Zee term with coefficient $1/4\pi$ mod$(1/2\pi)$. Hence, a $2\pi$ domain wall of the $\Theta$-term cannot be fully trivialized if $C_n$ rotation symmetry is also present.

\subsection{Charge and Geometry-Charge Responses in rTCIs with broken TRS}
Will will now argue, based on Eq.~\ref{eq:AppTRSEquiv}, that there exists a special type of spinless mirror symmetric rTCI  in a TRS-breaking context. To show this, we note that both the $RF$- and $\Theta$-terms are odd under mirror symmetry. Hence, for a mirror symmetric insulator $(\Phi,\Theta) = (-\Phi,-\Theta)$. This equation admits two non-trivial solutions $(\Phi,\Theta) = (\pi/2, \pi)$ and $(\pi/2, -\pi)$, each of which describes a non-trivial spinless rTCI with mirror symmetry. Since $\Theta$ is odd under time-reversal, the two rTCIs are related to each other by time-reversal and do not need to be considered individually. These insulators have both a non-trivial $RF$-term, and a non-trivial $\Theta$-term. The $\Theta$-term has the same quantized coefficient as the $\Theta$-term that describes time-reversal symmetric topological insulators and axion insulators~\cite{qi2008topological}. The coefficient of the $RF$-term is half of that which is allowed for mirror symmetric insulators with TRS (see Sec.~\ref{ssec:SymQuantMirror}). 

Since these rTCIs have both a non-vanishing $RF$-term and a non-vanishing $\Theta$ term, they exhibit topological charge responses as well as mixed geometry-charge responses. Individually, the charge responses should resemble those that have been previously studied in the contexts of $3$D topological insulators with $\Theta = \pi$~\cite{qi2008topological}. The geometry charge responses should be similar to those that we have discussed in previous sections, albeit with a different quantization. It is also possible that the combination of the $RF$-term and the $\Theta$ term may lead to fundamentally new phenomena,and we leave that to future work. 

Let us compare the spinless TRS-breaking system with mirror symmetry to that with PHS. Naively, for spinless insulators with PHS $\Phi = 0$ or $\pi$ (see second line of Eq.~\ref{eq:AppTRSEquiv}), and the value of $\Theta$ in unconstrained. However, a spinless $\Phi = \pi$  insulator with broken TRS can actually be adiabatically deformed into a $\Phi = 0$ insulator without breaking PHS. To show this, take a spinless insulator with PHS and broken TRS where $(\Phi, \Theta) = (\pi, 0)$. Since the value of $\Theta$ is not quantized by PHS, we can adiabatically increase $\Theta$ by $2\pi$ in this insulator, i.e., $(\Phi, \Theta) \rightarrow (\pi, 2\pi)$. However, from Eq.~\ref{eq:AppTRSEquiv} we have that $(\Phi, \Theta) = (\pi, 2\pi) \equiv (0,0)$. So, the $\Phi = \pi$  insulator can be adiabatically deformed into a trivial insulator without breaking PHS symmetry, and is therefore a trivial insulator itself. The fact that PHS alone cannot lead to a non-zero quantized value of $\Phi$ in spinless systems is a direct consequence of the shared periodicity of $\Phi$ and $\Theta$ in spinless systems without TRS.

For spin-1/2 insulators with broken TRS, $\Phi$ is $2\pi$ periodic (regardless of $\Theta$), and so $\Phi = 0$ or $\pi$ for systems with PHS or mirror symmetry.
As previously discussed, the $\Phi = \pi$ $RF$-term can also be realized in spinless insulators with TRS. So the geometric-charge responses of these insulators will match those already discussed in Sec.~\ref{sec:PHSrTCI} and Appendix~\ref{app:spinlessMirrorrTCI}.

\subsection{Models for the Mirror Symmetric rTCI with Broken TRS}
In this subsection, we present a lattice model for the $(\Phi,\Theta) = (\pi/2, \pi)$ rTCI with mirror symmetry and $C_4$ rotation symmetry. The minimal model for this rTCI is given by the following 4-band Bloch Hamiltonian:
\begin{equation}\begin{split}
\mathcal{H}(\bm{k}) &= \sin(k_x) \Gamma^x + \sin(k_y) \Gamma^y + \sin(k_z)\Gamma^z\\ 
&+ (M + \cos(k_x) + \cos(k_y)+ \cos(k_z) ) \Gamma^0.
\label{eq:AppMirrorRTCI}\end{split}\end{equation}
The $C_4$ rotation symmetry and mirror symmetry act on Eq.~\ref{eq:AppMirrorRTCI} as
\begin{equation}\begin{split}
&\hat{U}_4 = \exp(i\frac{\pi}{4}( \Gamma^{xy}  + \text{I}_4 )),\\
&\hat{\mathcal{M}}_z = \Gamma^{z5},
\end{split}\end{equation}
where $\text{I}_4$ is the $4\times 4$ identity matrix. This model also has inversion symmetry, which is the product of $\mathcal{M}_z$ and $C_2 = (C_4)^2$ symmetries. The spectrum of Eq.~\ref{eq:AppMirrorRTCI} is gapped except when $|M|= 1,3$, and below we show that this model realizes a mirror symmetric rTCI with $(\Phi,\Theta) = (\pi/2, \pi)$ when $1<|M|<3$.

We determine the response theory for this insulator using the same methods as in the main text. Specifically, for the band crossing near $M = -3$ the continuum Hamiltonian is
\begin{equation}\begin{split}
\mathcal{H} &= \Gamma^x i \partial_x + \Gamma^y  i\partial_y + \Gamma^z  i \partial_z+m \Gamma^0. 
\end{split}\end{equation}
We couple this system to the spin connection $\omega$ and U$(1)$ gauge field $A_\mu$ via the covariant derivative 
\begin{equation}\begin{split}
D_\mu = \partial_\mu - i A_\mu - i \frac{1}{2} \omega_\mu (\Gamma^{xy}  + \text{I}_4), 
\end{split}\end{equation}
and add the mirror symmetry breaking perturbation \begin{equation}\begin{split}
\mathcal{H}' = m' \Gamma^5.
\end{split}\end{equation}
If we set $m = - \bar{m}\cos(\phi)$, and $m' = - \bar{m}\sin(\phi)$,  we find that the effective response theory in terms of $\phi$, $A,$ and $\omega$ is given by
\begin{equation}\begin{split}
\mathcal{L}_{\text{eff}} &= \frac{\phi}{8\pi^2} \epsilon^{\mu\nu\rho\kappa} \partial_\mu \omega_\nu \partial_\rho A_\kappa +  \frac{\phi}{8\pi^2} \epsilon^{\mu\nu\rho\kappa} \partial_\mu A_\nu \partial_\rho A_\kappa\\ &+  \frac{\phi}{32\pi^2} \epsilon^{\mu\nu\rho\kappa} \partial_\mu \omega_\nu \partial_\rho \omega_\kappa. 
\end{split}\end{equation}
When $\phi = 0$ ($M<-3$ in the lattice model) the response theory vanishes. When $\phi = \pi$ ($-3<M<-1$ in the lattice model), the response theory is 
\begin{equation}\begin{split}
\mathcal{L}_{\text{eff}} &= \frac{1}{8\pi} \epsilon^{\mu\nu\rho\kappa} \partial_\mu \omega_\nu \partial_\rho A_\kappa +  \frac{1}{8\pi} \epsilon^{\mu\nu\rho\kappa} \partial_\mu A_\nu \partial_\rho A_\kappa\\ &+  \frac{1}{32\pi} \epsilon^{\mu\nu\rho\kappa} \partial_\mu \omega_\nu \partial_\rho \omega_\kappa. 
\label{eq:TRSBrokenResponse}\end{split}\end{equation}
The first two terms are the $RF$-term with $\Phi = \pi/2$ and the $\Theta$-term with $\Theta = \pi$, confirming that this model realizes the previously predicted mirror symmetric rTCI with broken TRS. We also find that there is an additional term that is quadratic in the spin connection. This term was not predicted by our earlier heuristic argument, but is not unexpected, and similar terms have been previously studied~\cite{ryu2012electromagnetic}.

If we ignore the $C_n$ rotation symmetry, Eq.~\ref{eq:AppMirrorRTCI} is simply a mirror symmetric axion insulator for $1<|M|<3$ insulator, as indicated by the non-vanishing $\Theta$-term in Eq.~\ref{eq:TRSBrokenResponse}. The spinless rTCI with $(\Phi,\Theta) = (\pi/2,\pi)$ is therefore equivalent to a mirror symmetric axion insulator with additional $C_n$ symmetry. Based on this, the $(\Phi,\Theta) = (\pi/2, \pi)$ rTCI with additional inversion symmetry is described by the same topological invariant as inversion symmetric axion insulators~\cite{hughes2011inversion}.

The surface theory of the model in Eq.~\ref{eq:AppMirrorRTCI} has been exhaustively analyzed elsewhere (see Ref. \onlinecite{varnava2018surfaces} for example). We would like to point out that the surface theory for Eq.~\ref{eq:AppMirrorRTCI} when $1<|M|<3$ consists of an odd number of 2-component Dirac fermions, with a Dirac mass term that is odd under mirror symmetry. So, for open boundary conditions, the surfaces on the top half and bottom half of the model can each be gapped out by adding opposite mass terms to each half. However, at the mirror invariant plane where the two halves meet, there will be a domain wall that hosts an odd number of one-dimensional chiral fermion modes. Since the fermions are chiral, they cannot acquire a mass without closing the bulk gap.

As mentioned above, there is also a spinless mirror symmetric rTCI with $(\Phi,\Theta) = (\pi/2, -\pi)$. This rTCI is related to the $(\Phi,\Theta) = (\pi/2, \pi)$ rTCI by time reversal symmetry. Hence, the minimal model for the $(\Phi,\Theta) = (\pi/2, -\pi)$ rTCI is found by acting on Eq.~\ref{eq:AppMirrorRTCI} with the TRS operator $\Gamma^y \mathcal{K}$, and redefining the $C_4$ rotation operator as
\begin{equation}\begin{split}
&\hat{U}_4 \rightarrow \exp(i\frac{\pi}{4}( \Gamma^{xy}  - \text{I}_4 )).
\end{split}\end{equation}
Using linear response theory we find that in the topological insulator phase the effective response theory is the same as in Eq.~\ref{eq:TRSBrokenResponse} but with opposite signs for the second and third terms, i.e., $\Phi = \pi/2$ and $\Theta = -\pi$ as expected.

\section{Conclusion and Outlook}\label{sec:Con}
In this work we analyzed how electromagnetic and geometric responses can be intertwined in $3$D rotation-invariant insulators. Our main focus was a mixed geometry-charge term, denoted the $RF$-term, that can occur in the effective response theories of such systems. The $RF$-term gives rise to a mixed Witten effect and imparts fractional statistics to magnetic flux lines and disclination lines. Additionally, Wen-Zee terms are bound to domain walls where the coefficient of the $RF$-term changes such as a surface. Using symmetry analyses and lattice models we show that a quantized $RF$-term occurs for a class of rotation-invariant topological crystalline insulators with either PHS or mirror symmetry. The coefficient of the $RF$-term depends on if the rTCI is composed of spinless fermions or spin-1/2 fermions. When a mass term is added to the surface of an rTCI, the resulting massive surface is described by a Wen-Zee response that has half the coefficient that is allowed in purely $2$D systems. 

Based on our results, there are several open questions for future work. First is the question of what mixed geometry-charge responses are exhibited by other $3$D topological crystalline insulators, and how to relate a given continuum response theory to a lattice model. We use linear response theory to accomplish the latter in this work, but this approach cannot be used on systems where the geometric effects are non-perturbative. Second is the question of what other anomalous symmetry-enriched topological orders can be realized  on the surface of only a topological crystalline insulator. A partial answer to this question would come from a set of anomaly indicators~\cite{wang2017anomaly} for topological orders that are enriched by crystalline symmetries. A final question is whether any physical systems can realize the $RF$-term constructed here. In this work, we find that for certain $C_n$ and mirror symmetric insulators without spin-orbit coupling, the $RF$-term is determined by the angular momentum and inversion eigenvalues of the occupied bands at TRIM. To consider more realistic materials, it is likely necessary to generalize this result to more generic band-structures and systems with spin-orbit coupling. In experiments, the $RF$-term could be observed by using scanning probes to detect the charge that is bound to surface disclinations. 
\section{Acknowledgements}
We thank Barry Bradlyn, Lei Gioia and Pranav Rao for helpful discussions and Abid Khan for assistance performing the numerics. JMM, MRH, and TLH thank ARO MURI W911NF2020166
for support. JMM is also supported by the National
Science Foundation Graduate Research Fellowship Program under Grant No. DGE - 1746047.
    
\bibliography{3D_TCI.bib}
\bibliographystyle{apsrev4-1}

\appendix

\section{Disclination Lines and Composite Disclination Lines With Embedded 1D Insulators}\label{app:RFOrigin}
As noted previously, the $RF$-term indicates that disclinations carry polarization. However, it is possible to change the local polarization of a disclination defect by embedding an additional isolated $1$D insulator at the disclination core. For a gapped $3$D system described by an $RF$-term with coefficient $\Phi$, embedding a $1$D insulator in the core of a $2\pi/n$ disclination will result in a composite defect (disclination + embedded $1$D insulator) with polarization $P_{\text{comp}} = \Phi/2\pi n + P_{1\text{D}}$. Note that a disclination and a composite defect differ only locally (i.e., near the defect core) and share the same Frank angle. When the composite defect terminates at a surface of the insulator,  a composite surface defect is formed (surface disclination + $0$D edge of the embedded $1$D insulator) with charge $\Phi/2\pi n + P_{1\text{D}}$. 

It is useful to consider embedding a $1$D insulator in the core of a disclination line in a $3$D insulator with either PHS or mirror symmetry. From the symmetry, we know the polarization of the $1$D insulator must be $0$ or $1/2 \mod (1)$ for spinless fermions (we discuss spin-1/2 fermions below). As noted in the main text, for spinless fermions with TRS, $\Phi$ is quantized to $0$ or $\pi \mod (2\pi)$, where the latter corresponds to an rTCI. Hence, for an rTCI,  a symmetric $1$D insulator embedded in the core of a $2\pi/n$ disclination line will result in a composite defect that carries polarization $\Phi/2\pi n + P_{1\text{D}} = 1/2n + 1/2$. This also leads to a composite surface defect with charge $1/2n + 1/2$. 

For $n = 2,4,6$, the difference in polarization bound to a $2\pi/n$ disclination and a $2\pi/n$ composite defect ($2\pi/n$ disclination + embedded $1$D insulator with polarization $1/2$) is inconsequential and does not change our conclusions. This is because $\Phi$ (which is $2\pi$ periodic)  determines the polarization of a disclination only mod $1/n$, and the polarization of the disclination and composite defect differs by $1/2 = 0 \mod (1/n)$ for $n = 2,4,6$. However, for $n = 3$ there is a distinction since the difference in polarization bound to a $2\pi/3$ disclination and a $2\pi/3$ composite defect is $1/6$ mod($1/3$). Hence, an $n = 3$ spinless rTCI hosts disclinations with polarization $1/6$ mod($1/3$) and composite defects with polarization $0$ mod($1/3$), while a trivial insulator hosts disclinations with polarization $0$ mod($1/3$) and composite defects with polarization $1/6$ mod($1/3$).  We find that for all $C_n$ symmetries, the polarization of a disclination of an rTCI and the polarization of a disclination of a trivial insulator differ by $1/2n$ mod$(1/n)$, and that the polarization of a composite defect of an rTCI and the polarization of a composite defect of a trivial insulator also differ by $1/2n$ mod$(1/n)$.  Similar arguments indicate that the difference between the surface charge bound to a surface disclination of an rTCI and the surface charge bound to a surface disclination of a trivial insulator is $1/2n$ mod$(1/n)$, as is the difference between the surface charge bound to a composite surface defects of an rTCI and the surface charge bound to a composite surface defects of a trivial insulator. For spin-1/2 fermions, the difference in the polarization (charge) of both disclinations (surface disclination) and composite defects (surface composite defects) between an rTCI and a trivial insulator is $1/n$ mod$(2/n)$ due to Kramers degeneracy.

\section{Coupling Lattice Dirac Fermions to the Spin Connection}\label{app:SpinConnectCoup}
In this appendix we discuss how to couple Dirac fermions to the spin connection. The spin connection is a gauge field whose flux distribution  encodes the configuration of  lattice disclinations. For our purposes, it will suffice to analyze a single 4-component Dirac fermion that is located near an $n$-fold high-symmetry point (HSPs) of the Brillouin zone of a $C_n$ symmetric lattice (i.e., the points of a Brillouin zone that are invariant under $C_n$ rotations).  One can describe the low-energy physics of generic systems that are Dirac-like near HSPs by combining multiple 4-component Dirac fermions.   

The Hamiltonian for a single Dirac fermion located at an $n$-fold HSP can be written as,
\begin{equation}\begin{split}
\hat{H} &= \psi^\dagger \mathcal{H} \psi \\
\mathcal{H} &= \Gamma^x i \partial_x + \Gamma^y  i\partial_y + \Gamma^z  i \partial_z+m \Gamma^0, 
\label{eq:AppDiracHamHSP}\end{split}\end{equation}
where the $\Gamma$ matrices are $4\times 4$ anti-commuting matrices. It is useful to write the $C_n$ rotation operator as
\begin{equation}\begin{split}
U_n = \exp(i \frac{2\pi}{n} L),
\label{eq:AppDiracRotDefHSP}\end{split}\end{equation}
where $L$ is the $C_n$ angular momentum operator, and $U_n$ satisfies
\begin{equation}\begin{split}
U^{\dagger}_n \mathcal{H}(\bm{k})U_n = \mathcal{H}(R_n \bm{k}),
\end{split}\end{equation} 
where $\mathcal{H}(\bm{k})$ is the Bloch Hamiltonian of Eq.~\ref{eq:AppDiracHamHSP}.
To proceed, we note that the continuum Dirac Hamiltonian in Eq.~\ref{eq:AppDiracHamHSP} has a continuous U$(1)$ rotation symmetry. In order for the continuum theory to be consistent, the $C_n$ lattice rotation symmetry should be embedded in this enlarged U$(1)$, i.e.
\begin{equation}\begin{split}
&U^{\dagger}(\theta) \mathcal{H}(\bm{k})U(\theta) = \mathcal{H}(R(\theta) \bm{k})\\
&U(\theta) \equiv \exp(i \theta L),
\end{split}\end{equation} 
where $R(\theta)$ is a rotation of the momentum by $\theta$. The most general, consistent definition of the $C_n$ angular momentum is 
\begin{equation}\begin{split}
L \equiv \frac{1}{2}\Gamma^{xy} + p \text{I}_4,
\label{eq:AppDiracAMDefHSP}\end{split}\end{equation}
where $\text{I}_4$ is the $4\times 4$ identity matrix, and $\Gamma^{xy} = -i \Gamma^x\Gamma^y$. For spinless fermions $(U_n)^n = +1$ and for spin-1/2 fermions $(U_n)^n = -1$. So, for spinless fermions $p$ must be a half-integer, while for spin-1/2 fermions, $p$ must be an integer. Here, the value of $p$ is defined only modulo $n$, and any physical quantity should only depend on the value of $p$ modulo $n$.

We now gauge the $C_n$ lattice rotation symmetry. To do this, we must introduce the frame-fields, $e^A_i$  (and inverses $E^i_A$) for $A = x,y,z$, and the spin connection $\omega$. Under a local $C_n$ transformation $\theta(x_\mu)$, the inverse frame-fields, and Dirac fermions transform as 
\begin{equation}\begin{split}
E^i_x &\rightarrow \cos(\theta) E^i_x + \sin(\theta)E^i_y, \\
E^i_y &\rightarrow  \cos(\theta) E^i_y - \sin(\theta)e^i_x +,\\
\omega_\mu &\rightarrow \omega_\mu - \partial_\mu \theta,\\
\psi &\rightarrow e^{i \theta L}\psi = e^{i \theta (\frac{1}{2} \Gamma^{xy} + p \text{I}_4)}\psi.
\label{eq:AppGaugeT}\end{split}\end{equation}
In terms of these fields, the minimally coupled Lagrangian is 
\begin{equation}\begin{split}
\mathcal{L} = \bar{\psi} [i \bar{\Gamma}^0 D_0 + i E^i_A \bar{\Gamma}^A D_i - m]\psi 
\label{eq:AppLagMin}\end{split}\end{equation}
where $\bar{\Gamma}^A = \Gamma^0 \Gamma^A$, $\bar{\Gamma}^5 = \Gamma^0 \Gamma^5$, and $\bar{\Gamma}^0 = \Gamma^0$. The covariant derivative is given by
\begin{equation}\begin{split}
D_\mu = \partial_\mu - i \omega_\mu[\frac{1}{2}\Gamma^{xy} + p \text{I}_4].
\label{eq:AppDiracMinC}\end{split}\end{equation}
The Lagrangian in Eq.~\ref{eq:AppLagMin} is invariant under the $C_n$ gauge transformation given in Eq.~\ref{eq:AppGaugeT}, as desired.

\section{Details of the Disclinated lattice}\label{app:latticeDisc}

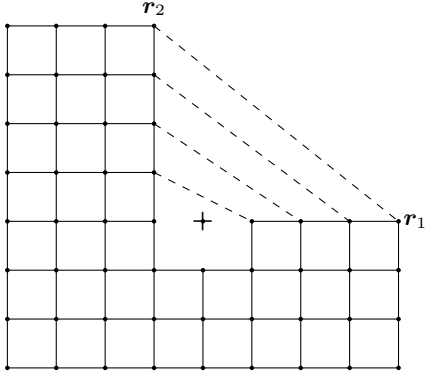
\begin{figure}
    \centering
    \begin{tikzpicture}[scale=0.65]
        \foreach \i in {0,1,...,3}{
            \foreach \j in {0,1,...,8}{
                \filldraw[black] (\j,\i) circle (1pt);
            }
        }
        \foreach \i in {0,1,...,2}{
            \draw (0,\i)--(8,\i);
        }
        \foreach \i in {3}{
            \draw (0,\i)--(3,\i);
        }
        \foreach \i in {3}{
            \draw (5,\i)--(8,\i);
        }
        \foreach \i in {0,1,...,3}{
            \draw (\i,0)--(\i,3);
        }
        \foreach \i in {4}{
            \draw (\i,0)--(\i,2);
        }
        \foreach \i in {5,6,7,8}{
            \draw (\i,0)--(\i,3);
        }

        \foreach \i in {4,5,...,7}{
            \foreach \j in {0,1,...,3}{
                \filldraw[black] (\j,\i) circle (1pt);
            }
        }{}
        \foreach \i in {0,1,...,3}{
            \draw (\i,3)--(\i,7);
        }
        \foreach \i in {4,5,...,7}{
            \draw (0,\i)--(3,\i);
        }
        \foreach \i in {4,5,...,7}{
            \draw[dashed] (3,\i)--(\i + 1,3);
        }
        \node at (8.35,3) {$\bm{r}_1$};
        \node at (3,7.35) {$\bm{r}_2$};
        \draw(4,3) node[]{$\bm{+}$};
    \end{tikzpicture}
    \caption{An xy-cross section of the disclinated lattice. The center of the disclination is marked as $\bm{+}$.}
    \label{fig:Disclattice}
\end{figure}
In this appendix we present details on how to define a tight-binding lattice model with disclinations. For a cubic lattice that is free of disclinations, a generic Hamiltonian can be written in real space as
\begin{equation}
\begin{aligned}
\hat{\mathcal{H}} &= \sum_{\bm{r}} \frac{1}{2}c^\dagger(\bm{r}) T_0 c(\bm{r}) + c^\dagger(\bm{r}+\hat{z}) T_{z} c(\bm{r}) \\ 
&+ c^\dagger(\bm{r}+\hat{x}) T_{x} c(\bm{r})+ c^\dagger(\bm{r}+\hat{y}) T_{y} c(\bm{r}) \\
&+ \text{H.c.}.\\
\label{eq:appLatticeRealSpace}
\end{aligned}
\end{equation}
where $c^\dagger$ is an $n$-component fermionic creation operator (in this work $n = 8$ for the spinless fermion models and $n = 16$ for the spin-1/2 fermion models). The matrices $T_0$ and $T_{x,y,z}$ are the on-site and nearest-neighbor hopping terms, respectively. The next-nearest-neighbor (NNN) hopping terms that appear in the models considered in the main text take the form
\begin{equation}\begin{aligned}
\hat{\mathcal{H}}_{\text{NNN}} = \sum_{\bm{r}} &c^\dagger(\bm{r}+\hat{x}+\hat{z}) T_{x+z} c(\bm{r})\\
+ &c^\dagger(\bm{r}+\hat{x}-\hat{z}) T_{x-z} c(\bm{r})\\
+ &c^\dagger(\bm{r}+\hat{y}+\hat{z}) T_{y+z} c(\bm{r})\\
+ &c^\dagger(\bm{r}+\hat{y}-\hat{z}) T_{y-z} c(\bm{r}) + \text{H.c.}\,.
\label{eq:appNNN}\end{aligned}\end{equation}

In this work we study $\pi/2$ site-centered disclinations, as depicted in Fig.~\ref{fig:Disclattice}. The important features of the disclination are the disclination core, indicated by the cross, and the hopping terms across the disclination cut, indicated by dashed lines. On this lattice, the Hamiltonian in Eq.~\ref{eq:appLatticeRealSpace} becomes: 
\begin{equation}
    \begin{aligned}
        \hat{\mathcal{H}}_{disc} &= \sum_{\bm{r}}  \frac{1}{2}c^\dagger(\bm{r}) T_0 c(\bm{r}) + c^\dagger(\bm{r}+\hat{z}) T_{+z} c(\bm{r})\\ 
        &+ \sum_{\substack{\langle \bm{r},\bm{r}'\rangle \\ \text{solid, }x}}  c^\dagger(\bm{r}') T_{x} c(\bm{r}) + \sum_{\substack{\langle \bm{r},\bm{r}'\rangle \\\text{solid, }y}}  c^\dagger(\bm{r}') T_{y} c(\bm{r}) \\ 
        &+ \sum_{\substack{\langle \bm{r},\bm{r}'\rangle \\ \text{dashed}}}  c^\dagger(\bm{r}') T_{\text{disc}} c(\bm{r}) \\
        &+ \text{H.c.},
        \label{eq:appLatticeRealSpaceDisc2}
    \end{aligned}
\end{equation}
where the first sum is over all sites of the lattice, the second sum is over sites connected by solid lines along the $+x$-direction in Fig.~\ref{fig:Disclattice}, the third sum is over sites connected by solid lines along the $+y$-direction, and the fourth sum is over sites connected by dashed lines. 
We exclude hopping terms between the disclination core and its nearest-neighbor sites as these terms are not determined by the bulk Hamiltonian. If included, these terms must be carefully chosen to respect the $C_3$ rotation symmetry of the disclinated lattice. The terms involving $T_{disc}$ are hopping terms across the disclination. Since crossing the disclination rotates the fermions, the disclination hopping terms are given by
 \begin{equation}\begin{split}
 T_{\text{disc}}=T_x U_4^{-1} = U_4^{-1}T_y,
 \end{split}\end{equation}
 where $U_4$ is the $C_4$ rotation matrix. 

The NNN hopping terms on the disclinated lattice are
\begin{equation}
    \begin{split}
        \hat{\mathcal{H}}_{\text{disc, NNN}} =
         &\sum_{\substack{\langle \bm{r},\bm{r}'\rangle \\ \text{solid, }x}}  c^\dagger(\bm{r}'+\hat{z}) T_{x+z} c(\bm{r}) \\
        + &\sum_{\substack{\langle \bm{r},\bm{r}'\rangle \\ \text{solid, }x}} c^\dagger(\bm{r}'-\hat{z}) T_{x-z} c(\bm{r})\\ 
        + &\sum_{\substack{\langle \bm{r},\bm{r}'\rangle \\ \text{solid, }y}}  c^\dagger(\bm{r}'+\hat{z}) T_{y+z} c(\bm{r}) \\
        + &\sum_{\substack{\langle \bm{r},\bm{r}'\rangle \\ \text{solid, }y}} c^\dagger(\bm{r}'-\hat{z}) T_{y-z} c(\bm{r})\\ 
        + &\sum_{\substack{\langle \bm{r},\bm{r}'\rangle \\ \text{dashed}}}  c^\dagger(\bm{r}'+\hat{z}) T_{\text{disc, }+z} c(\bm{r}) \\ 
        + &\sum_{\substack{\langle \bm{r},\bm{r}'\rangle \\ \text{dashed}}}  c^\dagger(\bm{r}'-\hat{z}) T_{\text{disc, }-z} c(\bm{r}) \\
        + &\text{H.c.}\,.  
        \label{eq:appLatticeRealSpaceDisc}
    \end{split}
\end{equation}
The first and second sums are over sites $\bm{r}$ and $\bm{r}'$ in the same xy-plane that are connected by a solid line along the $+x$-direction in Fig.~\ref{fig:Disclattice}. The third and fourth sums are over sites $\bm{r}$ and $\bm{r}'$ in the same xy-plane that are connected by a solid line along the $+y$-direction. The third and fourth sums are over sites $\bm{r}$ and $\bm{r}'$ in the same xy-plane that are connected by a dashed line with a clockwise orientation. 
The next-nearest neighbor disclination hopping terms $T_{\text{disc}, \pm z}$ are defined as
\begin{equation}\begin{split}
 T_{\text{disc, } +z} = -T_{\text{disc, } -z} = T_{x + z} U_4^{-1} = U_4^{-1}T_{y + z}.
\end{split}\end{equation}

\section{Zero-energy Surface Mode}\label{app:ZeroEnergy}
In this appendix we demonstrate the existence of a zero energy mode on the surface of the spinless rTCI when a mass vortex is added. The continuum surface Hamiltonian is
\begin{equation}
    H=\sigma^x\sigma^0(i\partial_x)-\sigma^y\sigma^0(i\partial_y),
\end{equation}
and the mass vortex term takes the form
\begin{equation}
    H_M= m_s(r)\sigma^z\left(\sigma^x\cos\theta + \sigma^y\sin\theta\right),
\end{equation} where $\theta$ is the polar-coordinate angle in the $xy$-plane, and $m_s(r)$ is a function of the polar-coordinate radius that vanishes at $r=0$. It is convenient to convert the Hamiltonian to polar coordinates to identify the zero energy eigenstate:
\begin{align}
    H=&i\begin{pmatrix}
        0&e^{i\theta}\\
        e^{-i\theta}&0
    \end{pmatrix}\otimes\sigma^0\frac{\partial}{\partial r}-
    \begin{pmatrix}
        0&e^{i\theta}\\
        -e^{-i\theta}&0
    \end{pmatrix}\otimes\sigma^0\frac{1}{r}\frac{\partial}{\partial \theta}
    \nonumber\\
    +&m_s(r)\sigma^z\otimes
    \begin{pmatrix}
        0&e^{-i\theta}\\
        e^{i\theta}&0
    \end{pmatrix}.
\end{align} Let us choose $m_s(r) = \bar{m}_s \Theta(r-R)$, where $\Theta$ is a step function and $R>0$. 
We make use of the following ansatz to obtain the zero mode:
\begin{equation}
    \Psi=\begin{pmatrix}
        u_1(r)e^{in\theta}\\
        u_2(r)e^{i(n+1)\theta}\\
        u_3(r)e^{i(n-1)\theta}\\
        u_4(r)e^{in\theta}
    \end{pmatrix}.
\end{equation}
Applying the Hamiltonian to this ansatz, we find that the zero energy solution must satisfy the following equations:
\begin{align}
    (\partial_r^2-\frac{n(n+1)}{r^2}-m_s^2)u_1&=0\\
    (\partial_r^2+\frac{2}{r}\partial_r-\frac{n(n+1)}{r^2}-m_s^2)u_2&=0\\
    (\partial_r^2+\frac{2}{r}\partial_r-\frac{n(n-1)}{r^2}-m_s^2)u_3&=0\\
    (\partial_r^2-\frac{n(n-1)}{r^2}-m_s^2)u_4&=0.
\end{align}
Making the substitution $u=r^{\frac{1}{2}}f$ in the first and last equations yields
\begin{equation}
    r^2\frac{d^2f}{dr^2}+r\frac{df}{dr}-[m_s^2r^2+n(n\pm1)+\frac{1}{4}]f=0,
\end{equation}
and making the substitution $u=r^{-\frac{1}{2}}f$ in the remaining two equations yields
\begin{equation}
    r^2\frac{d^2 f}{dr^2}+r\frac{df}{dr}-[m_s^2r^2+n(n\pm1)+\frac{1}{4}]f=0.
\end{equation}
The mass vanishes in the region $r<R$, $m_s(r)=0$, and the four equations have solutions
\begin{equation}
    u_1 = c_1r^{-n},\;u_2=c_2r^{-n-1},\;u_3=c_3r^{n-1},\;u_4=c_4r^{n}.
\end{equation}
For $r>R$ where the mass is finite, these equations can be solved with modified Bessel functions. 

As the potential is regular at the origin and at infinity, $\Psi$ must also be regular at $r=0$ and as $r\rightarrow \infty$. The wave function $\Psi$ must also be continuous at $r=R$. These conditions lead to the conclusion $u_2 = u_3 = 0$. First, suppose $u_2$ is finite for $r>R$. Then the regularity at $r\rightarrow \infty$ requires it to be a modified Bessel function of the second kind, which is non-vanishing at $r=R$. We therefore must have $n\leq -1$ to ensure continuity at $r=R$ and for $u_2$ to be regular at the origin. However, for $r<R$, the eigenvalue equations
\begin{align}
    u_2&=-\frac{i}{\bar{m}_s}e^{2i\theta} (\partial_r-\frac{1}{r}\partial_\theta)u_3 \\ 
    u_3&=\frac{i}{\bar{m}_s}e^{-2i\theta}(\partial_r+\frac{1}{r}\partial_\theta)u_2
\label{eq:appCoupled}\end{align}
indicates that $u_3$ must be non-zero for $r>R$ when $u_2$ is non-zero. Following the same logic as before, this indicates that $c_3\neq 0$, which in turn indicates that $n\geq 1$ for $u_3$ to be regular at the origin. Thus, there is a contradiction, and the only possibility is that $u_2=u_3 = 0$ for all r.

A similar argument applied to $u_1$ and $u_4$ leads to the conclusions $n=0$, and the solution must be
\begin{equation}
    u=\begin{pmatrix}
        1\\0\\0\\-i
    \end{pmatrix}\sqrt{\frac{\pi}{2\bar{m}_s}}
    \begin{cases}
        e^{-\bar{m}_sr}\;\;(r>R)\\
        e^{-\bar{m}_sR}\;\;(r<R)
    \end{cases}.
\end{equation}

The symmetries of the model are $\hat{ \mathcal{T}}=\sigma^y\otimes\sigma^y K$, $\hat C=\sigma^x\otimes\sigma^xK,$ and $\hat C_4 = \exp[i\frac{\pi}{4}(-\sigma^z\otimes\sigma^0+\sigma^0\otimes\sigma^z)].$ Under these symmetries the zero mode transforms as $\hat{ \mathcal{T}} u=-iu$, $\hat C u=iu$, and $\hat C_4u=u$.

\section{Topological crystalline insulator with PHS for spin-1/2 fermions}\label{app:Spin1/2PHS}
In this appendix we present a model for the spin-1/2 rTCI with TRS, PHS and $C_4$ rotation symmetry. The spin-1/2 rTCI is realized by the following 16-band model (8-bands per spin):
\begin{equation}\begin{split}
\mathcal{H}(\bm{k}) &= \Bigl[\sin(k_x)\Gamma^x\sigma^0 + \sin(k_y)\Gamma^y\sigma^0 + \sin(k_z)\Gamma^z\sigma^0\\
&+\sin(k_x)\sin(k_z)\Gamma^0\sigma^x + \sin(k_y)\sin(k_z)\Gamma^0\sigma^y\\
  &+(M+\cos(k_x)+\cos(k_y)+\cos(k_z))\Gamma^0\sigma^z\Bigr]\sigma^0,
\label{eq:appLatticeHamSpin1/2}\end{split}\end{equation}
where the spin of the fermions is given by $S^z = \frac{1}{2}\text{I}\sigma^0\sigma^z$. The spectrum for the lattice model is 4-fold degenerate and gapped for $|M|\neq 1,3$,
\begin{equation}\begin{split}
E_{\pm}(\bm{k}) &= \Bigl[\sin(k_x)^2 + \sin(k_y)^2 + \sin(k_z)^2\\
&+\sin(k_x)^2\sin(k_z)^2 + \sin(k_z)^2 \sin(k_z)^2\\
&+ (M+\cos(k_x)+\cos(k_y)+\cos(k_z))^2 \Bigr]^{1/2}.
\end{split}\end{equation}
Here we show that this lattice model realizes a spin-1/2 rTCI with a $\Phi = 2\pi$ $RF$-term for $1<|M|<3$. 

Eq.~\ref{eq:appLatticeHamSpin1/2} conserves charge and is invariant under TRS, PHS, and $C_4$. The TRS and PHS operations are defined as
\begin{equation}\begin{split}
&\hat{\mathcal{T}} = i\Gamma^y \sigma^y \sigma^y K, \\
&\hat{\mathcal{C}} = i\Gamma^{5y}\sigma^y \sigma^y K,
\end{split}\end{equation}
and $C_4$ rotation is defined as
\begin{equation}\begin{split}
\hat{U}_4 &= \exp(i \frac{\pi}{4}[ \Gamma^{yx}\sigma^0\sigma^0 + \text{I}\sigma^z\sigma^0 + \text{I}\sigma^0\sigma^z]).
\label{eq:appC4DefSpin1/2}\end{split}\end{equation}
Here $\hat{\mathcal{T}}^2 = (\hat{U}_4)^4= - 1 $ because the fermions have spin-1/2. 

\subsection{Response theory}\label{ssec:ResponseSpin1/2}
We follow the methodology used in Sec.~\ref{ssec:ResponseSpinless} to derive the response theory for the spin-1/2 model. We consider the system close to the band crossing at $M = -3$ where the low-energy degrees of freedom obtain the Dirac-like form
\begin{equation}\begin{split}
\mathcal{H} &= \left[\Gamma^x  \sigma^0  i \partial_x + \Gamma^y  \sigma^0  i\partial_y +\Gamma^z  \sigma^0  i \partial_z +m \Gamma^0  \sigma^z \right] \sigma^0, 
\label{eq:appLatticeHamLowEnergySpin1/2}\end{split}\end{equation}
with $m \sim M + 3$. To determine the effective response theory, we gauge the U$(1)$ charge and $C_4$ rotation symmetries and couple the fermions to the gauge field $A_\mu$ and spin connection $\omega_\mu$ via the covariant derivative (see Appendix~\ref{app:SpinConnectCoup}),
\begin{equation}\begin{split}
D_\mu &=   \partial_\mu - i A_\mu \\
&- i \frac{1}{2}\omega_\mu [\Gamma^{xy}\sigma^0\sigma^0 + \text{I}\sigma^z\sigma^0 + \text{I}\sigma^0\sigma^z ]. 
\label{eq:appCovariantDerivSpin1/2}\end{split}\end{equation}
Similar to before, the $C_4$ rotation symmetry of Eq.~\ref{eq:appLatticeHamLowEnergySpin1/2} is part of an enlarged U$(1)$ rotation symmetry. In addition to the gauge fields, we also include a PHS breaking perturbation
\begin{equation}\begin{split}
\mathcal{H}' =  m' \Gamma^5 \sigma^0\sigma^0,
\label{eq:appmass2spin1/2}\end{split}\end{equation}
and set $m = -\bar{m}\cos(\phi)$, and $m' = -\bar{m}\sin(\phi)$ , with $\bar{m}>0$, such that $m<0$ when $\phi = 0$, and $m>0$ when $\phi = \pi$. The effective response theory is obtained via a diagrammatic expansion in terms of $A_\mu$, $\omega_\mu$, and $\phi$. As before, we are primarily interested in the triangle diagrams shown in Fig.~\ref{fig:TDia}. 
The contribution from the triangle diagrams is
\begin{equation}\begin{split}
\mathcal{L}_{\text{eff}} = \frac{\phi}{2\pi^2} \epsilon^{\mu\nu\rho\kappa} \partial_\mu \omega_\nu \partial_\rho A_\kappa. 
\label{eq:appEffDerivedPhiSpin1/2}\end{split}\end{equation}
For $\phi = \pi$, the effective response is,
\begin{equation}\begin{split}
\mathcal{L}_{\text{eff}}= \frac{1}{2\pi} \epsilon^{\mu\nu\rho\kappa} \partial_\mu \omega_\nu \partial_\rho A_\kappa,
\label{eq:appEffDerivedPhiSpin2}\end{split}\end{equation}
which is exactly the $RF$-term with $\Phi = 2\pi$. We therefore find that the continuum model with $m>0$ (equiv. the lattice model with $-3<M<-1$) is a spin-1/2 rTCI with a $\Phi = 2\pi$ $RF$-term. Repeating this procedure for the band crossings at $M = \pm 1,3$ we conclude that $\Phi = 2\pi$ for $1<|M|<3$ and vanishes otherwise.

\subsection{Surface Theory}\label{ssec:SurfaceTheorySpin1/2}
Here we analyze the surface theory of the spin-1/2 rTCI. For a $C_4$ invariant surface with $-3<M < 1$ for $z<0$ and $M < -3$ for $z>0$, the surface theory consists of four 2-component Dirac fermions,
\begin{equation}\begin{split}
\hat{\mathcal{H}}_{\text{surf}} &= \psi^\dagger \mathcal{H}_{\text{surf}} \psi, \\
\mathcal{H}_{\text{surf}}& = \left[\sigma^x i\partial_x - \sigma^y i\partial_y\right] \sigma^0\sigma^0,
\label{eq:appSurfHamSpin1/2}\end{split}\end{equation}
where $\psi$ is an $8$-component spinor. The spin of the surface fermions is given by $S^z_{\text{surf}} = \frac{1}{2}\sigma^0\sigma^0 \sigma^z$ and the surface symmetry operations are
\begin{equation}\begin{split}
&\hat{\mathcal{T}}_{\text{surf}} = \sigma^y  \sigma^y \sigma^y K,\\
&\hat{\mathcal{C}}_{\text{surf}} = \sigma^x  \sigma^x \sigma^y K, \\
&\hat{U}_{4-\text{surf}} = \exp\left[i\frac{\pi}{4}\left(-\sigma^z  \sigma^0\sigma^0 + \sigma^0 \sigma^z \sigma^0 + \sigma^0 \sigma^0  \sigma^z\right)\right].
\label{eq:appspin1/2sym}\end{split}\end{equation}
The symmetry operations satisfy $\hat{\mathcal{T}}_{\text{surf}}^2 =  (\hat{U}_{4-\text{surf}})^4 =-1$ because the fermions have spin-1/2. 

As expected, the surface theory is gapped out by the PHS breaking surface mass term $m_s \sigma^z\sigma^z\sigma^0$. To find the response theory for the massive, PHS breaking surface, we once again introduce the gauge field $A_\mu$ and spin connection $\omega_\mu$ via the covariant derivative 
\begin{equation}\begin{split}
D_\mu &= \partial_\mu -i A_\mu \\
&- i \frac{1}{2} \omega_\mu [-\sigma^z  \sigma^0\sigma^0 + \sigma^0  \sigma^z\sigma^0 + \sigma^0\sigma^0\sigma^z].
\end{split}\end{equation}
The response theory for the spin-1/2 surface is given by the Wen-Zee term
\begin{equation}\begin{split}
\mathcal{L}_{\text{surf}} = \frac{\text{sgn}(m_s)}{2\pi} \epsilon^{\mu\nu\rho} \omega_\mu \partial_\nu A_\rho. 
\label{eq:appSurfaceResponseTh}\end{split}\end{equation}
This is exactly the anomalous surface term where $\Delta \Phi = 2\pi$, and indicates that charge $\pm \frac{1}{4}$ is bound to $\pi/2$ disclinations on the surface. The coefficient of the surface Wen-Zee term can be shifted by $1/\pi$ by purely surface effects, and, in general, a surface $\pi/2$ disclination binds charge $\frac{1}{4}+\frac{m}{2}$ for $m \in \mathbb{Z}$.

\subsection{Dimensional reduction to a 1+1D SPT}\label{sec:DimRedSpin1/2}
In this subsection, we use the logic of Ref. \onlinecite{song2017topological} and dimensionally reduce the $3$D spin-1/2 rTCI to a $1$D SPT. The resulting $1$D SPT is equivalent to the spin-1/2 SSH chain, with an additional $\mathbb{Z}_4$ symmetry that is inherited from the $C_4$ symmetry of the rTCI. The spin-1/2 SSH chain can be though of as a doubled version of the spinless SSH, one copy per spin. The edge of this system hosts two zero-energy modes that form a Kramers' pair. The edge also has charge $\pm 1$ when TRS is preserved on the edge. The $\pm 1$ edge charge is protected for the spin-1/2 system, unlike the spinless version, since TRS requires that particles are added in Kramers' pairs, which carry charge 2. The $\mathbb{Z}_4$ symmetry of the SSH chain can be interpreted as a discrete internal spin rotation symmetry along the z-axis. In this interpretation, the two zero-energy modes of the spin-1/2 SSH chain have spin $S^z = \pm \frac{1}{2}$ respectively. 

To show that the spin-1/2 rTCI can be dimensionally reduced to this system, we add a mass term of the form,
\begin{equation}\begin{split}
\mathcal{H}_{\text{surf-mass}} &= \left[m_x \sigma^z\sigma^x + m_y\sigma^z\sigma^y + m_z \sigma^z\sigma^z\right]\sigma^0,
\label{eq:appDimRedMassTermSpin1/2}\end{split}\end{equation}
to the surface theory in Eq.~\ref{eq:appSurfHamSpin1/2}, and set $m_z = 0$ and $m_x + im_y =  m_s(r) \exp(i\theta)$. Here, $(r,\theta)$ are polar coordinates on the surface and $m_s(r) \geq 0$ is a function of the radial coordinate that vanishes at $r = 0$ and goes to a non-vanishing constant value $\bar{m}_s > 0$ as $r\rightarrow \infty$. This mass term trivializes the surface, except for at the $C_4$ rotation center. 

At the rotation center, there are two localized zero-energy modes, $\psi_{0\uparrow}$, and $\psi_{0\downarrow}$ (see Appendix~\ref{app:ZeroEnergy}). These two zero-energy modes form a Kramers' pair under TRS, $\mathcal{T}: (\psi_{0\uparrow},\psi_{0\downarrow}) \rightarrow (\psi_{0\downarrow},-\psi_{0\uparrow})$. Under a $C_4$ rotation the zero modes transform as $C_4: (\psi_{0\uparrow},\psi_{0\downarrow})\rightarrow (\psi_{0\uparrow} e^{i\frac{\pi}{4}},\psi_{0\downarrow} e^{-i\frac{\pi}{4}})$. The $\psi_{0\uparrow}$ mode therefore has internal angular momentum (i.e. spin) $+1/2$ and $\psi_{0\downarrow}$ has spin $-1/2$. Since the zero modes have finite spin, both of the zero modes must either be empty or occupied in order to preserve TRS.


When the two zero-energy modes are empty, the effective response theory for the massive surface is
\begin{equation}\begin{split}
\mathcal{L}_{\text{eff-surf}} =& \frac{\epsilon^{\mu\nu\rho}}{4\pi} \bm{n}\cdot(\partial_\mu \bm{n}\times \partial_\nu \bm{n})A_\rho \\ 
&+\frac{n_z}{2\pi} \epsilon^{\mu\nu\rho} \omega_\mu \partial_\nu A_\rho,
\label{eq:appNLSMResponseSpin1/2}\end{split}\end{equation}
where $\bm{n} = \bm{m}/|\bm{m}|$ and $\bm{m} = (m_x,m_y,m_z)$.
Using the definition of the mass terms from above, we find there is a charge $Q= -1$ localized at $r = 0$. Due to the aforementioned gapless modes at $r=0$, this charge is only defined mod$(2)$ for a time-reversal invariant surface. 

Based on this, we can conclude that the surface physics of the mass-deformed spin-1/2 rTCI matches the surface physics of a spin-1/2 SSH chain with additional $\mathbb{Z}_4$ spin rotation symmetry. Namely, both surfaces have two zero modes, which form a Kramers' pair and have spin $\pm 1/2$. Additionally, when TRS is preserved, the charge at the surfaces is $1$ mod $(2)$. Using the bulk boundary correspondence, we conclude that the spin-1/2 rTCI and spin-1/2 SSH chain with additional $\mathbb{Z}_4$ spin-rotation symmetry are equivalent. 

\subsection{Surface Topological Order}\label{ssec:SurfTopoSpin1/2}
Similar to the spinless model, the spin-1/2 rTCI admits a symmetric gapped topologically ordered surface state. This surface state has anyon content $\{1,v,v^2,v^3,w,w^2,w^3,v^aw^b\}\times\{1,f\}$, for $a,b = 1,2,3$. Similar to before, the $f$ particle is a fermion and the $v$ and $w$ anyons are self-bosons with $\pi/2$ mutual statistics. The $v$ particle has charge $\frac{1}{2}$ and angular momentum $0$, and the $w$ particle has charge $0$ and angular momentum $\frac{1}{2}$. This topological order can be viewed as the spinless topological order described in Sec.~\ref{ssec:SurfTopo}, except that the $m$ particle has angular momentum $\frac{1}{2}$ instead of $\frac{1}{4}$. Nevertheless, we shall label the anyons of the spin-1/2 surface topological order as $v$ and $w$ instead of $e$ and $m$ to avoid confusion. 

The spin-1/2 topological order can be constructed using a vortex proliferation argument similar to that in Sec.~\ref{ssec:SurfTopo}. As before, the starting point is to add a superconducting term to the surface theory. If we divide the 8-component spinor in Eq.~\ref{eq:appSurfHamSpin1/2} into four 2-component Dirac fermions, $\psi = (\psi_1,\psi_2,\psi_3,\psi_4)$, the superconducting surface can be written as 
\begin{equation}\begin{split}
\hat{\mathcal{H}}_{\text{SC}} =& i \Delta_1 \psi_1 \sigma^y \psi_1 + i \Delta_2 \psi_2 \sigma^y \psi_2\\ &  + i \Delta_3 \psi_3 \sigma^y \psi_3 + i \Delta_4 \psi_4 \sigma^y \psi_4 + \text{H.c.}
\end{split}\end{equation}
Under $U(1)$ and $C_4$ rotations, the $\Delta_i$'s transform as,
\begin{equation}\begin{split}
 &\text{U}(1): (\Delta_1,\Delta_2, \Delta_3,\Delta_4)\\ &\phantom{====}\rightarrow (\Delta_1e^{i 2\theta},\Delta_2e^{i 2\theta},\Delta_3e^{i 2\theta},\Delta_4e^{i 2\theta})\\
 &C_4:  (\Delta_1,\Delta_2, \Delta_3,\Delta_4)\\ &\phantom{====} \rightarrow (\Delta_1e^{i \pi},\Delta_2, \Delta_3 , \Delta_4e^{-i \pi}).
 \label{eq:appSCSymmetryTransSpin1/2}\end{split}\end{equation}
Therefore $\Delta_1$ describes a Cooper pair with charge $2$ and angular momentum $2$, $\Delta_2$ and $\Delta_3$ describe Cooper pairs with charge $2$ and angular momentum $0$, and $\Delta_4$ describes a Cooper pair with charge $2$ and angular momentum $-2$. By extension, there also exist composite Cooper pairs with charge $0$ and angular momentum $\pm 2$. The TRS and PHS operations act via
\begin{equation}\begin{split}
&T: (\Delta_1,\Delta_2, \Delta_3,\Delta_4)\rightarrow -(\Delta^*_4,\Delta^*_3,\Delta^*_2,\Delta^*_1)\\
&C: (\Delta_1,\Delta_2, \Delta_3,\Delta_4)\rightarrow (\Delta^*_4,\Delta^*_3,\Delta^*_2,\Delta^*_1).
\label{eq:appSCSymmetryTransSpin1/2OS}\end{split}\end{equation}

 We can identify 2-types of vortices that must be proliferated in order to restore all symmetries. First are $2\pi n$ $w$-vortices, where all $\Delta_{i}$ wind by $2\pi n$. Second are $2\pi n$ $v$-vortices where $\Delta_1$ winds by $2\pi n$, $\Delta_4$ winds by $-2\pi n$, and $\Delta_{2}$ and $\Delta_3$ are left invariant. Based on Eq.~\ref{eq:appSCSymmetryTransSpin1/2}, a $2\pi$ $w$-vortex is generated by a $\pi$ electromagnetic flux, while a $2\pi$ $v$-vortex is generated by a $\pi$ disclination. 

Using the effective response theory, we find that a $-2\pi$ $w$-vortex has charge $0$ and angular momentum $\frac{1}{2}$, and a $-2\pi$ $v$-vortex has charge $\frac{1}{2}$ and angular momentum $0$. Both the $w$ and $v$-vortices are self-bosons, and the $2\pi$ $w$ and $v$-vortices have $\pi/2$ mutual statistics. Additionally, a $2\pi$ $w$-vortex binds 4 Majorana fermions (2-complex fermions) and a $2\pi$ $v$-vortex binds $2$ Majorana fermions (1-complex fermion). Following the same logic used in Sec.~\ref{ssec:SurfTopo}, the following two types of vortices can be simultaneously condensed: first, an $8\pi$ $w$-vortex and composite Cooper pair with charge $0$ angular momentum $-2$, and second, an $8\pi$ $v$-vortex along and a Cooper pair with charge $2$ angular momentum $0$. The Majorana zero modes of these vortices can all be gapped while preserving symmetry. Based on Eqs.~\ref{eq:appSCSymmetryTransSpin1/2} and~\ref{eq:appSCSymmetryTransSpin1/2OS}, proliferating these two types of vortices restores the symmetry of the surface. 

The anyon content of the gapped surface theory corresponds to the vortices that have trivial braiding statistics with the condensate. These are the $2\pi n$ $w$-vortices, $2\pi n$ $v$-vortices, and their combinations. There is also a fermion $f$, which is the remnant of the gapped fermionic zero modes. The $-2\pi$ $w$-vortex is a self-boson with charge $0$ and angular momentum $\frac{1}{2}$, and it constitutes the $w$ anyon. The $-2\pi$ $v$-vortex is a self-boson with charge $\frac{1}{2}$ and angular momentum $0$, and it constitutes the $v$ anyon. The $v$ and $w$ anyons have $\pi/2$ mutual statistics. The $w^4$ and $v^4$ anyons have trivial braiding statistics and unfractionalized quantum numbers, so they can be regarded as local particles that do not enter into the anyonic data. 

We conclude that the topological order described at the beginning of the section can be realized on the surface of the spin-1/2 rTCI. Due to the same logic used in Sec.~\ref{ssec:SurfTopo}, this topological order cannot be realized in a purely $2$D system with PHS, but can be realized on the surface of the particle-hole symmetric spin-1/2 rTCI.

\section{Topological crystalline insulator with mirror symmetry for spinless fermions}\label{app:spinlessMirrorrTCI}
In this subsection, we present a model for the spinless rTCI with TRS, $C_4$ rotation symmetry and mirror symmetry. Our starting point is the 8-band lattice model in Eq.~\ref{eq:LatticeHam},
\begin{equation}\begin{split}
\mathcal{H}(\bm{k}) &= \left[\sin(k_x)\Gamma^x + \sin(k_y)\Gamma^y + \sin(k_z)\Gamma^z\right]\sigma^0\\ 
&+\sin(k_x)\sin(k_z)\Gamma^0\sigma^x + \sin(k_y)\sin(k_z)\Gamma^0\sigma^y\\
&+(M+\cos(k_x)+\cos(k_y)+\cos(k_z))\Gamma^0\sigma^z,
\end{split}\end{equation}
In Sec.~\ref{sec:SpinlessLattice} we were primarily interested in the topological features of Eq.~\ref{eq:LatticeHam} associated with PHS. Here, we are interested in the topological features associated with the mirror symmetry. The mirror symmetry operator is given by
\begin{equation}\begin{split}
\hat{\mathcal{M}}_z = \Gamma^{5z} \sigma^0,
\end{split}\end{equation}
and satisfies the relation $\hat{\mathcal{M}}_z^{-1} \mathcal{H}(k_x,k_y,k_z) \hat{\mathcal{M}}_z = \mathcal{H}(k_x,k_y,-k_z)$. Since mirror reflection is equivalent to the combination of a $\pi$ rotation and inversion, $(\hat{\mathcal{M}}_z)^2 = +1$ for spinless fermions. 

The lattice model also has PHS,
\begin{equation}\begin{split}
\hat{\mathcal{C}} = \Gamma^{5y} \sigma^y \mathcal{K},
\end{split}\end{equation}
but the PHS should be regarded as an ``accidental'' symmetry of the lattice model and we explicitly break this symmetry throughout this section. 

\subsection{Response theory}
The response theory for the rTCI with mirror symmetry is with the same technique as in Sec.~\ref{ssec:ResponseSpinless}. The continuum theory near the band crossing at $M = -3$ is the same as in Eq.~\ref{eq:LatticeHamLowEnergy}. In the continuum limit, the effective response theory is found by coupling the Dirac fermions to the spin connection $\omega$ and U$(1)$ gauge field $A$ (see Eq.~\ref{eq:CovariantDeriv}). We also include the perturbation in Eq.~\ref{eq:ResponsePert} and set $m=-\bar{m}\cos(\phi)$ and $m'=-\bar{m}\sin(\phi)$. Here, if $\phi$ is a function of $z$, mirror symmetry is preserved only when $\phi(z) = -\phi(-z)$ mod$(2\pi)$. After integrating out the massive fermions, the response theory as a function of $\omega$, $A$, and $\phi$ is again given by Eq.~\ref{eq:EffDerivedphi}. 

When $\phi$ is constant, mirror symmetry requires that $\phi = 0$ or $\pi$. The former corresponds to the $m<0$ insulator ($M < -3$ in the lattice model), which has a trivial $RF$-term. The latter corresponds to the $m > 0$ insulator ($-3<M<-1$ in the lattice model), which is a mirror symmetric rTCI with a $\Phi = \pi$ $RF$-term. This analysis is much the same as that of the rTCI with PHS. However, as noted before, the coefficient of the $RF$-term can fluctuate while preserving mirror symmetry, provided that $\Phi(z) = -\Phi(-z)$ mod$(2\pi)$. Because of this, it is possible to have mirror symmetry preserving domain walls between the rTCI and a trivial insulator (see Sec.~\ref{ssec:SymQuantMirror}). We show this explicitly in the next subsection.

\subsection{Surface Theory}
To analyze the surface theory of the rTCI with mirror symmetry, we consider a pair of domain walls that are related to one another by mirror symmetry. Specifically, we use a geometry where $-3 < M < 1$ for $|z| < z_{\text{dw}}$ and $M < -3$ for $|z| > z_{\text{dw}}$, which corresponds to a pair of symmetry related domain walls at $z = \pm z_{\text{dw}}$ ($z_{\text{dw}}$ is taken to be large compared to the correlation length of the insulators). 

The Hamiltonians for the two surfaces are,
\begin{equation}\begin{split}
&\mathcal{H}_{\text{t}} = \left[\sigma^x  i\partial_x - \sigma^y i\partial_y \right]\sigma^0,\\ &\mathcal{H}_{\text{b}} = \left[\sigma^x  i\partial_x - \sigma^y i\partial_y\right]\sigma_0,\\ 
\label{eq:appSurfHamMirror}\end{split}\end{equation}
where the t and b subscripts indicate the top and bottom surfaces, respectively. The two surface theories can be combined as
\begin{equation}\begin{split}
\mathcal{H}_{\text{t-b}} &= \left[\sigma^x i\partial_x - \sigma^y i\partial_y\right]\sigma^0\sigma^0,
\label{eq:appSurfHamMirror2}\end{split}\end{equation}
where the two domain walls are indexed by $\sigma^0\sigma^0\sigma^z$. 
Mirror symmetry acts on Eq.~\ref{eq:appSurfHamMirror2} as
\begin{equation}\begin{split}
&\hat{M}_{z-\text{surf}} = \sigma^0 \sigma^0 \sigma^x.
\end{split}\end{equation}

Using the $8$-band description of the surfaces in Eq.~\ref{eq:appSurfHamMirror2}, there are two surface mass terms of note. First is the mass term promotional to $\sigma^z \sigma^z \sigma^z$ that preserves TRS and breaks mirror symmetry. Second is the mass term proportional to $\sigma^z \sigma^z \sigma^0$ that which preserves both TRS and mirror symmetry. Hence, in agreement with our discussion from Sec.~\ref{ssec:SymQuantMirror}, we find that the surface Dirac fermions are not protected by mirror symmetry. 

For the mirror symmetry breaking surface masses, the surface response theory consists of two Wen-Zee terms, one per surface. The Wen-Zee terms have coefficients of the form $1/4\pi$ mod$(1/2\pi)$, half the amount allowed in $2$D systems. In general, the two surfaces will have different coefficients. Following the same logic used before, a $\pi/2$ disclination of the rTCI with mirror symmetry breaking surfaces binds charge $\pm \frac{1}{8}$ mod$(\frac{1}{4})$ on one surface and charge $\mp \frac{1}{8}$ mod$(\frac{1}{4})$ on the other surface. 

For the mirror symmetry preserving surface masses, the response theory consists of a Wen-Zee term defined on each surface. Due to mirror symmetry, these Wen-Zee terms have the same coefficient of $1/4\pi$ mod$(1/2\pi)$. A $\pi/2$ disclination of the rTCI with mirror symmetry preserving surfaces will therefore bind charge $\pm \frac{1}{8}$ mod$(\frac{1}{4})$ on \textit{both} surfaces.

\subsection{Dimensional Reduction to $1$D SPT}
Here we dimensionally reduce the rTCI with mirror symmetry to a $1$D SPT, as in Sec.~\ref{sec:DimRed}. The resulting SPT is the SSH chain protected by mirror symmetry with a trivial on-site $\mathbb{Z}_4$ symmetry. Much like the SSH chain with PHS, the SSH chain with mirror symmetry has half-integer quantized charges at its boundaries. However, the SSH chain with mirror symmetry does not have protected zero edge modes. This is because mirror symmetry only requires that the energies of the two edge modes are equal to each other, (under PHS, the energies of any edge modes must be exactly zero). However, the fractional charge localized at a pair of mirror symmetry related edges must be the same for the SSH chain, which leads to a filling anomaly~\cite{benalcazar2019quantization}. 

With this in mind, consider the two surface Hamiltonians in Eq.~\ref{eq:appSurfHamMirror} with additional mass perturbations of the form
\begin{equation}\begin{split}
&\mathcal{H}_{\text{t-mass}} =  m_{x,\text{t}} \sigma^z\sigma^x + m_{y,\text{t}} \sigma^z\sigma^y + m_{z,\text{t}} \sigma^z \sigma^z,\\ &\mathcal{H}_{\text{b-mass}} = m_{x,\text{b}} \sigma^z\sigma^x + m_{y,\text{b}} \sigma^z\sigma^y + m_{z,\text{b}} \sigma^z \sigma^z.\\ 
\label{eq:appSurfHamMirrorMass}\end{split}\end{equation}
Under mirror symmetry, $m_{i,\text{t}} \rightarrow m_{i,\text{b}}$, for $i = x,y,z$. As discussed in Sec.~\ref{sec:DimRed}, the rTCI can be dimensionally reduced to an SSH chain with gapless edge modes by setting $m_{z,\text{t}} = m_{z,\text{b}} = 0$, $m_{x,\text{t}} + im_{y,\text{t}} = m_{x,\text{b}} + im_{y,\text{b}} = m_s(r) \exp(i\theta)$, where $(r,\theta)$ are polar coordinates on the surface, and $m_s(r) \geq 0$ is a function of the radial coordinate that vanishes at $r = 0$, and goes to a non-zero constant $\bar{m}_s$, as $r\rightarrow \infty$. This mass configuration preserves TRS, mirror symmetry and $C_4$ rotation symmetry. Similar to before, there is a zero energy mode on each surface located near $r = 0$, which transforms trivially under $C_4$ symmetry.

It is possible to gap out the edge modes of the SSH chain by setting $m_{z,\text{t}} = m_{z,\text{b}} = \sqrt{\bar{m}^2_s - m_s(r)^2}$, such that $m_{z,\text{t}}$ and $m_{z,\text{b}}$ take on the same non-zero value at $r = 0$. This perturbation preserves all symmetries of the model, and gaps out the zero modes located at $r = 0$ on each surface (see Appendix~\ref{app:ZeroEnergy}).

We integrate out the massive fermions to determine the charge that is bound at $r = 0$, leading to the effective response theory
\begin{equation}\begin{split}
\mathcal{L}_{\text{eff-t}} =& \frac{\epsilon^{\mu\nu\rho}}{8\pi} \bm{n}_{\text{t}}\cdot(\partial_\mu \bm{n}_{\text{t}}\times \partial_\nu \bm{n}_{\text{t}})A_\rho\\
&+\frac{n_z}{4\pi} \epsilon^{\mu\nu\rho} \omega_\mu \partial_\nu A_\rho,\\
\mathcal{L}_{\text{eff-b}} =& \frac{\epsilon^{\mu\nu\rho}}{8\pi} \bm{n}_{\text{b}}\cdot(\partial_\mu \bm{n}_{\text{b}}\times \partial_\nu \bm{n}_{\text{b}})A_\rho\\
&+\frac{n_z}{4\pi} \epsilon^{\mu\nu\rho} \omega_\mu \partial_\nu A_\rho,\\
\bm{n}_{\text{t}/\text{b}} =& \frac{\bm{m}_{\text{t}/\text{b}}}{|\bm{m}_{\text{t}/\text{b}}|},\phantom{=} \bm{m}_{\text{t}/\text{b}} = (m_{x,\text{t}/\text{b}},m_{y,\text{t}/\text{b}},m_{z,\text{t}/\text{b}}). 
\label{eq:appNLSMResponseMirror}\end{split}\end{equation}
For the mass configurations discussed above, the response theory indicates that charge $1/2$ is localized near $r = 0$ on both the top and bottom surfaces (this charge is only defined modulo $1$ due to surface effects).

Viewed as two $0$D systems, the rotation centers of the top and bottom surfaces each have an unprotected mode and carry the same half-integer of charge. These are exactly the characteristic features of the $0$D edges of a $1$D SSH chain with mirror symmetry. Using the bulk-boundary correspondence, we conclude that the deformed rTCI and SSH chain with mirror symmetry are adiabatically connected. 

It is worth noting that the $1$D SSH chain with mirror symmetry can be further dimensionally reduced to a non-trivial $0$D system with on-site $\mathbb{Z}_2$ symmetry, which is inherited from the mirror symmetry (see Ref. \onlinecite{khalaf2021boundary} for further discussion). By extension, the rTCI can also be dimensionally reduced to the same $0$D system, with an additional trivial $\mathbb{Z}_4$ symmetry. 

\section{Topological invariant and $RF$-term for Dirac-like insulators}\label{app:InvariantEx}

In this appendix we calculate the $RF$-term and topological invariant for insulators that have a Dirac-like band structure at the time-reversal invariant momentum (TRIM). We show that $\nu_{RF} = \Phi/\pi$ for spinless Dirac-like insulators with TRS, $C_n$ rotation, mirror, and inversion symmetry, where $\nu_{RF}$ is defined as in Eq.~\ref{eq:NuRF}. Similarly, we show that $\nu_{RF\uparrow} = \Phi/2\pi$ for spin-1/2  Dirac-like insulators with TRS, $C_n$ rotation, mirror, and inversion symmetry, and additional spin conservation, where $\nu_{RF \uparrow}$ is defined as in Eq.~\ref{eq:NuRFUp}.

The analysis we present here is simplified by the following observations. First, the combination of mirror and inversion symmetry leads to a $C_2$ symmetry, and so we need to consider $C_n$ symmetry only for $n = 2,4,6$, as combining $C_3$ and $C_2$ symmetry leads to $C_6$ symmetry. Second, for $C_n$-invariant systems ($n = 2,4,6$), the $RF$-term can be determined by gauging only the $C_2$ subgroup of the full rotation symmetry. All the TRIM are invariant under $C_2$ rotations, so these considerations greatly simplify our analysis. Physically, gauging only the $C_2$ rotation symmetry is equivalent to considering responses only to the $\pi$ disclinations of a $C_4$ or $C_6$ symmetric system. Since a $\pi$ disclination is the fusion of two $\pi/2$ disclinations or three $\pi/3$ disclination, the response of a system to either $\pi/2$ or $\pi/3$ disclinations can be deduced from the response of the system to $\pi$ disclinations and the disclination fusion rules. 

First, we consider the $RF$-term for a lattice model of spinless fermions, where the band-structure is Dirac-like near the TRIM. We take this model to have TRS (with $\mathcal{T}^2 = 1$), U$(1)$ charge conservation, $C_n$ symmetry, $M_z$ mirror symmetry, and inversion symmetry. As noted before, the rotation symmetry has a $C_2$ subgroup, and we need to consider only this $C_2$ subgroup to determine the $RF$-term of this system. A 3D spinless Dirac-fermion with TRS requires a minimum of 8-bands, and the lattice models therefore have $N_{\text{band}} \in 8\mathbb{Z}$ bands.

Our goal is to show that $\nu_{RF} = \Phi/\pi$ for these systems, where $\nu_{RF}$ is defined as in Eq.~\ref{eq:NuRF}. To do this, we note that any two band insulators with the same symmetries (and representations) can be symmetrically evolved into one another via a sequence of band crossings. Since the $RF$-term is quantized in mirror symmetric insulators, the difference in the $RF$-term between two mirror symmetric insulators is equal to the total change in the $RF$-term that occurs during the aforementioned gapless band crossings. Based on this, we prove that $\nu_{RF} = \Phi/\pi$ by first proving that $\nu_{RF} = 0$ for a trivial symmetric insulator (where $\Phi = 0$ by definition), and that any band crossing that generates a non-trivial $RF$-term also changes the value of $\nu_{RF}$.

To this end, take a trivial (atomic) insulator, where the lattice Hamiltonian contains only a constant on-site potential $\mathcal{H}_{\text{Triv}} = \sigma^z \otimes \text{I}_{N_{\text{band}}/2}$.  Importantly, the band structure is constant throughout the Brillouin zone, which implies that $\eta_{0,+1} = \eta_{\pi, +1}$ (see Eq.~\ref{eq:etaDef}). Because of this, $\nu_{RF}=0$ for such a spinless trivial insulator.

Now consider a generic symmetric band crossing. This band crossing can either occur at a TRIM or at an arbitrary point in the Brillouin zone. We begin with the latter case. Due to the $C_2$ and $M_z$ symmetries of the lattice Hamiltonian, such a band crossing must be accompanied by an odd number of other symmetry related band crossings. In general, it is possible to adiabatically and symmetrically evolve the lattice Hamiltonian such that the momentum space distance between the multiple band crossings is taken to zero (modulo a reciprocal lattice vector). After this evolution, an even number of band crossings will occur at a single TRIM, $\Lambda_n$. Because of this, any two band insulators that are related by band crossings at arbitrary momenta are also related by an even number of band crossings that only occur at TRIM. Since moving the location of the band crossings can be done symmetrically and adiabatically, it does not affect the change in the $RF$-term that is generated by the band-crossing. 

We now consider the case where the band crossing occurs at a TRIM, $\Lambda_n$ (recall that we assume the band-structure is Dirac-like at $\Lambda_n$). To begin, we analyze a band crossing that involves 8-bands (the minimal number of bands for a spinless Dirac fermion with TRS). In an appropriate basis, the Hamiltonian for the low energy bands can be written as
 \begin{equation}\begin{split}
\mathcal{H}_{\text{Dirac}} &= \Gamma^x \sigma^0 i \partial_x  + \Gamma^y\sigma^0 i\partial_y  +  \Gamma^z  \sigma^0 i \partial_z\\ &+ m \Gamma^0 \sigma^z,
\label{eq:GenHam}\end{split}\end{equation}
where $m$ parameterizes the band-crossing and the $\Gamma$-matrices are a set of $4\times 4$ anti-commuting matrices. The TRS, $C_2$, inversion, and mirror symmetries act on Eq.~\ref{eq:GenHam} via 
\begin{equation}\begin{split}
&\hat{\mathcal{T}} = \Gamma^y \sigma^y \mathcal{K},\\
&\hat{U}_2 = (\hat{U}_n)^{\frac{n}{2}} = \exp(i \frac{\pi}{2} [\Gamma^{xy} \sigma^0  +  \text{I}_4 \sigma^z ]\text{I}_N),\\
&\hat{P} = \Gamma^0 \sigma^z,\\
&\hat{\mathcal{M}}_z = \hat{U}_2\hat{P}.
\label{eq:SymmetryDefGen}\end{split}\end{equation}
We now determine how the $RF$-term for Eq.~\ref{eq:GenHam} changes during the band crossing where $m$ changes from a negative to a positive value. As before, we do this by gauging the U$(1)$ and $C_2$ symmetries, and coupling the Dirac fermions to the electromagnetic gauge field $A_\mu$ and spin connection $\omega_\mu$, via the covariant derivative 
\begin{equation}\begin{split}
D_\mu = \partial_\mu - iA_\mu - i \omega_\mu \frac{1}{2}[\Gamma^{xy}\sigma^0  + \text{I}_4 \sigma^z].
\end{split}\end{equation}
We also add an additional mass term,
\begin{equation}\begin{split}
\mathcal{H}'= m' \Gamma^5 \sigma^0 ,
\end{split}\end{equation}
that preserves TRS but breaks mirror symmetry. Upon setting $m = -\bar{m}\cos(\phi),$ $m' = -\bar{m}\sin(\phi)$ and integrating out the massive fermions, we find the following $RF$-term:
\begin{equation}\begin{split}
\mathcal{L}_{\text{eff}}[A_\mu, \omega_\mu, \phi]= \frac{\phi}{2\pi^2} \epsilon^{\mu\nu\rho\kappa} \partial_\mu \omega_\nu \partial_\rho A_\kappa.
\label{eq:EffDerivedPhiSpin}\end{split}\end{equation}
As we can see, the coefficient of the $RF$-term $\Phi$ shifts by $\pi$ when $m$ changes sign (since the $RF$-term is a total derivative, it is only possible to determine the \textit{change} in the coefficient of the $RF$-term during this process). It is straightforward to generalize this result to a Dirac band crossing at $\Lambda_n$ that involves $8N$ bands, in which case $\Phi$ shifts by $N\pi$ during the band crossing. Taking into account that $\Phi$ is defined mod$(2\pi)$, the $8N$ band crossing changes the physical value of $\Phi$ by $0$ when $N$ is even and by $\pi$ when $N$ is odd. 

Let us now consider the 8-band Dirac Hamiltonian restricted to the mirror invariant plane that contains $\Lambda_n$. On this mirror plane, we can divide the Hamiltonian into sectors with mirror eigenvalue $\hat{\mathcal{M}}_z = \pm1$. The Hamiltonian for the $\hat{\mathcal{M}}_z = +1$ sector can be written as
 \begin{equation}\begin{split}
\mathcal{H}_{\text{mirror}} &= \Gamma^x i \partial_x  - \Gamma^y i\partial_y  - m  \Gamma^z,
\label{eq:GenMirrorHam}\end{split}\end{equation}
where the TRS, and $C_2$ (equivalently 2D inversion) 
symmetry act as
\begin{equation}\begin{split}
&\hat{\mathcal{T}}_{\text{mirror}} = \sigma^y \sigma^y \mathcal{K},\\
&\hat{U}_2 = \hat{P} =\exp(i \frac{\pi}{2} [-\Gamma^{xy}  +  \sigma^0 \sigma^z ]).
\label{eq:SymmetryDef2D}\end{split}\end{equation}
Eq.~\ref{eq:GenMirrorHam} therefore describes a pair of TRS related $2$D Dirac fermions. Using Eq.~\ref{eq:ChernNumberC2} we find that the Chern number parities of the two occupied bands with $\hat{\mathcal{M}}_z = +1$ at this mirror invariant plane change sign when $m$ changes. Based on this, the band crossing sends $\eta_{k_z,+1}\rightarrow -\eta_{k_z,+1},$ where $k_z = 0,\pi$ is the mirror invariant plane that contains $\Lambda_n$ (see Eq.~\ref{eq:etaDef}). It is again straightforward to extend this analysis to an $8N$-band crossing at $\Lambda_n$, where the band crossing sends $\eta_{k_z,+1} \rightarrow +\eta_{k_z,+1}$ when $N$ is even and $\eta_{k_z,+1} \rightarrow -\eta_{k_z,+1}$ when $N$ is odd. Based on this, $\nu_{RF} \rightarrow \nu_{RF}+1$ mod$(2)$ and $\Phi\rightarrow \Phi+\pi$ mod($\pi$) if $N$ is odd for an $8N$-band, Dirac-like crossing at a TRIM. The values of both $\nu_{RF}$ and $\Phi$ do not change if $N$ is even. 

We can therefore conclude that when the coefficient of the $RF$-term changes due to \textit{any} symmetric band crossing, the value of $\nu_{RF}$ will also change and vice versa. Combined with the fact that $\nu_{RF} = 0$ for a trivial insulator, and that any two symmetric insulators (with the same representations) can be evolved into one another via a sequence of symmetric band crossings, we conclude that $\nu_{RF} = \Phi/\pi$ for the insulators considered here. 

We now turn our attention to Dirac-like lattice models of spin-1/2 fermions with TRS (with $\mathcal{T}^2 = -1$), U$(1)$ charge conservation, $C_n$ rotation symmetry, mirror symmetry, inversion symmetry, and spin conservation. To show that $\nu_{RF,\uparrow} = \Phi/2\pi$, where $\nu_{RF,\uparrow}$ is defined as in Eq.~\ref{eq:NuRFUp}, we note that such a system can be divided into $S^z = +1/2$ and $S^z = -1/2$ sectors. Furthermore, as shown in Sec.~\ref{ssec:spin1/2invariant}, each sector can be treated as a system of spinless fermions. In this framework, the invariant $\nu_{RF,\uparrow}$ is simply the invariant $\nu_{RF}$ evaluated for the effectively spinless $S_z = +1/2$ sector. Using our previous analysis, we find the effective response theory of the $S_z = +1/2$ sector contains an $RF$-term with coefficient  $\Phi_{\uparrow} =  \pi\nu_{RF,\uparrow}$. The total $RF$-term of the spin-1/2 insulator contains contributions from the both the $S_z = \pm 1/2$ sectors, $\Phi = \Phi_{\uparrow} + \Phi_{\downarrow}$, and by TRS $\Phi_{\uparrow} = \Phi_{\downarrow}$. Hence, $\nu_{RF,\uparrow} = \Phi/2\pi$ for these spin-1/2 insulators.

\section{Linear response for the 3D Wen-Zee term}\label{app:LinRespWZ}
In this appendix we use linear response theory to determine the topological response theory that describes the $3$D layered Hamiltonian
\begin{equation}\begin{split}
\mathcal{H}(\bm{k}) &= \sin(k_x)\sigma^x\sigma^0 + \sin(k_y)\sigma^y\sigma^0\\
&-(m+\cos(k_x)+\cos(k_y))\sigma^z\sigma^z.
\label{eq:app2DHamwithTRS}\end{split}\end{equation} The TRS, $C_2$ rotation symmetry (which is part of a large $C_4$ symmetry), and mirror symmetry act as
\begin{equation}\begin{split}
&\hat{\mathcal{T}} = \sigma^y\sigma^y \mathcal{K},\\
&\hat{C}_2 = \sigma^z\sigma^z,\\
&\hat{M}_z = \sigma^0 \sigma^0.
\end{split}\end{equation}
Note that this Hamiltonian is independent of $k_z$, as it has a decoupled-layer structure.

The Hamiltonian is gapped except when $|m|=0,2$. The $2<|m|$ phase is a trivial insulator that is adiabatically connected to an atomic insulator when $|m|\rightarrow \infty.$ To find the effective response theory, we consider the system near the $m = -2$ band crossing, where the low energy physics near momentum $k_x=k_y=0$ takes on the Dirac form
\begin{equation}\begin{split}
\mathcal{H} = i\partial_x \sigma^x\sigma^0 + i \partial_y \sigma^y \sigma^0 - m' \sigma^z \sigma^z
\end{split}\end{equation}
where $m' \propto m +2$ for $m \sim -2$ controls the transition between the $m<-2$ and $-2<m<0$ phases. When introducing the spin connection $\omega$, and U$(1)$ gauge field $A$, we should replace ordinary derivatives with the covariant derivative 
\begin{equation}\begin{split}
D_\mu = \partial_\mu - i A_\mu - i \omega_\mu \frac{1}{2}[\sigma^z \sigma^0 + \sigma^0 \sigma^2],
\end{split}\end{equation}
where $\omega_\mu$ couples via the $C_2$ angular momentum. 

Let us first consider the Hamiltonian for a single layer at a fixed $z$-coordinate. The topological term in the linear response theory for the 2D system is the Wen-Zee term
\begin{equation}\begin{split}
\mathcal{L}_{\text{eff},2D} = \frac{\text{sgn}(m')+1}{4\pi} \epsilon^{\mu\nu\rho}\omega_\mu \partial_\nu A_\rho,
\label{eq:app2DResponse}\end{split}\end{equation}
where $\mu$, and $\nu$, and $\rho$ run over $x$, $y$, and $t$. The term proportional to $\frac{\text{sgn}(m')}{4\pi}$ is contributed by the low-energy fermions while the term proportional to $\frac{1}{4\pi}$ arises from massive fermions at the other $C_2$ invariant momenta. This second term ensures that the Wen-Zee term vanishes for the trivial insulator with $m'<0$ ($m < -2$ in the lattice model).  Since the Hamiltonian is independent of the z-direction, the full topological response theory is given by
\begin{equation}\begin{split}
\mathcal{L}_{\text{eff},3D} = \frac{\text{sgn}(m')+1}{8\pi^2} G_z \epsilon^{\mu\nu\rho}\omega_\mu \partial_\nu A_\rho.
\end{split}\end{equation}
where $G_z$ is the reciprocal lattice vector along the z-direction, which results from stacking the response in Eq.~\ref{eq:app2DResponse}. The coefficient of the response theory is therefore $-\frac{1}{4\pi^2}$ for $-2<m<0$ and zero for $m<-2$ in the lattice model. Similar calculations show that the coefficient is $\frac{1}{4\pi^2}$ for $0<m<2$ and zero for $2<m$. These conclusions are not modified if we couple the layers as long as the rotation symmetry and bulk gap are maintained.

\end{document}